\documentclass[11pt, a4paper]{article}
\pdfoutput=1
\usepackage{jheppub}
\usepackage{amssymb,amsmath}
\usepackage{graphicx}
\usepackage{booktabs}
\usepackage{subfigure}
\usepackage{enumerate}
\allowdisplaybreaks

\newcommand{\beq}{\begin{equation}}
\newcommand{\eeq}{\end{equation}}
\newcommand{\beqa}{\begin{eqnarray}}
\newcommand{\eeqa}{\end{eqnarray}}
\newcommand{\cm}{{\cal M}^0}

\newcommand{\cmb}{{\cal M}}
\newcommand{\ds}{{\rm d}\hat{\sigma}}
\newcommand{\dsigma}{{\rm d}\hat{\sigma}}
\newcommand{\dphi}{{\rm d}\Phi}
\newcommand{\wt}{\widetilde}
\newcommand{\re}{{\rm{Re}}}
\newcommand{\norm}{{\cal N}}
\newcommand{\ssoft}[3]{{\cal S}_{#1 #2 #3}}

\newcommand{\order}[1]{{\cal O}(#1)}

\newcommand{\qp}[1]{#1_{q'}}
\newcommand{\qpb}[1]{#1_{\bar{q}'}}
\newcommand{\Q}[1]{#1_Q}
\newcommand{\Qb}[1]{#1_{\bar{Q}}}
\newcommand{\qi}[1]{\hat{#1}_q}
\newcommand{\qbi}[1]{\hat{#1}_{\bar{q}}}

\newcommand{\gl}[1]{#1_g}

\newcommand{\ph}[1]{#1_{\gamma}}

\newcommand{\ione}[2]{{\bf I}^{(1)}_{#1 #2}}

\newcommand{\cep}{C(\epsilon)}
\newcommand{\cepb}{\bar{C}(\epsilon)}
\newcommand{\asmu}{\alpha_s(\mu)}
\newcommand{\poles}{{\cal P}oles}

\newcommand{\LO}{\mathrm{LO}}

\newcommand{\NNLO}{\mathrm{NNLO}}

\def\e{\epsilon}

\title{Light fermionic NNLO QCD corrections to top-antitop production in the quark-antiquark channel}

\author[a]{Gabriel Abelof,}
\author[b,c]{Aude Gehrmann-De Ridder}

\affiliation[a]{Department of Physics \& Astronomy, Northwestern University, Evanston, IL 60208, USA}
\affiliation[b]{Institute for Theoretical Physics, ETH, CH-8093 Z\"urich, Switzerland}
\affiliation[c]{Physics Institute
University of Z\"urich, Winterthurerstrasse 190, CH-8057, Z\"urich, Switzerland}

\emailAdd{gabriel.abelof@northwestern.edu}
\emailAdd{gehra@itp.phys.ethz.ch}


\abstract{
We present the NNLO corrections to top pair production in the quark-antiquark channel 
proportional to the number of light quark flavors $N_l$. While the double real corrections were 
derived previously, here we compute the real-virtual and virtual-virtual contributions in this partonic channel. 
Using the antenna subtraction formalism, we show that the subtraction terms correctly approximate the 
real-virtual contributions in all their infrared limits. Combined with the integrated forms of the double real 
and real-virtual subtraction terms, we show analytically that the explicit infrared poles cancel at the 
real-virtual and virtual-virtual levels respectively, thereby demonstrating the validity of the massive extension 
of the NNLO antenna formalism. These NNLO corrections are implemented in a Monte Carlo parton level 
generator providing full kinematical information on an event-by event basis. With this program, NNLO 
differential distributions in the form of binned histograms are obtained and presented here.}
   
\preprint{
\hfill\begin{minipage}[t]{9em}
\today\\    
ZU-TH 32/14\\
\end{minipage}}

\begin{document}
\bibliographystyle{JHEP-2}

\maketitle

%%%%%%%%%%%%%%%%%%%%%%%%%%%%%%%%%%%%%%%%%%%%%%%%%%%%%%
%%%%%%%%%%%%%%%%%%%%%%%%%%%%%%%%%%%%%%%%%%%%%%%%%%%%%%
%%%%%%%%%%%%%%%%%%%%%%%%%%%%%%%%%%%%%%%%%%%%%%%%%%%%%%
%                                                          SECTION: Introduction
%%%%%%%%%%%%%%%%%%%%%%%%%%%%%%%%%%%%%%%%%%%%%%%%%%%%%%

\section{Introduction}
\label{sec:intro} 
With the large number of top quark pairs produced at the LHC, the study of the properties of top quarks is becoming precision physics. Recently, the ATLAS and CMS collaborations at CERN have reported first measurements of differential observables in top-quark pair production, such as the transverse momentum and rapidity of the top quark as well as the transverse momentum, rapidity, and invariant mass of the $t\bar t$ system \cite{Aad:2012hg,CMS:fxa,CMS:cxa}. These measurements will become more and more accurate with higher statistics and will allow for a much more detailed probe of the top quark production mechanism than what can obtained from the total cross section. To reliably interpret the data, those precise measurements have to be matched by equally accurate theoretical predictions. Those can be obtained by computing these hadron collider observables at the next-to-next-to leading order (NNLO) in perturbative QCD. At present, a fully differential NNLO calculation of the cross section for top pair production including all partonic channels is still missing. Intermediate results have recently become available in \cite{Abelof:2011jv,Abelof:2011ap,Abelof:2012he,Abelof:2012rv,Abelof:2014fza,Anastasiou:2008vd,Baernreuther:2012ws,Bernreuther:2011jt,Bernreuther:2013uma,Bierenbaum:2011gg,Bonciani:2008az,Bonciani:2009nb,Bonciani:2010mn,Czakon:2008zk,Czakon:2011ve,Czakon:2012zr,Kniehl:2008fd,Korner:2008bn,Bonciani:2013ywa}, and new developments were achieved in \cite{Catani:2014qha,Gao:2014nva,Li:2013mia,vonManteuffel:2014mva,Zhu:2012ts}. Most notably, the total hadronic cross section for $t\bar{t}$ production has been presented in \cite{Czakon:2013goa}. It has been obtained using a combination of two well-known and tested procedures: the NLO so-called FKS subtraction formalism \cite{Frixione:1997np} and the sector-decomposition approach \cite{Anastasiou:2003gr,Binoth:2004jv}, leading to a numerical extraction and cancellation of the infrared $\e$-poles. 

In this paper we consider the hadro-production of a top-antitop pair and present those ${\cal O}(\alpha_s^4)$ corrections to the partonic process $q\bar{q}\to t\bar{t}$ which are proportional to the number of light quark flavours $N_l$. While at ${\cal O}(\alpha_s^3)$ this partonic channel only contains a $N_l$ dependence in the finite part of the virtual contributions, at ${\cal O}(\alpha_s^4)$ we must consider three different ingredients: virtual-virtual, real-virtual and double real, with two, three and four particles in the final state respectively. Individually, these contributions can contain both ultraviolet and infrared singularities. While the ultraviolet divergencies are removed by ultraviolet renormalisation, the infrared singularities can be made explicit using a subtraction formalism in which subtraction terms that approximate the infrared behaviour of the real radiation matrix elements are added and subtracted. At NNLO, a systematic procedure to construct these terms is provided by the antenna formalism \cite{Abelof:2011ap,Abelof:2012rv,Abelof:2014fza,Currie:2013dwa,Currie:2013vh,GehrmannDeRidder:2005cm,GehrmannDeRidder:2011aa,GehrmannDeRidder:2012dg,Glover:2010im}. 

The double real contributions for the process considered in this paper were treated in \cite{Abelof:2011ap}, where antenna subtraction terms were derived and their convergence to the real radiation matrix elements in all single and double unresolved limits was shown. In this paper we shall present the real-virtual and virtual-virtual contributions, employing the massive extension of the antenna formalism to construct their corresponding subtraction terms. We will show that the real-virtual subtraction terms correctly approximate the corresponding matrix elements in all singular limits, and that all explicit infrared singularities are cancelled analytically. Furthermore we also provide differential cross sections obtained with a fully differential parton-level event generator containing the real-virtual and virtual-virtual contributions presented in this paper together with the double real contributions of \cite{Abelof:2011ap}.

Employing a subtraction method, the NNLO partonic cross section to the $t\bar{t}$ production cross section in a given partonic channel reads, 
\beqa
\label{eq.sigNNLO}
{\rm d}\hat\sigma_{NNLO}&=&\int_{{\rm{d}}\Phi_{4}}\left({\rm{d}}\hat\sigma_{NNLO}^{RR}-{\rm{d}}\hat\sigma_{NNLO}^{S}\right)+\int_{{\rm{d}}\Phi_{4}}{\rm{d}}\hat\sigma_{NNLO}^S\nonumber\\
&+&\int_{{\rm{d}}\Phi_{3}}\left({\rm d}\hat\sigma_{NNLO}^{RV}-{\rm{d}}\hat\sigma_{NNLO}^{V S}\right)+\int_{{\rm{d}}\Phi_{3}}{\rm{d}}\hat\sigma_{NNLO}^{V S}+\int_{{\rm{d}}\Phi_{3}}{\rm{d}}\hat\sigma_{NNLO}^{MF,1}\nonumber\\
&+&\int_{{\rm{d}}\Phi_2}{\rm{d}}\hat\sigma_{NNLO}^{VV}+\int_{{\rm{d}}\Phi_2}{\rm{d}}\hat\sigma_{NNLO}^{MF,2},
\eeqa
where $\int_{{\rm{d}}\Phi_m}$ represents the phase space integration for an m-parton final state. ${\rm d} \hat\sigma^{S}_{NNLO}$ denotes the subtraction term for the four-parton final state which, by construction, behaves like the double real radiation contribution ${\rm d}\hat\sigma^{RR}_{NNLO}$ in all single and double unresolved infrared limits. Likewise, ${\rm d} \hat\sigma^{VS}_{NNLO}$ is the one-loop virtual subtraction term coinciding with the one-loop three-parton final state ${\rm d} \hat\sigma^{RV}_{NNLO}$ in all single unresolved singular limits. ${\rm d} \hat\sigma^{VV}_{NNLO}$ denotes the virtual-virtual two-parton final state contribution, where no particles can become unresolved and therefore no subtraction is needed. In addition, due to the presence of initial-state partons that can emit collinear radiation, there are also two mass factorisation counter-terms $\ds_{NNLO}^{MF,1}$ and $\ds_{NNLO}^{MF,2}$, for the three and two-particle final state respectively. Both are constructed as convolutions of Altarelli-Parisi splitting kernels with lower order partonic cross sections: $\ds_{NNLO}^{MF,1}$ involves the next-to-leading order (NLO) real radiation cross sections, while $\ds_{NNLO}^{MF,2}$ contains the leading order (LO) and NLO virtual partonic cross sections. Since there are no NLO real contributions to $t\bar{t}$ production proportional to $N_l$, in this paper we do not need to consider $(MF,1)$-type counter-terms. As mentioned above, the quark-antiquark channel contains a piece proportional to $N_l$ in the NLO virtual contributions, and therefore $\int_{{\rm{d}}\Phi_2}{\rm{d}}\hat\sigma_{NNLO}^{MF,2}$ is present and will be derived below.

The numerical implementation of the NNLO cross section requires the rearrangement of the different terms in eq.(\ref{eq.sigNNLO}) according to the multiplicity of the final states. After this rearrangement is performed we have
\beqa\label{eq.subnnlo}
\ds_{NNLO}
&=&\int_{{\rm{d}}\Phi_{4}}\left[\ds_{NNLO}^{RR}-\ds_{NNLO}^S\right]\nonumber \\
&+& \int_{{\rm{d}}\Phi_{3}}\left[\ds_{NNLO}^{RV}-\ds_{NNLO}^{T}\right] \nonumber \\
&+&\int_{{\rm{d}}\Phi_{2}}\left[\ds_{NNLO}^{VV}-\ds_{NNLO}^{U}\right],
\eeqa
where the terms in each of the square brackets is finite, well behaved in the infrared singular regions and can be implemented in a parton-level event generator computing the NNLO cross section and related differential distributions numerically. 

The double real subtraction term $\ds_{NNLO}^S$ contains distinct pieces corresponding to different limits in different colour-ordered configurations. Some of these pieces ought to be integrated analytically over the unresolved phase space of one particle and combined with the three-parton final state integral, while the remaining terms are to be integrated over the unresolved phase space of two particles and combined with the two-parton contribution. This separation amounts to splitting the integrated form of $\ds_{NNLO}^S$ as explained in \cite{Abelof:2014fza,Currie:2013dwa,Currie:2013vh,GehrmannDeRidder:2011aa,GehrmannDeRidder:2012dg}
\beq
\int_{{\rm{d}}\Phi_{4}}\ds_{NNLO}^S=\int_{{\rm{d}}\Phi_{3}} \int_1 \ds_{NNLO}^{S,1}+\int_{{\rm{d}}\Phi_{2}} \int_2 \ds_{NNLO}^{S,2}.
\eeq
After performing this separation, the counterterms for the partonic channel considered in this paper which must be added at the three and two-parton final states are
\beqa
\label{eq.Tdef}  \ds_{q \bar{q},NNLO,N_l}^{T} &=& \phantom{ -\int_1 }\ds_{q \bar{q},NNLO,N_l}^{VS}- \int_1 \ds_{q \bar{q},NNLO,N_l}^{S,1},\\
\label{eq.Udef}  \ds_{q \bar{q},NNLO,N_l}^{U}&=& -\int_1 \ds_{q \bar{q},NNLO,N_l}^{VS}-\int_2 \ds_{q \bar{q},NNLO,N_l}^{S,2}-\ds_{q \bar{q},NNLO,N_l}^{MF,2}.\\ \nonumber
\eeqa
While the construction of  $\ds_{q \bar{q},NNLO,N_l}^{S}$ has been presented in \cite{Abelof:2011ap}, the explicit derivation of $\ds_{q \bar{q},NNLO,N_l}^{T}$ and $ \ds_{q \bar{q},NNLO,N_l}^{U}$ will be given here for the first time.  

The NNLO antenna subtraction method, which we shall employ in this paper, can be viewed as belonging to the class of so-called classical subtraction methods in which the matrix elements, subtraction terms and especially the integrated subtraction terms are evaluated analytically.  As a result, at the real-virtual and virtual-virtual levels one obtains an exact and analytic pole cancellation. The finite remainders at each of those levels can be implemented in a parton-level Monte Carlo event generator containing the full kinematical information on the final states on an event-by-event basis and enabling the description of differential NNLO observables. 

The general structure of the NNLO antenna subtraction terms that are required for massless hadronic observables has been detailedly explained in \cite{Currie:2013dwa,Currie:2013vh,GehrmannDeRidder:2011aa,GehrmannDeRidder:2012dg,Glover:2010im}. For the treatment of observables involving massive final states, further ingredients are needed, namely, massive antennae and different phase space factorisations and mappings \cite{Abelof:2011ap,Abelof:2014fza}. The general structure of the subtraction terms, however, remains unchanged it will not be repeated in this paper. We shall instead restrict ourselves to presenting those ingredients of our computation which are new at the real-virtual and virtual-virtual levels focusing on challenging aspects such as the correct subtraction in the subleading colour contributions at each partonic level. 

The plan of the paper is as follows: In section \ref{sec:structure} we present the general structure of the individual NNLO contributions related to the process $q \bar{q}\to t\bar{t}$. Section \ref{sec:doublereal} deals with the double real contributions proportional to the colour factors $N_l N_c$ and $N_l/N_c$, and it specifies how the integrated form of the double real subtraction term ${\rm{d}}\hat\sigma_{NNLO}^S$ ought to be added back to the three and two-parton phase space integrals. In section \ref{sec:realvirtual}, we derive the real-virtual contributions and explicitly construct their corresponding subtraction term ${\rm{d}}\hat\sigma_{NNLO}^T$, while in section \ref{sec:virtualvirtual} the two-parton final state contributions involving the two-loop matrix elements and the counterterm ${\rm{d}}\hat\sigma_{NNLO}^U$ are derived. Section \ref{sec:realvirtual} also presents some numerical tests checking the validity of the real-virtual subtraction term ${\rm{d}}\hat\sigma_{NNLO}^T$, while in section \ref{sec:virtualvirtual} we show that the explicit poles present at the two-parton level cancel analytically. In section \ref{sec:numerics} we present differential distributions at NNLO in  kinematical variables of the top quark and the top-antitop system. Finally, section \ref{sec:conclusions} contains our conclusions and an outlook. 

Included are also different appendices: appendix \ref{sec:dss} presents the double real subtraction term ${\rm{d}}\hat\sigma_{NNLO}^S$ derived in \cite{Abelof:2011ap}. In appendix \ref{sec.iones} we define the massive and massless infrared singularity operators required to express the explicit infrared poles of several ingredients in our calculation. Appendix \ref{sec:infraredfact} summarises the infrared factorisation of tree-level and one-loop amplitudes, while appendix \ref{sec:a13} presents the infrared limits of the one-loop antennae present in the real-virtual subtraction term. Finally, appendix \ref{sec.b04} contains the expression for the integrated initial-final four-parton B-type massive antenna required in our calculation at the virtual-virtual level. 

%%%%%%%%%%%%%%%%%%%%%%%%%%%%%%%%%%%%%%%%%%%%%%%%%%%%%%
%%%%%%%%%%%%%%%%%%%%%%%%%%%%%%%%%%%%%%%%%%%%%%%%%%%%%%
%%%%%%%%%%%%%%%%%%%%%%%%%%%%%%%%%%%%%%%%%%%%%%%%%%%%%%
%   SECTION: General structure of the NNLO contributions to top pair production in the $q\bar{q}$ channel
%%%%%%%%%%%%%%%%%%%%%%%%%%%%%%%%%%%%%%%%%%%%%%%%%%%%%%

\section{General structure of the NNLO contributions to top pair production in the $q\bar{q}$ channel}
\label{sec:structure}
At leading order (${\cal O}(\alpha_s^2$)), the partonic cross section for top-antitop pair production via the 
quark-antiquark channel reads 
\beq\label{eq.qqblo}
\ds_{q\bar{q},LO}=
\norm_{\LO}^{\:q\bar{q}}\dphi_2(p_3,p_4;p_1,p_2)\:|\cm_4(\Q{3},\Qb{4},\qbi{2},\qi{1})|^2 J^{(2)}_2(p_3,p_4).
\eeq
In this equation, $\dphi_2(p_3,p_4;p_1,p_2)$ is a $2 \to 2$ phase space with $p_1$ and $p_2$ denoting the momenta of the incoming massless quark and antiquark respectively, and $p_3$ and $p_4$ those of the top and antitop. $J^{(2)}_2(p_3,p_4)$ represents the measurement function that constructs an experimental observable with a top-antitop quark pair in the final state. For the total cross section, we have $J^{(2)}_2(p_3,p_4)=1$ while for a generic differential distribution in a given variable $X$, which is defined in terms of the momenta $p_3$  and $p_4$ through the function $\hat{X}(p_3,p_4)$ we take $J^{(2)}_2(p_3,p_4)=\delta(X -\hat{X}(p_3,p_4))$. 

The tree-level matrix element $\cm_4(...)$ in eq.(\ref{eq.qqblo}) is colour and coupling-stripped, and it is related to the full tree-level amplitude for $q_1\bar{q}_2\to Q_3 \bar{Q}_4$ through the (trivial) colour decomposition
\beq\label{eq.coldecqqblo}
{\cal M}^0_{q_1\bar{q}_2\rightarrow Q_3\bar{Q}_4}=g_s^2\left( \delta_{i_3i_1}\delta_{i_2i_4}-\frac{1}{N_c}\delta_{i_3i_4}\delta_{i_2i_1}\right)\cm_4(\Q{3},\Qb{4},\qbi{2},\qi{1}).
\eeq
The hats in $\hat{1}_q$ and $\hat{2}_{\bar{q}}$ indicate that $p_1$ and $p_2$ are incoming momenta.

The normalisation factor $\norm_{\LO}^{\:q\bar{q}}$ is given by
\beq\label{eq.normlo}
\norm_{\LO}^{\:q\bar{q}}=\frac{1}{2s}\:\left( \frac{\asmu}{2\pi}\right)^2\:\frac{\cepb^2}{\cep^2}\:\frac{(N_c^2-1)}{4N_c^2},
\eeq
where $s$ is the energy squared in the hadronic center-of-mass frame. Included in this normalisation factor are the flux factor, as well as the sum and average over colour and spin. The constants $\cep$ and $\cepb$ are defined as:
\beq\label{eq.Ce}
\cep=\frac{(4\pi)^{\e}}{8\pi^2}e^{-\e \gamma_E} \hspace{1.5in}\cepb=(4\pi)^{\e} e^{-\e \gamma_E},
\eeq
providing the useful relation
\beq
g_s^2=4\pi\alpha_s=\left( \frac{\alpha_s}{2\pi}\right)\frac{\cepb}{\cep}.
\eeq

The NNLO corrections to the leading-order cross section in eq.(\ref{eq.qqblo}) receive contributions from partonic processes at tree, one-loop and two-loop levels. The double real NNLO corrections that are proportional to the number of light quark flavors are due to the partonic process $q \bar q \to Q \bar{Q}  q' \bar{q'}$ at tree level. They read
\beq\label{eq.RR}
\ds_{q\bar{q},NNLO,N_l}^{RR}=
\norm_{NNLO}^{RR,q\bar{q}}\,N_l\,\dphi_4(p_3,p_4,p_5,p_6;p_1,p_2)
|{\cal M}^0_{q_1\bar{q}_2\rightarrow Q_3\bar{Q}_4q'_5\bar{q}'_6}|^2 J_2^{(4)}(p_3,p_4,p_5,p_6),
\eeq
where $\dphi_4$ is the $2 \to 4$ phase space, and $|{\cal M}^0_{q_1\bar{q}_2\rightarrow Q_3\bar{Q}_4q'_5\bar{q}'_6}|^2$ is the square of the full coupling-stripped tree-level amplitude normalized to $(N_c^2-1)$. This factor is included in the overall normalisation $\norm_{NNLO}^{RR,q\bar{q}}$, which is given by
\beq\label{eq.Nrrnnlo}
\norm_{NNLO}^{RR,q\bar{q}}=\norm_{LO}^{q\bar{q}}\left(\frac{\asmu}{2\pi}\right)^2\frac{\cepb^2}{\cep^2}.
\eeq

Given the fact that the measurement function $J_2^{(4)}$ allows the massless final state fermions with momentum $p_5$ and $p_6$ to become unresolved, i.e. soft or collinear, eq.(\ref{eq.RR}) contains infrared divergences. These divergences are implicit, in the sense that they only become explicit as poles in the dimensional regulator $\e$ after the integration over the phase space is performed. The infrared behaviour of $\ds_{q\bar{q},NNLO,N_l}^{RR}$ can be captured with the antenna subtraction term presented in \cite{Abelof:2011ap}, which is also recalled in appendix \ref{sec:dss}. A more detailed discussion on the structure of the double real contributions and their infrared behaviour will be presented in section \ref{sec:doublereal}.

The mixed real-virtual contributions are given by the phase space integral of the one-loop and tree-level amplitudes for the $2\to 3$ process $ q \bar{q} \to Q \bar{Q} g$. They read
\beqa\label{eq.RV}
&&\hspace{-0.3in}\ds_{q \bar{q},NNLO,N_l}^{RV}=\norm_{NNLO}^{q\bar{q},RV}\,N_l\, \int\frac{{\rm d}x_1}{x_1}\frac{{\rm d}x_2}{x_2}\dphi_3(p_3,p_4,p_5;x_1p_1,x_2p_2)\delta(1-x_1)\delta(1-x_2)\nonumber\\
&&\hspace{1in}\times\bigg[2\re \left({\cal M}^1_{q_1 \bar{q}_2\rightarrow Q_3 \bar{Q}_4 g_5} {\cal M}^{0\,\dagger}_{q_1\bar{q}_2\rightarrow Q_3 \bar{Q}_4 g_5}\right)\bigg]\Bigg|_{N_l}J_2^{(3)}(p_3,p_4,p_5),\nonumber\\
\eeqa
with the normalisation factor $\norm_{NNLO}^{q\bar{q},RV}$ given by
\beq
\label{eq.NnnloRV}
\norm_{NNLO}^{q\bar{q},RV}=\norm_{LO}^{q\bar{q}}\left(\frac{\asmu}{2\pi}\right)^2\cepb=\norm_{NNLO}^{q\bar{q},RR}\:\cep.
\eeq
The subscript $N_l$ on the interference of the tree-level and one-loop amplitudes indicates that only those terms proportional to $N_l$ are to be kept.

The real-virtual contributions in eq.(\ref{eq.RV}) contain explicit ultraviolet and infrared divergences as well as implicit infrared ones. The explicit poles in $\e$ originate from the loop integration in ${\cal M}^1_{q_1\bar{q}_2\rightarrow Q_3 \bar{Q}_4 g_5}$, whereas the implicit singularities are due to the phase space integration over regions where the matrix elements diverge: the soft limit $p_5\rightarrow 0$ and the collinear limits $p_1||p_5$, $p_2||p_5$. While the ultraviolet poles are cancelled upon renormalisation, we employ a subtraction term $\ds_{q \bar{q},NNLO,N_l}^T$ to deal with the infrared ones. This subtraction term has the twofold purpose of canceling the explicit poles, whose structure is well known \cite{Catani:2000ef}, while simultaneously regularising the phase space integrand in the soft and collinear limits. We shall present the explicit construction of $\ds_{q \bar{q},NNLO,N_l}^T$ in section \ref{sec:realvirtual}.

Finally, the double virtual contributions contain two terms: the interference of a two-loop $2\rightarrow 2$ matrix element with its tree-level counterpart, and a one-loop $2\rightarrow 2$ amplitude squared. Those can be written as, 
\beqa\label{eq.setup.sigmannloVV}
&&\hspace{-0.3in}\ds_{q \bar{q},NNLO,N_l}^{VV}=\norm_{NNLO}^{q\bar{q},VV}\,N_l\,\int\frac{{\rm d}x_1}{x_1}\frac{{\rm d}x_2}{x_2}\dphi_2(p_3,p_4;x_1p_1,x_2p_2)\delta(1-x_1)\delta(1-x_2)\nonumber\\
&&\hspace{1in}\times\bigg[ 2\re \left({\cal M}^2_{q_1\bar{q}_2\rightarrow Q_3 \bar{Q}_4} {\cal M}^{0\,\dagger}_{q_1\bar{q}_2\rightarrow Q_3 \bar{Q}_4}\right)+|{\cal M}^1_{q_1\bar{q}_2\rightarrow Q_3 \bar{Q}_4}|^2\bigg]\Bigg|_{N_l}\hspace{-0.075in}J_2^{(2)}(p_3,p_4)\nonumber\\
\eeqa
where the normalisation factor  $\norm_{NNLO}^{q\bar{q},VV}$ is given by
\beq
\label{eq:NnnloVV}
\norm_{NNLO}^{q\bar{q},VV}=\norm_{LO}^{q\bar{q}}\left(\frac{\asmu}{2\pi}\right)^2\cepb^2=\norm_{NNLO}^{q\bar{q},RR}\:\cep^2.
\eeq
 
After ultraviolet renormalisation, these double virtual contributions have explicit infrared poles coming from the loop integration, but no implicit poles, since the measurement function $J_2^{(2)}$, does not allow any final state particle to be unresolved. The construction of $\ds_{q \bar{q},NNLO,N_l}^U$ to be associated with these virtual-virtual contributions and dealing with the explicit infrared divergencies present in those will be presented in section \ref{sec:virtualvirtual}.

In the expressions of the real-virtual and virtual-virtual contributions defined above as $\ds^{RV}_{q\bar{q},NNLO,N_l}$ and of $\ds^{VV}_{q\bar{q},NNLO,N_l}$, we have introduced a dependence on $x_1$ and $ x_2$, the momentum fractions carried by the initial state quark-antiquark pair, through delta functions. This trivial dependence on $x_1$ and $x_2$ is introduced in order to facilitate the combination with the counterterms $\ds_{q \bar{q},NNLO,N_l}^{T}$ and $\ds_{q \bar{q},NNLO,N_l}^{U}$ respectively, whose dependance on these two variables is not trivial. As we will see in sections \ref {sec:realvirtual} and \ref{sec:virtualvirtual}, this dependance on $x_1$ and $x_2$ is related to the fact that in the integrated subtraction terms and mass factorisation counterterms there are contributions where the partons that enter the hard scattering carry a fraction $x_i$ of the incoming momenta. In general, there are three regions: the soft ($x_1=x_2=1$), the collinear ($x_1=1,\,x_2\neq 1$ and $x_1\neq 1\,x_2=1$) and the hard ($x_1\neq 1\,x_2\neq 1$). The two delta functions in the above equations imply that the real-virtual and the virtual-virtual corrections only contribute in the soft region. Their corresponding subtraction terms denoted respectively as $\ds_{q \bar{q},NNLO,N_l}^{T}$ and $\ds_{q \bar{q},NNLO,N_l}^{U}$, on the other hand, will contribute in all three regions.

%%%%%%%%%%%%%%%%%%%%%%%%%%%%%%%%%%%%%%%%%%%%%%%%%%%%%%
%%%%%%%%%%%%%%%%%%%%%%%%%%%%%%%%%%%%%%%%%%%%%%%%%%%%%%
%%%%%%%%%%%%%%%%%%%%%%%%%%%%%%%%%%%%%%%%%%%%%%%%%%%%%%
%                SECTION: Real-real contributions to $q\bar{q} \rightarrow t \bar{t}$: the $N_l$ part
%%%%%%%%%%%%%%%%%%%%%%%%%%%%%%%%%%%%%%%%%%%%%%%%%%%%%%

\section{Real-real contributions to $q\bar{q} \rightarrow t \bar{t}$: the $N_l$ part} 
\label{sec:doublereal}
The double real contributions to $q\bar{q} \rightarrow t\bar{t}$ that are proportional to the number of light quark flavours and their corresponding subtraction terms were derived in \cite{Abelof:2011ap}. It is the purpose of this section to identify the parts of the integrated form of the subtraction term that must be included in the counter-terms at the two and three parton levels.

Using the colour decomposition of \cite{Abelof:2011ap} for the tree-level matrix element 
${\cal M}^0_{q_1\bar{q}_2\rightarrow Q_3\bar{Q}_4q'_5\bar{q}'_6}$, eq.(\ref{eq.RR}) can be re-written as
\beqa
&&\hspace{-0.1in}\ds^{RR}_{q\bar{q},NNLO,N_l}=\norm_{NNLO}^{q\bar{q},RR}\,N_l\,\dphi_4(p_3,p_4,p_5,p_6;p_1,p_2)\phantom{\bigg[}\nonumber\\
&&\times\bigg\{ N_c\bigg[|\cm_6(\Q{3},\qi{1};;\qbi{2},\qpb{6};;\qp{5},\Qb{4})|^2+|\cm_6(\Q{3},\qpb{6};;\qbi{2},\Qb{4};;\qp{5},\qi{1})|^2\bigg]\nonumber\\
&&\hspace{0.075in}+\frac{1}{N_c}\bigg[|\cm_6(\Q{3},\Qb{4};;\qbi{2},\qpb{6};;\qp{5},\qi{1})|^2+|\cm_6(\Q{3},\qpb{6};;\qbi{2},\qi{1};;\qp{5},\Qb{4})|^2\nonumber\\
&&\hspace{0.15in}+|\cm_6(\Q{3},\qi{1};;\qbi{2},\Qb{4};;\qp{5},\qpb{6})|^2-3|\cm_6(\Q{3},\Qb{4};;\qbi{2},\qi{1};;\qp{5},\qpb{6})|^2\bigg]\bigg\} J_2^{(4)}(p_3,p_4,p_5,p_6).\nonumber\\
\eeqa
The different colour-ordered amplitudes in this equation can become singular in the $\qp{5}||\qpb{6}$ single collinear limit, as well as in the $\qp{5}$,$\qpb{6}$ double soft limit and the $\qi{4}||\qp{5}||\qpb{6}$ and $\qbi{3}||\qp{5}||\qpb{6}$ triple collinear limits. The subtraction term required to capture these limits is recalled in the appendix \ref{sec:dss} for completeness. 

Following the labelling of the subtraction terms employed in the computation of hadronic jet observables with the antenna formalism \cite{GehrmannDeRidder:2011aa,GehrmannDeRidder:2012dg,Glover:2010im}, we find that our double real subtraction term can be split as
\beq
\ds^S_{q\bar{q},NNLO,N_l}=
\ds^{S,a}_{q\bar{q},NNLO,N_l}+\ds^{S,b,4}_{q\bar{q},NNLO,N_l}+\ds^{S,b,3\times3}_{q\bar{q},NNLO,N_l}.
\eeq

The $(S,a)$-type subtraction term, denoted by $\ds_{q\bar{q},NNLO,N_l}^{S,a}$, subtracts the single unresolved limits of $\ds^{RR}_{q\bar{q},NNLO,N_l}$ and it is built with products of a tree-level three-parton antenna, generally denoted as  $X_3^0$ and five-parton reduced matrix elements with remapped momenta. 

The $(S,b)$-type subtraction term, denoted by $\ds_{q\bar{q},NNLO,N_l}^{S,b}$, takes care of the double unresolved limits. Two different kinds of structures are involved in this subtraction term: $\ds_{q\bar{q},NNLO,N_l}^{S,b,4}$ which has the form  $X_4^0\times |\cm_4|^2$, with $X_4^0$ being a general four parton tree-level antenna, and $\ds_{q\bar{q},NNLO,N_l}^{S,b,3\times 3}$ which has the form $X_3^0\times X_3^0\times |\cm_4|^2$. The former subtracts the double unresolved limits while introducing spurious single unresolved singularities, whereas the latter removes these spurious limits ensuring that the four-parton antenna is only active in the double unresolved regions. 

The integrated form of $\ds_{q\bar{q},NNLO,N_l}^{S,b,4}$ is obtained by integrating the four-parton antennae $X_4^0$ inclusively over their corresponding antenna phase space. These integrated subtraction terms are added back at the two-parton level, and therefore, recalling eq.(\ref{eq.Udef}), for the present calculation we have
\beq
\label{eq:S2}
\int_2 \ds_{q\bar{q},NNLO,N_l}^{S,2}= \int_2 \ds_{q\bar{q},NNLO,N_l}^{S,b,4}.
\eeq

On the other hand, the subtraction terms $\ds_{q\bar{q},NNLO,N_l}^{S,a}$ and $\ds_{q\bar{q},NNLO,N_l}^{S,b,3\times 3}$ are added back in integrated form at the three-parton level. The integrated forms are obtained by integrating the three-parton antennae $X_3^0$ over the corresponding antenna phase space (in the case of $\ds_{q\bar{q},NNLO,N_l}^{S,b,3\times 3}$ the ``outer'' antenna is integrated). We therefore have
\beq
\label{eq:S1}
\int_1 \ds_{q\bar{q},NNLO,N_l}^{S,1}= \int_1 \ds_{q\bar{q},NNLO,N_l}^{S,a}+\int_1 \ds_{q\bar{q},NNLO,N_l}^{S,b,3\times3} .
\eeq

Before concluding this section we would like to make a few more remarks about the double real subtraction terms. 

As was pointed out in \cite{Abelof:2011ap}, the construction of $\ds^{S,b,4}_{q\bar{q},NNLO,N_l}$, which is given in appendix \ref{sec:dss}, requires three different B-type antennae: a massive final-final, a flavour-violating initial-final, and an (massless) initial-initial antenna. While the leading-colour piece, that is, the part proportional to $N_l\,N_c$, only requires the initial-final antenna, a combination of all three of them is needed in the subleading-colour part, propotional to $N_l/N_c$. The form of this combination is non-trivial and it is derived in such a way that the subtraction term correctly matches the double soft limit of the double real radiation matrix elements. The integrated forms of the three different B-type antennae, which shall be used in section \ref{sec:virtualvirtual} for the construction of the double virtual counter-term, were derived in \cite{Abelof:2012he,Bernreuther:2011jt,Boughezal:2010mc,GehrmannDeRidder:2012ja}.

For the construction of $\ds^{S,a}_{q\bar{q},NNLO,N_l}$, only one antenna function is needed: a massive final-final E-type antenna. The same antenna function is employed in $\ds^{S,b,3\times3}_{q\bar{q},NNLO,N_l}$ in order to subtract the single collinear limits of the B-type four-parton antennae. The unintegrated and integrated form of this massive E-type three-parton antenna has been derived in \cite{Abelof:2011jv,GehrmannDeRidder:2009fz}.

%%%%%%%%%%%%%%%%%%%%%%%%%%%%%%%%%%%%%%%%%%%%%%%%%%%%%%
%               SECTION: Real-virtual contributions to $q\bar{q} \rightarrow t \bar{t}$: the $N_l$ part
%%%%%%%%%%%%%%%%%%%%%%%%%%%%%%%%%%%%%%%%%%%%%%%%%%%%%%

\section{Real-virtual contributions to $q\bar{q} \rightarrow t \bar{t}$: the $N_l$ part} 
\label{sec:realvirtual}
In this section we shall present the real-virtual contributions to $q\bar{q} \rightarrow t \bar{t}$ proportional to the number of light quark flavours together with their corresponding antenna subtraction terms. In other words we will construct the phase space integrand of the following three-parton contribution 
\beq
\int_{\dphi_3}\Big[\ds_{q \bar{q},NNLO,N_l}^{RV}-\ds_{q \bar{q},NNLO,N_l}^{T}\Big]. 
\eeq

%%%%%%%%%%%%%%%%%%%%%%%%%%%%%%%%%%%%%%%%%%%%%%%%%%%%%%
%%%%%%%%%%%%%%%%%%%%%%%%%%%%%%%%%%%%%%%%%%%%%%%%%%%%%%
%%%%%%%%%%%%%%%%%%%%%%%%%%%%%%%%%%%%%%%%%%%%%%%%%%%%%%
%                                              SUBSECTION: Real-virtual contributions
%%%%%%%%%%%%%%%%%%%%%%%%%%%%%%%%%%%%%%%%%%%%%%%%%%%%%%

\subsection{Real-virtual contributions} 
The real-virtual corrections to top pair production in the $q\bar{q}$ channel are obtained from the interference of the one-loop and tree-level amplitudes for the partonic process $q\bar{q} \to t\bar{t}g$. The colour decomposition of the one-loop matrix-element reads, 
\beqa
\label{eq.colourdecqqbRV}
&&\hspace{-0.2in}{\cal M}^1_{q_1 \bar{q}_2\rightarrow Q_3 \bar{Q}_4 g_5}=
\sqrt{2}\,g_s^5\,\cep \nonumber\\
&&\times\bigg\{\bigg[ (T^{a_5})_{i_3i_1}\delta_{i_2i_4}\cmb_5^1(\Q{3},\gl{5},\qi{1};;\qbi{2},\Qb{4})
+(T^{a_5})_{i_2i_4}\delta_{i_3i_1}\cmb_5^1(\Q{3},\qi{1};;\qbi{2},\gl{5},\Qb{4})\bigg]\nonumber\\
&&\hspace{-0.1in}-\frac{1}{N_c}\bigg[(T^{a_5})_{i_3i_4}\delta_{i_2i_1}\cmb_5^1(\Q{3},\gl{5},\Qb{4};;\qbi{2},\qi{1})
+(T^{a_5})_{i_2i_1}\delta_{i_3i_4}\cmb_5^1(\Q{3},\Qb{4};;\qbi{2},\gl{5},\qi{1})\bigg]\bigg\},\nonumber\\
\eeqa
where each of the sub-amplitudes has the following decomposition into primitives:
\beq
\cmb_5^1(...)=N_c\cmb_5^{[lc]}(...)+N_l\cmb_5^{[l]}(...)+ N_h\cmb_5^{[h]}(...)-\frac{1}{N_c}\cmb_5^{[slc]}(...).
\eeq

For the tree-level amplitude, the colour decomposition reads, 
\beqa
\label{eq.colourdecqqbR}
&&\hspace{-0.2in}{\cal M}^0_{q_1 \bar{q}_2\rightarrow t_3 \bar{t}_4 g_5}=\sqrt{2}\,g_s^3\nonumber\\
&&\times\bigg\{\bigg[ (T^{a_5})_{i_3i_1}\delta_{i_2i_4}\cm_5(\Q{3},\gl{5},\qi{1};;\qbi{2},\Qb{4})+(T^{a_5})_{i_2i_4}\delta_{i_3i_1}\cm_5(\Q{3},\qi{1};;\qbi{2},\gl{5},\Qb{4})\bigg]\nonumber\\
&&\hspace{-0.1in}-\frac{1}{N_c}\bigg[(T^{a_5})_{i_3i_4}\delta_{i_2i_1}\cm_5(\Q{3},\gl{5},\Qb{4};;\qbi{2},\qi{1})+(T^{a_5})_{i_2i_1}\delta_{i_3i_4}\cm_5(\Q{3},\Qb{4};;\qbi{2},\gl{5},\qi{1})\bigg]\bigg\}.\nonumber\\
\eeqa

Interfering these one-loop and tree-level amplitudes given respectively in eq.(\ref{eq.colourdecqqbRV}) and eq.(\ref{eq.colourdecqqbR}), combining the result with the phase space and the measurement function, and retaining only the terms multiplied by $N_l$, we have
\beqa\label{eq.qqbNlRV}
&&\hspace{-0.1in}\ds_{q\bar{q},NNLO,N_l}^{RV}=\norm_{NNLO}^{q\bar{q},RV}\:N_l\int\frac{{\rm d}x_1}{x_1}\frac{{\rm d}x_2}{x_2}\dphi_3(p_3,p_4,p_5;x_1p_1,x_2p_2)\delta(1-x_1)\delta(1-x_2)\nonumber\\
&&\hspace{0.2in}\phantom{\ds_{q\bar{q},NNLO,N_l}^{RV}}\times\bigg\{ N_c\bigg[ |\cmb_5^{[l]}(\Q{3},\gl{5},\qi{1};;\qbi{2},\Qb{4})|^2 +|\cmb_5^{[l]}(\Q{3},\qi{1};;\qbi{2},\gl{5},\Qb{4})|^2 \bigg]\nonumber\\
&& \hspace{0.275in}\phantom{\ds_{q\bar{q},NNLO,N_l}^{RV}}+\frac{1}{N_c}\bigg[ |\cmb_5^{[l]}(\Q{3},\gl{5},\Qb{4};;\qbi{2},\qi{1})|^2 +|\cmb_5^{[l]}(\Q{3},\Qb{4};;\qbi{2},\gl{5},\qi{1})|^2\nonumber\\
&&\hspace{0.525in}\phantom{\ds_{q\bar{q},NNLO,N_l}^{RV}}-2|\cmb_5^{[l]}(\Q{3},\Qb{4},\qbi{2},\qi{1},\ph{5})|^2\bigg]\bigg\}J^{(3)}_2(p_3,p_4,p_5),
\eeqa
where the overall factor $\norm_{NNLO}^{q\bar{q},RV}$ was given in eq.(\ref{eq.NnnloRV}). Furthermore, in eq.(\ref{eq.qqbNlRV}) we have defined the following amplitude
\beqa
\cmb_5^{[l]}(\Q{3},\Qb{4},\qbi{2},\qi{1},\ph{5})&=&\cmb_5^{[l]}(\Q{3},\gl{5},\qi{1};;\qbi{2},\Qb{4})+\cmb_5^{[l]}(\Q{3},\qi{1};;\qbi{2},\gl{5},\Qb{4})\nonumber\\
&=&\cmb_5^{[l]}(\Q{3},\gl{5},\Qb{4};;\qbi{2},\qi{1})+\cmb_5^{[l]}(\Q{3},\Qb{4};;\qbi{2},\gl{5},\qi{1}),
\eeqa
where the gluon is photon-like.

\begin{figure}[t]
\begin{center}
\label{fig.RVFeynDiag}
\includegraphics[width=0.8\textwidth]{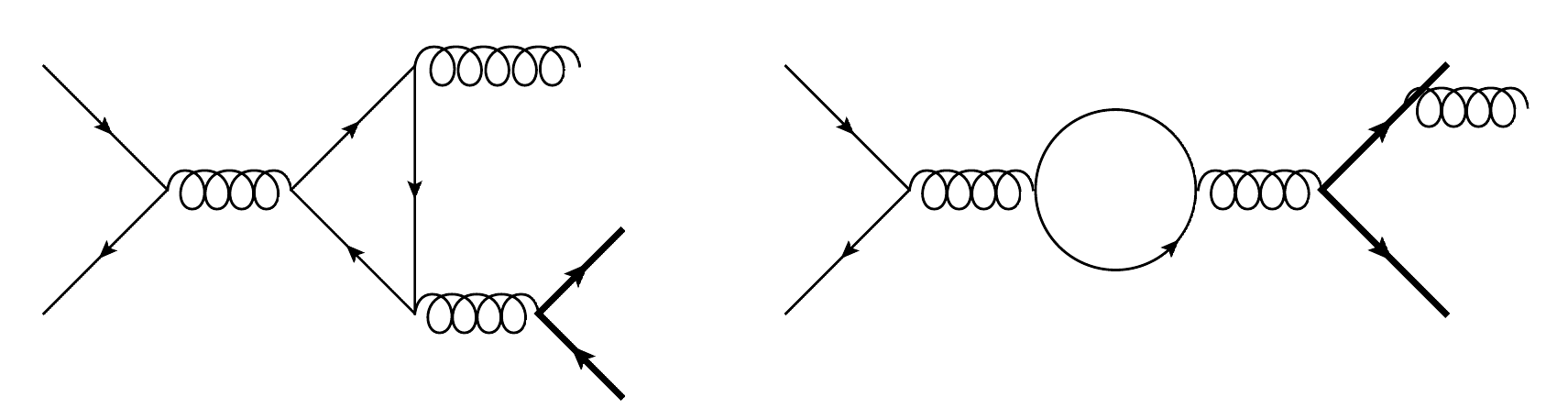}
\caption{Sample Feynman diagrams for the $N_l$ part of the one-loop amplitude ${\cal M}^1_{q_1 \bar{q}_2\rightarrow t_3 \bar{t}_4 g_5}$. The thick solid lines represent massive fermions.}
\end{center}
\end{figure}

The matrix elements in eq.(\ref{eq.qqbNlRV}) were not available in analytic form in the literature. We computed them ourselves with an in-house program based on {\tt Qgraf} \cite{Nogueira:1991ex}. Only eight Feynman diagrams of the full one-loop amplitude ${\cal M}^1_{q_1 \bar{q}_2\rightarrow Q_3 \bar{Q}_4 g_5}$ contribute to the $N_l$ piece, with topologies involving up to three massless loop propagators. Two sample Feynman diagrams are presented in figure 1. After the reduction of all tensor integrals to scalars, we find that all the one-loop amplitudes in $\ds_{q\bar{q},NNLO,N_l}^{RV}$ can be written entirely in terms of two massless bubbles with virtualities $(p_1+p_2)^2$ and $(p_3+p_4)^2$.

These matrix elements contain explicit infrared and ultraviolet poles, and because of their singular behavior in soft and collinear limits, they also yield implicit infrared singularities upon phase space integration. The ultraviolet poles are removed after renormalisation, for which we follow the scheme of \cite{Bonciani:2008az,Bonciani:2009nb,Bonciani:2010mn,Bonciani:2013ywa}: the heavy quark mass and wave function are renormalised on-shell, while for the strong coupling, the $\overline{\rm{MS}}$ scheme is employed.

In general, the explicit infrared pole structure of real-virtual contributions can be written in terms of colour-ordered infrared singularity operators denoted as $\ione{i}{j}$, with $i,j$ the partons involved in this operators which can be either massless or massive. For completeness those are recalled in appendix \ref{sec.iones}. In the context of this computation, it turns out to be more convenient to write the pole parts explicitly, since the $\ione{i}{j}$ operators involved are $\ione{Q}{g,F}(\e,s_{ij})$ and $\ione{q}{g,F}(\e,s_{ij})$, all of which yield the same explicit poles proportional to the colour-ordered renormalisation constant $b_{0,F}=-1/3$. For all colour orderings we have:
\beq
\poles\left( |\cmb_5^{[l]}(\ldots)|^2 \right)=-\frac{1}{3\e}|\cm_5(\ldots)|^2.
\eeq
With this, we can write the pole part of the real-virtual contributions as, 
\beqa\label{eq.polesNlRV}
&&\hspace{-0.3in}{\cal P}oles\left( \ds_{q\bar{q},NNLO,N_l}^{RV} \right)=\nonumber\\
&&\norm_{NNLO}^{q\bar{q},RV}\,N_l\hspace{-0.03in}\int\frac{{\rm d}x_1}{x_1}\hspace{-0.03in} \frac{{\rm d}x_2}{x_2}\,\dphi_3(p_3,p_4,p_5; x_1p_1,x_2p_2)\delta(1-x_1)\delta(1-x_2)\nonumber\\
&&\hspace{0.2in}\times \frac{b_{0,F}}{\e}\,\bigg\{ N_c\bigg[ |\cm_5(\Q{3},\gl{5},\qi{1};;\qbi{2},\Qb{4})|^2 + |\cm_5(\Q{3},\qi{1};;\qbi{2},\gl{5},\Qb{4})|^2 \bigg]\nonumber\\
&& \hspace{0.5925in}+\frac{1}{N_c}\bigg[ |\cm_5(\Q{3},\gl{5},\Qb{4};;\qbi{2},\qi{1})|^2 + |\cm_5(\Q{3},\Qb{4};;\qbi{2},\gl{5},\qi{1})|^2\nonumber\\
&& \hspace{0.9in}-2|\cm_5(\Q{3},\Qb{4},\qbi{2},\qi{1},\ph{5})|^2\bigg]\bigg\}J^{(3)}_2(p_3,p_4,p_5).
\eeqa

The implicit infrared singularities in eq.(\ref{eq.qqbNlRV}) are related to the fact the final state gluon can become either soft or collinear to one of incoming particles. These implicit divergences as well as the explicit poles are captured by the real-virtual subtraction term $\dsigma^T_{q\bar{q},NNLO,N_l}$ which we shall construct below. 

%%%%%%%%%%%%%%%%%%%%%%%%%%%%%%%%%%%%%%%%%%%%%%%%%%%%%%
%                                     SUBSECTION: The real-virtual subtraction term: $\dsigma^T$
%%%%%%%%%%%%%%%%%%%%%%%%%%%%%%%%%%%%%%%%%%%%%%%%%%%%%%

\subsection{The real-virtual subtraction term: $\dsigma^T$ }
The purpose of the real-virtual counter term $\ds^{T}_{q\bar{q},NNLO,N_l}$ is to cancel the explicit $\e$-poles of the real-virtual contributions $\ds^{RV}_{q\bar{q},NNLO,N_l }$ and to simultaneously subtract their infrared limits in such a way that the difference $\ds^{RV}_{q\bar{q},NNLO,N_l}-\ds^{T}_{q\bar{q},NNLO,N_l}$ can be safely integrated numerically in four dimensions. The general structure of this subtraction term was developed for hadronic (massless) jet observables in \cite{GehrmannDeRidder:2011aa}, and it remains unchanged for processes involving massive fermions. For the leading-colour contributions to $q\bar{q}\to t\bar{t}$ it has been applied in \cite{Abelof:2014fza}. We shall not repeat it here.

As shown in eq.(\ref{eq.Tdef}), the real-virtual subtraction term for the present calculation contains singly integrated double real subtraction terms, as well as a pure virtual subtraction term $\ds^{VS}_{q \bar{q}, NNLO,N_l}$. The pieces of the integrated double real subtraction terms that must be added back in this three-parton final state are specified in eq.(\ref{eq:S1}). Regarding the purely virtual subtraction term $\ds^{VS}_{q \bar{q}, NNLO,N_l}$, only two of the four contributions discussed in \cite{GehrmannDeRidder:2011aa} are needed in the context of this paper: a term denoted as $\ds^{VS,a}_{q \bar{q}, NNLO,N_l}$, which approximates the behaviour of the real-virtual contributions in their infrared limits, and another term that we denote as $\ds^{VS,d}_{q\bar{q}, NNLO,N_l}$, which is related to the ultraviolet renormalisation of one-loop antennae. With all these ingredients put together, the real-virtual subtraction term that will be presented in the remainder of this section takes the following form
\beq
\ds^T_{q\bar{q},NNLO,N_l}=-\int_1 \ds^{S,a}_{q\bar{q},NNLO,N_l}+\left[ \ds^{VS,a}_{q\bar{q},NNLO,N_l} +\ds^{VS,d}_{q\bar{q},NNLO,N_l} - \int_1 \ds^{S,b\:3\times 3}_{q\bar{q},NNLO,N_l}\right].
\eeq

%%%%%%%%%%%%%%%%%%%%%%%%%%%%%%%%%%%%%%%%%%%%%%%%%%%%%%
%                                   SUBSUBSECTION: Explicit infrared singularity subtraction
%%%%%%%%%%%%%%%%%%%%%%%%%%%%%%%%%%%%%%%%%%%%%%%%%%%%%%

\subsubsection{Explicit infrared singularity subtraction}
We start the construction of $\dsigma^T_{q \bar{q},NNLO,N_l}$ by showing that the explicit poles of the real-virtual contributions are cancelled as
\beq\label{eq.exppolesrv}
\poles\left( \ds_{q\bar{q},NNLO,N_l}^{RV}+ \int_1 \ds^{S,a}_{q\bar{q},NNLO,N_l} \right)=0.
\eeq
The explicit poles of the real-virtual matrix elements were given in eq.(\ref{eq.polesNlRV}). Those of the singly integrated subtraction term $\int_1\dsigma^{S,a}_{q\bar{q},NNLO,N_l}$ can be obtained from its corresponding unintegrated form given in eq.(\ref{eq.subtermqqttqq,a}) by replacing in that expression the unintegrated antennae by their integrated forms and relabeling the remapped momenta. We find
\beqa
\label{eq.dssaint}
&&\hspace{-0.1in}\int_1 \ds^{S,a}_{q\bar{q},NNLO,N_l}=\norm_{NNLO}^{q\bar{q},RV}\,N_l\int\frac{{\rm d}x_1}{x_1}\frac{{\rm d}x_2}{x_2}\,\dphi_3(p_3,p_4,p_5; x_1p_1,x_2p_2)\nonumber\\
&&\hspace{0.1in}\times\bigg\{ \frac{N_c}{2}\bigg({\cal E}^0_{Qq\bar{q}}(\e,s_{35},x_1,x_2)+{\cal E}^0_{Qq\bar{q}}(\e,s_{45},x_1,x_2) \bigg)\bigg( |\cm_5(\Q{3},\gl{5},\qi{1};;\qbi{2},\Qb{4})|^2\nonumber\\
&&\hspace{0.9in} + |\cm_5(\Q{3},\qi{1};;\qbi{2},\gl{5},\Qb{4})|^2 \bigg)\nonumber\\
&& \hspace{0.175in}+\frac{1}{2N_c}\bigg({\cal E}^0_{Qq\bar{q}}(\e,s_{35},x_1,x_2)+{\cal E}^0_{Qq\bar{q}}(\e,s_{45},x_1,x_2) \bigg)\bigg( |\cm_5(\Q{3},\gl{5},\Qb{4};;\qbi{2},\qi{1})|^2\nonumber\\
&&\hspace{0.9in} + |\cm_5(\Q{3},\Qb{4};;\qbi{2},\gl{5},\qi{1})|^2-2|\cm_5(\Q{3},\Qb{4},\qbi{2},\qi{1},\ph{5})|^2\bigg)\bigg\}J^{(3)}_2(p_3,p_4,p_5).\nonumber\\
\eeqa

The integrated final-final massive three-parton E-type antenna function denoted by ${\cal E}^0_{Qq\bar{q}}$ was derived in \cite{Abelof:2011jv}. It is given by
\beqa
\label{eq.E03Qqqint}
\lefteqn{{\cal E}_{Qq\bar{q}}^0(\e,s_{Qq\bar{q}},x_1,x_2)=-4\ione{Q}{g,F}(\e,s_{Qq\bar{q}})\delta(1-x_1)\delta(1-x_2)}\nonumber\\
&&+\delta(1-x_1)\delta(1-x_2)\left( \frac{s_{Qq\bar{q}}}{\mu^2} \right)^{-\e}\bigg[-\frac{6+3\rho-8\rho^2+3\rho^3+6\rho^4}{6(1-\rho^2)^2}-\frac{\rho^3(3+3\rho-\rho^3)}{3(1-\rho^2)^3}\ln(\rho^2)\nonumber\\
&&\hspace{0.25in}+\frac{1}{3}\ln(1-\rho^2)\bigg]+\order{\e},
\eeqa
with
\beq
\rho=\frac{m_Q}{E_{cm}}=\frac{m_Q}{\sqrt{s_{Qq\bar{q}}+m_Q^2}}.
\eeq
We omit the explicit functional dependence of the integrated antennae on $\rho$ for conciseness. 

Combining eqs.(\ref{eq.polesNlRV}), (\ref{eq.dssaint}) and (\ref{eq.E03Qqqint}) we find that eq.(\ref{eq.exppolesrv}) is satisfied.

%%%%%%%%%%%%%%%%%%%%%%%%%%%%%%%%%%%%%%%%%%%%%%%%%%%%%%
%                             SUBSUBSECTION: The implicit infrared substraction term: $\ds^{VS}$
%%%%%%%%%%%%%%%%%%%%%%%%%%%%%%%%%%%%%%%%%%%%%%%%%%%%%%

\subsubsection{The implicit infrared substraction term: $\ds^{VS}$}
As we mentioned before, the real-virtual subtraction term $\ds^{VS}_{q\bar{q},NNLO,N_l}$ receives two different contributions. It can be written as
\beq
\ds_{q\bar{q},NNLO,N_l}^{VS}=\ds_{q\bar{q},NNLO,N_l}^{VS,a}+\ds_{q\bar{q},NNLO,N_l}^{VS,d}.
\eeq
Since $\ds_{q\bar{q},NNLO,N_l}^{VS,d}$ does not have any explicit infrared poles, and no $\ds_{q\bar{q},NNLO,N_l}^{MF,1}$ mass factorisation counterterms terms are present, a way to see that no $(VS,b)$-type subtraction terms are needed, unlike in the general case \cite{GehrmannDeRidder:2011aa}, is by showing that
\beq\label{eq.conditionsNl}
{\cal P}oles\left(\ds_{q\bar{q},NNLO,N_l}^{VS,a}- \int_1 \ds^{S,b\:3\times3}_{q\bar{q},NNLO,N_l} \right)=0,
\eeq
and that the implicit infrared singularities of $\int_1 \ds^{S,a}_{q\bar{q},NNLO,N_l}$ are compensated by \linebreak
$\int_1 \ds^{S,b\:3\times3}_{q\bar{q},NNLO,N_l}$.

%%%%%%%%%%%%%%%%%%%%%%%%%%%%%%%%%%%%%%%%%%%%%%%%%%%%%%
%                             SUBSUBSECTION: Construction of $\ds^{VS,a}$
%%%%%%%%%%%%%%%%%%%%%%%%%%%%%%%%%%%%%%%%%%%%%%%%%%%%%%

\subsubsection*{Construction of $\ds^{VS,a}$}
Following the general framework described in \cite{GehrmannDeRidder:2011aa}, in order to subtract the single unresolved limits of the real-virtual contributions given in eq.(\ref{eq.qqbNlRV}) we construct our subtraction terms of the type $(VS,a)$ with tree-level and one-loop antennae multiplied by reduced matrix elements at one-loop and tree level respectively. For the process under consideration we find
\beqa\label{eq.dsqqbvsanl}
&&\hspace{-0.4in}\ds_{q\bar{q},NNLO,N_l}^{VS,a}=\norm_{NNLO}^{q\bar{q},RV}\,N_l\int\frac{{\rm d}x_1}{x_1}\frac{{\rm d}x_2}{x_2}\,\dphi_3(p_3,p_4,p_5; x_1p_1,x_2p_2)\delta(1-x_1)\delta(1-x_2)\nonumber\\
&&\hspace{0.2in}\times\bigg\{ N_c\bigg[ A_3^0(\Q{3},\gl{5},\qi{\bar{1}})|\cmb_{4,1}^{[l]}(\Q{(\wt{35})},\Qb{4},\qbi{\bar{2}},\qi{\bar{\bar{1}}})|^2 J_2^{(2)}(p_{\wt{35}},p_4)\nonumber\\
&&\hspace{0.56in}+A_3^{1,l}(\Q{3},\gl{5},\qi{\bar{1}})|\cm_4(\Q{(\wt{35})},\Qb{4},\qbi{\bar{2}},\qi{\bar{\bar{1}}})|^2 J_2^{(2)}(p_{\wt{35}},p_4)\phantom{\bigg[}\nonumber\\
&&\hspace{0.56in}+A_3^0(\Qb{4},\gl{5},\qbi{\bar{2}})|\cmb_{4,1}^{[l]}(\Q{3},\Qb{(\wt{45})},\qbi{\bar{\bar{2}}},\qi{\bar{1}})|^2 J_2^{(2)}(p_3,p_{\wt{45}})\phantom{\bigg[}\nonumber\\
&&\hspace{0.56in}+A_3^{1,l}(\Qb{4},\gl{5},\qbi{\bar{2}})|\cm_4(\Q{3},\Qb{(\wt{45})},\qbi{\bar{\bar{2}}},\qi{\bar{1}})|^2 J_2^{(2)}(p_3,p_{\wt{45}})\bigg]\phantom{\bigg[}\nonumber\\
&&\hspace{0.25in}+\frac{1}{N_c}\bigg[ 2A_3^0(\Q{3},\gl{5},\qbi{\bar{2}})|\cmb_{4,1}^{[l]}(\Q{(\wt{35})},\Qb{4},\qbi{\bar{\bar{2}}},\qi{\bar{1}})|^2 J_2^{(2)}(p_{\wt{35}},p_4)\phantom{\bigg[}\nonumber\\
&&\hspace{0.56in}+2A_3^{1,l}(\Q{3},\gl{5},\qbi{\bar{2}})|\cm_4(\Q{(\wt{35})},\Qb{4},\qbi{\bar{\bar{2}}},\qi{\bar{1}})|^2 J_2^{(2)}(p_{\wt{35}},p_4)\phantom{\bigg[}\nonumber\\
&&\hspace{0.56in}+2A_3^0(\Qb{4},\gl{5},\qi{\bar{1}})|\cmb_{4,1}^{[l]}(\Q{3},\Qb{(\wt{45})},\qbi{\bar{2}},\qi{\bar{\bar{1}}})|^2 J_2^{(2)}(p_3,p_{\wt{45}})\phantom{\bigg[}\nonumber\\
&&\hspace{0.56in}+2A_3^{1,l}(\Qb{4},\gl{5},\qi{\bar{1}})|\cm_4(\Q{3},\Qb{(\wt{45})},\qbi{\bar{2}},\qi{\bar{\bar{1}}})|^2 J_2^{(2)}(p_3,p_{\wt{45}})\phantom{\bigg[}\nonumber\\
&&\hspace{0.56in}-2A_3^0(\Q{3},\gl{5},\qi{\bar{1}})|\cmb_{4,1}^{[l]}(\Q{(\wt{35})},\Qb{4},\qbi{\bar{2}},\qi{\bar{\bar{1}}})|^2 J_2^{(2)}(p_{\wt{35}},p_4)\phantom{\bigg[}\nonumber\\
&&\hspace{0.56in}-2A_3^{1,l}(\Q{3},\gl{5},\qi{\bar{1}})|\cm_4(\Q{(\wt{35})},\Qb{4},\qbi{\bar{2}},\qi{\bar{\bar{1}}})|^2 J_2^{(2)}(p_{\wt{35}},p_4)\phantom{\bigg[}\nonumber\\
&&\hspace{0.56in}-2A_3^0(\Qb{4},\gl{5},\qbi{\bar{2}})|\cmb_{4,1}^{[l]}(\Q{3},\Qb{(\wt{45})},\qbi{\bar{\bar{2}}},\qi{\bar{1}})|^2 J_2^{(2)}(p_3,p_{\wt{45}})\phantom{\bigg[}\nonumber\\
&&\hspace{0.56in}-2A_3^{1,l}(\Qb{4},\gl{5},\qbi{\bar{2}})|\cm_4(\Q{3},\Qb{(\wt{45})},\qbi{\bar{\bar{2}}},\qi{\bar{1}})|^2 J_2^{(2)}(p_3,p_{\wt{45}})\phantom{\bigg[}\nonumber\\
&&\hspace{0.56in}-A_3^0(\Q{3},\gl{5},\Qb{4})|\cmb_{4,1}^{[l]}(\Q{(\wt{35})},\Qb{(\wt{45})},\qbi{\bar{2}},\qi{\bar{1}})|^2 J_2^{(2)}(p_{\wt{35}},p_{\wt{45}})\phantom{\bigg[}\nonumber\\
&&\hspace{0.56in}-A_3^{1,l}(\Q{3},\gl{5},\Qb{4})|\cm_4(\Q{(\wt{35})},\Qb{(\wt{45})},\qbi{\bar{2}},\qi{\bar{1}})|^2 J_2^{(2)}(p_{\wt{35}},p_{\wt{45}})\phantom{\bigg[}\nonumber\\
&&\hspace{0.56in}-A_3^0(\qbi{\bar{2}},\gl{5},\qi{\bar{1}})|\cmb_{4,1}^{[l]}(\Q{\tilde{3}},\Qb{\tilde{4}},\qbi{\bar{\bar{2}}},\qi{\bar{\bar{1}}})|^2 J_2^{(2)}(\wt{p}_3,\wt{p}_4)\phantom{\bigg[}\nonumber\\
&&\hspace{0.56in}-A_3^{1,l}(\qbi{\bar{2}},\gl{5},\qi{\bar{1}})|\cm_4(\Q{\tilde{3}},\Qb{\tilde{4}},\qbi{\bar{\bar{2}}},\qi{\bar{\bar{1}}})|^2 J_2^{(2)}(\wt{p}_3,\wt{p}_4)\bigg]\bigg\}.
\eeqa
This subtraction term contains three different three-parton tree-level A-type antennae: a (massless) initial-initial $A_3^0(\hat{q},g,\hat{\bar{q}})$, a massive flavour-violating initial-final $A_3^0(Q,g,\hat{q})$ and a massive final-final $A_3^0(Q,g,\bar{Q})$. All these antennae were derived and integrated in \cite{Abelof:2011jv,Daleo:2006xa,GehrmannDeRidder:2009fz}. The one-loop antennae in eq.(\ref{eq.dsqqbvsanl}) are $A_3^{1,l}(\hat{q},g,\hat{\bar{q}})$, $A_3^{1,l}(Q,g,\hat{q})$ and $A_3^{1,l}(Q,g,\bar{Q})$. Only the first of them has been already derived \cite{Gehrmann:2011wi}. The remaining two are massive antennae which we present for the first time together with their singular limits and infrared structure in appendix \ref{sec:a13}.

For the construction of $\ds_{q\bar{q},NNLO,N_l}^{VS,a}$, the U(1)-like sub-amplitude squared \linebreak
$|\cmb_5^{[l]}(\Q{3},\Qb{4},\qbi{2},\qi{1},\ph{5})|^2$ in eq.(\ref{eq.qqbNlRV}) requires a special treatment, since the hard radiations cannot be simply identified as the ``colour neighbours'' of the unresolved gluon. It is crucial to notice here that the unresolved behaviour of the one-loop matrix elements $\cmb_5^{[l]}$ is ``tree-level-like''. This can be seen in the fact that the $N_l$ parts of the one-loop colour-ordered soft factor and splitting functions are proportional to their tree-level counterparts as shown in appendix \ref{sec:infraredfact}. Therefore, the subtraction terms for the one-loop matrix-element $|\cmb_5^{[l]}(\Q{3},\Qb{4},\qbi{2},\qi{1},\ph{5})|^2$ can be constructed following eq.(\ref{eq.factabeliang2}) which characterises the soft behaviour of tree-level subleading-colour contributions. 

Because the renormalized one-loop reduced matrix elements $|\cmb_{4,1}^{[l]}(\Q{3},\Qb{4},\qbi{2},\qi{1})|^2$ in eq.(\ref{eq.dsqqbvsanl}) are finite, all explicit poles present in $\ds_{q\bar{q},NNLO,N_l}^{VS,a}$ come from the one-loop $A_3^{1,l}$ antennae, which can be found in appendix \ref{sec:a13}. We find
\beqa\label{eq.polesqqbvsNl}
&&\hspace{-0.1in}\poles\left( \ds_{q\bar{q},NNLO,N_l}^{VS,a} \right)=\nonumber\\
&&\hspace{0.2in}-\norm_{NNLO}^{q\bar{q},RV}\,N_l\,\frac{b_{0,F}}{\e}\int\frac{{\rm d}x_1}{x_1}\frac{{\rm d}x_2}{x_2}\,\dphi_3(p_3,p_4,p_5; x_1p_1,x_2p_2)\delta(1-x_1)\delta(1-x_2)\nonumber\\
&&\hspace{0.8in}\times\bigg\{ N_c\bigg[ A_3^0(\Q{3},\gl{5},\qi{\bar{1}})|\cm_4(\Q{(\wt{35})},\Qb{4},\qbi{\bar{2}},\qi{\bar{\bar{1}}})|^2 J^{(2)}_2(p_{\wt{35}},p_4)\nonumber\\
&&\hspace{1.175in}+A_3^0(\Qb{4},\gl{5},\qbi{2})|\cm_4(\Q{3},\Qb{(\wt{45})},\qbi{\bar{\bar{2}}},\qi{\bar{1}})|^2 J^{(2)}_2(p_3,p_{\wt{45}})\bigg]\nonumber\\
&&\hspace{0.88in}+\frac{1}{N_c}\bigg[ 2 A_3^0(\Q{3},\gl{5},\qbi{\bar{2}})|\cm_4(\Q{(\wt{35})},\Qb{4},\qbi{\bar{\bar{2}}},\qi{\bar{1}})|^2 J^{(2)}_2(p_{\wt{35}},p_4)\phantom{\bigg(}\nonumber\\
&&\hspace{1.175in}+2 A_3^0(\Qb{4},\gl{5},\qi{\bar{1}})|\cm_4(\Q{3},\Qb{(\wt{45})},\qbi{\bar{2}},\qi{\bar{\bar{1}}})|^2 J^{(2)}_2(p_3,p_{\wt{45}})\phantom{\bigg(}\nonumber\\
&&\hspace{1.175in} -2A_3^0(\Q{3},\gl{5},\qi{\bar{1}})|\cm_4(\Q{(\wt{35})},\Qb{4},\qbi{\bar{2}},\qi{\bar{\bar{1}}})|^2 J^{(2)}_2(p_{\wt{35}},p_4)\phantom{\bigg(}\nonumber\\
&&\hspace{1.175in}-2A_3^0(\Qb{4},\gl{5},\qbi{\bar{2}})|\cm_4(\Q{3},\Qb{(\wt{45})},\qbi{\bar{\bar{2}}},\qi{\bar{1}})|^2 J^{(2)}_2(p_3,p_{\wt{45}})\phantom{\bigg(}\nonumber\\
&&\hspace{1.175in}- A_3^0(\Q{3},\gl{5},\Qb{4})|\cm_4(\Q{(\wt{35})},\Qb{(\wt{45})},\qbi{\bar{2}},\qi{\bar{1}})|^2 J^{(2)}_2(p_{\wt{35}},p_{\wt{45}})\phantom{\bigg(}\nonumber\\
&&\hspace{1.175in}- A_3^0(\qi{\bar{1}},\gl{5},\qbi{\bar{2}})|\cm_4(\Q{\tilde{3}},\Qb{\tilde{4}},\qbi{\bar{2}},\qi{\bar{1}})|^2 J^{(2)}_2(\wt{p}_3,\wt{p}_4) \bigg]\bigg\}.
\eeqa 

Furthermore, the integrated form of the double real subtraction term $\ds^{S,b,3\times3}_{q\bar{q},NNLO,N_l}$ can be obtained from its unintegrated form by integrating the ``outer'' antennae in those terms in eq.(\ref{eq.subtermqqttqq,b}) which contain a product of two three-parton antennae. It reads
\beqa
&&\hspace{-0.275in}\int_1 \ds^{S,b\:3\times3}_{q\bar{q},NNLO,N_l}=\nonumber\\
&&-\norm_{NNLO}^{q\bar{q},RV}\,N_l\int\frac{{\rm d}x_1}{x_1}\frac{{\rm d}x_2}{x_2}\,\dphi_3(p_3,p_4,p_5; x_1p_1,x_2p_2)\delta(1-x_1)\delta(1-x_2)\nonumber\\
&&\hspace{0.45in}\times\frac{1}{2}\bigg({\cal E}^0_{Qq\bar{q}}(\e,s_{35},x_1,x_2)+{\cal E}^0_{Qq\bar{q}}(\e,s_{45},x_1,x_2) \bigg)\nonumber\\
&&\hspace{1in}\times\bigg\{ N_c\bigg[ A_3^0(\Q{3},\gl{5},\qi{\bar{1}})|\cm_4(\Q{(\wt{35})},\Qb{4},\qbi{\bar{2}},\qi{\bar{\bar{1}}})|^2 J^{(2)}_2(p_{\wt{35}},p_4)\nonumber\\
&&\hspace{1.36in}+A_3^0(\Qb{4},\gl{5},\qbi{\bar{2}})|\cm_4(\Q{3},\Qb{(\wt{45})},\qbi{\bar{\bar{2}}},\qi{\bar{1}})|^2 J^{(2)}_2(p_3,p_{\wt{45}})\bigg]\nonumber\\
&&\hspace{1.085in}+\frac{1}{N_c}\bigg[ 2 A_3^0(\Q{3},\gl{5},\qbi{\bar{2}})|\cm_4(\Q{(\wt{35})},\Qb{4},\qbi{\bar{\bar{2}}},\qi{\bar{1}})|^2 J^{(2)}_2(p_{\wt{35}},p_4)\nonumber\\
&&\hspace{1.37in}+2 A_3^0(\Qb{4},\gl{5},\qi{\bar{1}})|\cm_4(\Q{3},\Qb{(\wt{45})},\qbi{\bar{2}},\qi{\bar{\bar{1}}})|^2 J^{(2)}_2(p_3,p_{\wt{45}})\phantom{\bigg(}\nonumber\\
&&\hspace{1.37in} -2A_3^0(\Q{3},\gl{5},\qi{\bar{1}})|\cm_4(\Q{(\wt{35})},\Qb{4},\qbi{\bar{2}},\qi{\bar{\bar{1}}})|^2 J^{(2)}_2(p_{\wt{35}},p_4)\phantom{\bigg(}\nonumber\\
&&\hspace{1.37in}-2A_3^0(\Qb{4},\gl{5},\qbi{\bar{2}})|\cm_4(\Q{3},\Qb{(\wt{45})},\qbi{\bar{\bar{2}}},\qi{\bar{1}})|^2 J^{(2)}_2(p_3,p_{\wt{45}})\phantom{\bigg(}\nonumber\\
&&\hspace{1.37in}- A_3^0(\Q{3},\gl{5},\Qb{4})|\cm_4(\Q{(\wt{35})},\Qb{(\wt{45})},\qbi{\bar{2}},\qi{\bar{1}})|^2 J^{(2)}_2(p_{\wt{35}},p_{\wt{45}})\phantom{\bigg(}\nonumber\\
&&\hspace{1.37in}- A_3^0(\qi{\bar{1}},\gl{5},\qbi{\bar{2}})|\cm_4(\Q{\tilde{3}},\Qb{\tilde{4}},\qbi{\bar{\bar{2}}},\qi{\bar{\bar{1}}})|^2 J^{(2)}_2(\wt{p}_3,\wt{p}_4) \bigg]\bigg\}.
\eeqa

Using the pole part of ${\cal E}^0_{Q q\bar{q}} $ as given above in eq.(\ref{eq.E03Qqqint}), one can show that eq.(\ref{eq.conditionsNl}) holds and that no further real virtual subtraction term (except $\dsigma^{VS,d}_{q\bar{q},NNLO,N_l}$ derived below) is necessary in the context of the calculation presented in this paper. In section \ref{subsec:num} we shall furthermore show that all implicit infrared singularities present in $\int_1\ds^{S,a}_{q\bar{q},NNLO,N_l}$ are captured by $\int_1\ds^{S,b,3\times 3}_{q\bar{q},NNLO,N_l}$ .

%%%%%%%%%%%%%%%%%%%%%%%%%%%%%%%%%%%%%%%%%%%%%%%%%%%%%%
%                             SUBSUBSECTION: Renormalisation scale dependent subtraction
%%%%%%%%%%%%%%%%%%%%%%%%%%%%%%%%%%%%%%%%%%%%%%%%%%%%%%

\subsubsection{Renormalisation scale dependent subtraction} 
In general, one-loop antennae are evaluated  at the renormalisation scale $|s_{ijk}|$ (see Appendix C) while all one-loop matrix elements are evaluated at scale $\mu^2$. To ensure the proper renormalisation of the real-virtual subtraction terms, a contribution denoted as $\ds^{VS,d}_{q\bar{q},NNLO,N_l}$ is required. It is given by
\beqa\label{eq.dsqqbvsdnl}
&&\hspace{-0.1in}\ds_{q\bar{q},NNLO,N_l}^{VS,d}=\norm_{NNLO}^{q\bar{q},RV}\,N_l\int\frac{{\rm d}x_1}{x_1} \frac{{\rm d}x_2}{x_2}\,\dphi_3(p_3,p_4,p_5; x_1p_1,x_2p_2)\delta(1-x_1)\delta(1-x_2)\nonumber\\
&&\hspace{0.2in}\times\bigg\{ N_c\bigg[b_{0,F}\log\left( \frac{\mu^2}{|s_{135}|}\right)A_3^0(\Q{3},\gl{5},\qi{\bar{1}})|\cm_4(\Q{(\wt{35})},\Qb{4},\qbi{\bar{2}},\qi{\bar{\bar{1}}})|^2 J_2^{(2)}(p_{\wt{35}},p_4)\nonumber\\
&&\hspace{0.56in}+b_{0,F}\log\left( \frac{\mu^2}{|s_{245}|}\right)A_3^0(\Qb{4},\gl{5},\qbi{\bar{2}})|\cm_4(\Q{3},\Qb{(\wt{45})},\qbi{\bar{\bar{2}}},\qi{\bar{1}})|^2 J_2^{(2)}(p_3,p_{\wt{45}})\bigg]\nonumber\\
&&\hspace{0.25in}+\frac{1}{N_c}\bigg[2b_{0,F}\log\left( \frac{\mu^2}{|s_{235}|}\right)A_3^0(\Q{3},\gl{5},\qbi{\bar{2}})|\cm_4(\Q{(\wt{35})},\Qb{4},\qbi{\bar{\bar{2}}},\qi{\bar{1}})|^2 J_2^{(2)}(p_{\wt{35}},p_4)\nonumber\\
&&\hspace{0.56in}+2b_{0,F}\log\left( \frac{\mu^2}{|s_{145}|}\right)A_3^0(\Qb{4},\gl{5},\qi{\bar{1}})|\cm_4(\Q{3},\Qb{(\wt{45})},\qbi{\bar{2}},\qi{\bar{\bar{1}}})|^2 J_2^{(2)}(p_3,p_{\wt{45}})\nonumber\\
&&\hspace{0.56in}-2b_{0,F}\log\left( \frac{\mu^2}{|s_{135}|}\right)A_3^0(\Q{3},\gl{5},\qi{\bar{1}})|\cm_4(\Q{(\wt{35})},\Qb{4},\qbi{\bar{2}},\qi{\bar{\bar{1}}})|^2 J_2^{(2)}(p_{\wt{35}},p_4)\nonumber\\
&&\hspace{0.56in}-2b_{0,F}log\left( \frac{\mu^2}{|s_{245}|}\right)A_3^0(\Qb{4},\gl{5},\qbi{\bar{2}})|\cm_4(\Q{3},\Qb{(\wt{45})},\qbi{\bar{\bar{2}}},\qi{\bar{1}})|^2 J_2^{(2)}(p_3,p_{\wt{45}})\nonumber\\
&&\hspace{0.56in}-b_{0,F}\log\left( \frac{\mu^2}{|s_{345}|}\right)A_3^0(\Q{3},\gl{5},\Qb{4})|\cm_4(\Q{(\wt{35})},\Qb{(\wt{45})},\qbi{\bar{2}},\qi{\bar{1}})|^2 J_2^{(2)}(p_{\wt{35}},p_{\wt{45}})\nonumber\\
&&\hspace{0.56in}-b_{0,F}\log\left( \frac{\mu^2}{|s_{125}|}\right)A_3^0(\qbi{\bar{2}},\gl{5},\qi{\bar{1}})|\cm_4(\Q{\tilde{3}},\Qb{\tilde{4}},\qbi{\bar{\bar{2}}},\qi{\bar{\bar{1}}})|^2 J_2^{(2)}(\wt{p}_3,\wt{p}_4)\bigg]\bigg\}.
\eeqa

%%%%%%%%%%%%%%%%%%%%%%%%%%%%%%%%%%%%%%%%%%%%%%%%%%%%%%
%                             SUBSUBSECTION: The real-virtual subtraction term $\dsigma^T$
%%%%%%%%%%%%%%%%%%%%%%%%%%%%%%%%%%%%%%%%%%%%%%%%%%%%%%

\subsubsection{The real-virtual subtraction term $\dsigma^T$}
Putting everything together, the real-virtual counterterm term $\dsigma^T_{q\bar{q},NNLO,N_l}$ takes the following form
\beqa\label{eq.dsTnl}
&&\hspace{-0.05in}\ds_{q\bar{q},NNLO,N_l}^{T}=\norm_{NNLO}^{q\bar{q},RV}\,N_l\int\frac{{\rm d}x_1}{x_1} \frac{{\rm d}x_2}{x_2}\,\dphi_3(p_3,p_4,p_5; x_1p_1,x_2p_2)\nonumber\\
&&\times\bigg\{N_c\bigg\{ - \frac{1}{2}\bigg({\cal E}^0_{Qq\bar{q}}(\e,s_{35},x_1,x_2)+{\cal E}^0_{Qq\bar{q}}(\e,s_{45},x_1,x_2) \bigg)\bigg(|\cm_5(\Q{3},\gl{5},\qi{\bar{1}};;\qbi{\bar{2}},\Qb{4})|^2\nonumber\\
&&\hspace{1.25in} + |\cm_5(\Q{3},\qi{\bar{1}};;\qbi{\bar{2}},\gl{5},\Qb{4})|^2\bigg)J_2^{(3)}(p_3,p_4,p_5)\nonumber\\
&&\hspace{0.25in}+\bigg[ A_3^{1,l}(\Q{3},\gl{5},\qi{\bar{1}})\delta(1-x_1)\delta(1-x_2)|\cm_4(\Q{(\wt{35})},\Qb{4},\qbi{\bar{2}},\qi{\bar{\bar{1}}})|^2 \nonumber\\
&& \hspace{0.25in}+\frac{1}{2}\bigg({\cal E}^0_{Qq\bar{q}}(\e,s_{35},x_1,x_2)+{\cal E}^0_{Qq\bar{q}}(\e,s_{45},x_1,x_2)\bigg)A_3^0(\Q{3},\gl{5},\qi{\bar{1}})|\cm_4(\Q{(\wt{35})},\Qb{4},\qbi{\bar{2}},\qi{\bar{\bar{1}}})|^2\nonumber\\
&&\hspace{0.25in}+A_3^0(\Q{3},\gl{5},\qi{\bar{1}})\delta(1-x_1)\delta(1-x_2)|\cmb_4^{[l]}(\Q{(\wt{35})},\Qb{4},\qbi{\bar{2}},\qi{\bar{\bar{1}}})|^2\bigg]J_2^{(2)}(p_{\wt{35}},p_4)\nonumber\\
&&\hspace{0.25in}+\bigg[ A_3^{1,l}(\Qb{4},\gl{5},\qbi{\bar{2}})\delta(1-x_1)\delta(1-x_2)|\cm_4(\Q{3},\Qb{(\wt{45})},\qbi{\bar{\bar{2}}},\qi{\bar{1}})|^2 \nonumber\\
&& \hspace{0.35in}+\frac{1}{2}\bigg({\cal E}^0_{Qq\bar{q}}(\e,s_{35},x_1,x_2)+{\cal E}^0_{Qq\bar{q}}(\e,s_{45},x_1,x_2)\bigg)A_3^0(\Qb{4},\gl{5},\qbi{\bar{2}})|\cm_4(\Q{3},\Qb{(\wt{45})},\qbi{\bar{\bar{2}}},\qi{\bar{1}})|^2\nonumber\\
&&\hspace{0.35in}+A_3^0(\Qb{4},\gl{5},\qbi{\bar{2}})\delta(1-x_1)\delta(1-x_2)|\cmb_4^{[l]}(\Q{3},\Qb{(\wt{45})},\qbi{\bar{\bar{2}}},\qi{\bar{1}})|^2\bigg]J_2^{(2)}(p_3,p_{\wt{45}})\nonumber\\
&&\hspace{0.25in}+b_{0,F}\log\left( \frac{\mu^2}{|s_{135}|}\right)A_3^0(\Q{3},\gl{5},\qi{\bar{1}})\delta(1-x_2)\delta(1-x_2)|\cm_4(\Q{(\wt{35})},\Qb{4},\qbi{\bar{2}},\qi{\bar{\bar{1}}})|^2 J_2^{(2)}(p_{\wt{35}},p_4)\nonumber\\
&&\hspace{0.25in}+b_{0,F}\log\left( \frac{\mu^2}{|s_{245}|}\right)A_3^0(\Qb{4},\gl{5},\qbi{\bar{2}})\delta(1-x_1)\delta(1-x_2)|\cm_4(\Q{3},\Qb{(\wt{45})},\qbi{\bar{\bar{2}}},\qi{\bar{1}})|^2 J_2^{(2)}(p_3,p_{\wt{45}})\bigg\}\nonumber\\
&&\hspace{0.075in}+\frac{1}{N_c}\bigg\{ - \frac{1}{2}\bigg({\cal E}^0_{Qq\bar{q}}(\e,s_{35},x_1,x_2)+{\cal E}^0_{Qq\bar{q}}(\e,s_{45},x_1,x_2) \bigg)\bigg(|\cm_5(\Q{3},\gl{5},\Qb{4};;\qbi{\bar{2}},\qi{\bar{1}})|^2\nonumber\\
&&\hspace{1in} + |\cm_5(\Q{3},\Qb{4};;\qbi{\bar{2}},\gl{5},\qi{\bar{1}})|^2-2 |\cm_5(\Q{3},\Qb{4}\qbi{\bar{2}},\qi{\bar{1}},\ph{5})|^2\bigg)J_2^{(3)}(p_3,p_4,p_5)\nonumber\\
&&\hspace{0.25in}+2\bigg[ A_3^{1,l}(\Q{3},\gl{5},\qbi{\bar{2}})\delta(1-x_1)\delta(1-x_2)|\cm_4(\Q{(\wt{35})},\Qb{4},\qbi{\bar{\bar{2}}},\qi{\bar{1}})|^2 \nonumber\\
&& \hspace{0.4in}+\frac{1}{2}\bigg({\cal E}^0_{Qq\bar{q}}(\e,s_{35},x_1,x_2)+{\cal E}^0_{Qq\bar{q}}(\e,s_{45},x_1,x_2)\bigg)A_3^0(\Q{3},\gl{5},\qbi{\bar{2}})|\cm_4(\Q{(\wt{35})},\Qb{4},\qbi{\bar{2}},\qi{1})|^2\nonumber\\
&&\hspace{0.4in}+A_3^0(\Q{3},\gl{5},\qbi{\bar{2}})\delta(1-x_1)\delta(1-x_2)|\cmb_4^{[l]}(\Q{(\wt{35})},\Qb{4},\qbi{\bar{\bar{2}}},\qi{\bar{1}})|^2\bigg]J_2^{(2)}(p_{\wt{35}},p_4)\nonumber\\
&&\hspace{0.25in}+2\bigg[ A_3^{1,l}(\Qb{4},\gl{5},\qi{\bar{1}})|\cm_4(\Q{3},\Qb{(\wt{45})},\qbi{\bar{2}},\qi{\bar{\bar{1}}})|^2 \nonumber\\
&& \hspace{0.4in}+\frac{1}{2}\bigg({\cal E}^0_{Qq\bar{q}}(\e,s_{35},x_1,x_2)+{\cal E}^0_{Qq\bar{q}}(\e,s_{45},x_1,x_2)\bigg)A_3^0(\Qb{4},\gl{5},\qi{\bar{1}})|\cm_4(\Q{3},\Qb{(\wt{45})},\qbi{\bar{2}},\qi{\bar{\bar{1}}})|^2\nonumber\\
&&\hspace{0.4in}+A_3^0(\Qb{4},\gl{5},\qi{\bar{1}})\delta(1-x_1)\delta(1-x_2)|\cmb_4^{[l]}((\Q{3},\Qb{(\wt{45})},\qbi{\bar{2}},\qi{\bar{\bar{1}}})|^2\bigg]J_2^{(2)}(p_3,p_{\wt{45}})\nonumber\\
&&\hspace{0.25in}-2\bigg[ A_3^{1,l}(\Q{3},\gl{5},\qi{\bar{1}})\delta(1-x_1)\delta(1-x_2)|\cm_4(\Q{(\wt{35})},\Qb{4},\qbi{\bar{2}},\qi{\bar{\bar{1}}})|^2 \nonumber\\
&& \hspace{0.4in}+\frac{1}{2}\bigg({\cal E}^0_{Qq\bar{q}}(\e,s_{35},x_1,x_2)+{\cal E}^0_{Qq\bar{q}}(\e,s_{45},x_1,x_2)\bigg)A_3^0(\Q{3},\gl{5},\qi{\bar{1}})|\cm_4(\Q{(\wt{35})},\Qb{4},\qbi{\bar{2}},\qi{\bar{\bar{1}}})|^2\nonumber\\
&&\hspace{0.4in}+A_3^0(\Q{3},\gl{5},\qi{\bar{1}})\delta(1-x_1)\delta(1-x_2)|\cmb_4^{[l]}(\Q{(\wt{35})},\Qb{4},\qbi{\bar{2}},\qi{\bar{\bar{1}}})|^2\bigg]J_2^{(2)}(p_{\wt{35}},p_4)\nonumber\\
&&\hspace{0.25in}-2\bigg[ A_3^{1,l}(\Qb{4},\gl{5},\qbi{\bar{2}})\delta(1-x_1)\delta(1-x_2)|\cm_4(\Q{3},\Qb{(\wt{45})},\qbi{\bar{\bar{2}}},\qi{\bar{1}})|^2 \nonumber\\
&& \hspace{0.4in}+\frac{1}{2}\bigg({\cal E}^0_{Qq\bar{q}}(\e,s_{35},x_1,x_2)+{\cal E}^0_{Qq\bar{q}}(\e,s_{45},x_1,x_2)\bigg)A_3^0(\Qb{4},\gl{5},\qbi{\bar{2}})|\cm_4(\Q{3},\Qb{(\wt{45})},\qbi{\bar{2}},\qi{1})|^2\nonumber\\
&&\hspace{0.4in}+A_3^0(\Qb{4},\gl{5},\qbi{\bar{2}})\delta(1-x_1)\delta(1-x_2)|\cmb_4^{[l]}(\Q{3},\Qb{(\wt{45})},\qbi{\bar{\bar{2}}},\qi{\bar{1}})|^2\bigg]J_2^{(2)}(p_3,p_{\wt{45}})\nonumber\\
&&\hspace{0.25in}-\bigg[ A_3^{1,l}(\Q{3},\gl{5},\Qb{4})\delta(1-x_1)\delta(1-x_2)|\cm_4(\Q{(\wt{35})},\Qb{(\wt{45})},\qbi{\bar{2}},\qi{\bar{1}})|^2 \nonumber\\
&& \hspace{0.35in}+\frac{1}{2}\bigg({\cal E}^0_{Qq\bar{q}}(\e,s_{35},x_1,x_2)+{\cal E}^0_{Qq\bar{q}}(\e,s_{45},x_1,x_2)\bigg)A_3^0(\Q{3},\gl{5},\Qb{4})|\cm_4(\Q{(\wt{35})},\Qb{(\wt{45})},\qbi{\bar{2}},\qi{\bar{1}})|^2\nonumber\\
&&\hspace{0.35in}+A_3^0(\Q{3},\gl{5},\Qb{4})\delta(1-x_1)\delta(1-x_2)|\cmb_4^{[l]}(\Q{(\wt{35})},\Qb{(\wt{45})},\qbi{\bar{2}},\qi{\bar{1}})|^2\bigg]J_2^{(2)}(p_{\wt{35}},p_{\wt{45}})\nonumber\\
&&\hspace{0.25in}-\bigg[ A_3^{1,l}(\qi{\bar{1}},\gl{5},\qbi{\bar{2}})\delta(1-x_1)\delta(1-x_2)|\cm_4(\Q{\tilde{3}},\Qb{\tilde{4}},\qbi{\bar{\bar{2}}},\qi{\bar{\bar{1}}})|^2 \nonumber\\
&& \hspace{0.35in}+\frac{1}{2}\bigg({\cal E}^0_{Qq\bar{q}}(\e,s_{35},x_1,x_2)+{\cal E}^0_{Qq\bar{q}}(\e,s_{45},x_1,x_2)\bigg)A_3^0(\qi{\bar{1}},\gl{5},\qbi{\bar{2}})|\cm_4(\Q{\tilde{3}},\Qb{\tilde{4}},\qbi{\bar{2}},\qi{\bar{1}})|^2\nonumber\\
&&\hspace{0.35in}+A_3^0(\qi{\bar{1}},\gl{5},\qbi{\bar{2}})\delta(1-x_1)\delta(1-x_2)|\cmb_4^{[l]}(\Q{\tilde{3}},\Qb{\tilde{4}},\qbi{\bar{\bar{2}}},\qi{\bar{\bar{1}}})|^2\bigg]J_2^{(2)}(\wt{p}_3,\wt{p}_4)\nonumber\\
&&\hspace{0.25in}+2b_{0,F}\log\left( \frac{\mu^2}{|s_{235}|}\right)\delta(1-x_1)\delta(1-x_2)A_3^0(\Q{3},\gl{5},\qbi{\bar{2}})|\cm_4(\Q{(\wt{35})},\Qb{4},\qbi{\bar{\bar{2}}},\qi{\bar{1}})|^2 J_2^{(2)}(p_{\wt{35}},p_4)\nonumber\\
&&\hspace{0.25in}+2b_{0,F}\log\left( \frac{\mu^2}{|s_{145}|}\right)A_3^0(\Qb{4},\gl{5},\qi{\bar{1}})\delta(1-x_1)\delta(1-x_2)|\cm_4(\Q{3},\Qb{(\wt{45})},\qbi{\bar{2}},\qi{\bar{\bar{1}}})|^2 J_2^{(2)}(p_3,p_{\wt{45}})\bigg]\nonumber\\
&&\hspace{0.25in}-2b_{0,F}\log\left( \frac{\mu^2}{|s_{135}|}\right)A_3^0(\Q{3},\gl{5},\qi{\bar{1}})\delta(1-x_1)\delta(1-x_2)|\cm_4(\Q{(\wt{35})},\Qb{4},\qbi{\bar{2}},\qi{\bar{\bar{1}}})|^2 J_2^{(2)}(p_{\wt{35}},p_4)\nonumber\\
&&\hspace{0.25in}-2b_{0,F}\log\left( \frac{\mu^2}{|s_{245}|}\right)A_3^0(\Qb{4},\gl{5},\qbi{\bar{2}})\delta(1-x_1)\delta(1-x_2)|\cm_4(\Q{3},\Qb{(\wt{45})},\qbi{\bar{\bar{2}}},\qi{\bar{1}})|^2 J_2^{(2)}(p_3,p_{\wt{45}})\bigg]\nonumber\\
&&\hspace{0.25in}-b_{0,F}\log\left( \frac{\mu^2}{|s_{345}|}\right)A_3^0(\Q{3},\gl{5},\Qb{4})\delta(1-x_1)\delta(1-x_2)|\cm_4(\Q{(\wt{35})},\Qb{(\wt{45})},\qbi{\bar{2}},\qi{\bar{1}})|^2 J_2^{(2)}(p_{\wt{35}},p_{\wt{45}})\nonumber\\
&&\hspace{0.25in}-b_{0,F}\log\left( \frac{\mu^2}{|s_{125}|}\right)A_3^0(\qbi{\bar{2}},\gl{5},\qi{\bar{1}})\delta(1-x_1)\delta(1-x_2)|\cm_4(\Q{\tilde{3}},\Qb{\tilde{4}},\qbi{\bar{\bar{2}}},\qi{\bar{\bar{1}}})|^2 J_2^{(2)}(\wt{p}_3,\wt{p}_4)\bigg\}\bigg\}.\nonumber\\
\eeqa
The pole part of the terms proportional to the tree-level five-parton matrix elements squared exactly cancels the explicit poles in $\e$ present in the real-virtual contributions $\ds^{RV}_{q\bar{q},NNLO,N_l}$. Furthermore, the content of the square brackets $[\ldots]$ is free of poles in $\e$. 

From all the terms in $\ds_{q\bar{q},NNLO,N_l}^{T}$, only those coming from $\ds_{q\bar{q},NNLO,N_l}^{VS,a}$ (eq.(\ref{eq.dsqqbvsanl})) and $\ds_{q\bar{q},NNLO,N_l}^{VS,d}$ (eq.(\ref{eq.dsqqbvsdnl})) must be integrated and combined with the double virtual contributions. 

In the following section we shall show that the counter-term in eq.(\ref{eq.dsTnl}) approximates the real-virtual contributions $\ds_{q\bar{q},NNLO,Nl}^{RV}$ given in eq.(\ref{eq.qqbNlRV}) in all single soft and collinear limits. 

%%%%%%%%%%%%%%%%%%%%%%%%%%%%%%%%%%%%%%%%%%%%%%%%%%%%%%
%                              SUBSECTION: Numerical tests of soft and collinear cancellations
%%%%%%%%%%%%%%%%%%%%%%%%%%%%%%%%%%%%%%%%%%%%%%%%%%%%%%

\subsection{Numerical tests of soft and collinear cancellations} 
\label{subsec:num}
The real-virtual contributions and their related subtraction terms presented above have been implemented in a {\tt Fortran} code. In this section we show the results of a series of numerical tests devised in order to test how well the subtraction terms fulfill their purpose of approximating the real-virtual contributions in all single unresolved regions. Since the heavy quark mass regulates all final state collinear limits, we must consider only soft and initial state collinear limits.

The tests were performed by using {\tt RAMBO} \cite{Kleiss:1985gy} to generate phase space points in the singular regions, with the exact distance between each event and the singularity parametrized with a control variable $x$. We quantify the level of real-virtual cancellations as
\beq 
\label{eq:RVcanc}
\delta_{RV}=\left|\frac{\ds^{RV}_{q\bar{q},NNLO,N_l}}{\ds^{T}_{q\bar{q},NNLO,N_l}}-1\right|.
\eeq 
To demonstrate the consistency and stability of the subtraction terms we will show that the $\delta_{RV}$ distributions converge to zero in all relevant $x \to 0$ limits. On the right-hand-side of eq.(\ref{eq:RVcanc}) the consistent subtraction of explicit infrared singularities in the numerator and denominator is implicitly understood. Each of the employed samples consists of about $10^4$ points with $\sqrt{\hat{s}}=1$\,TeV 
\footnote{For simplicity, $\hat{s}$ will be denoted by $s$ in this section.} and $m_Q=173.5$ GeV.

Figure 2(b) shows the degree of cancellation $\delta_{RV}$ in the soft region for samples of $10^4$ phase space points for several values of the control variable $x=(s-s_{34}-2m_Q^2)/s$, which describes the softness of the phase space points.  As the singularity is approached with smaller values of $x$, the subtraction term $\ds^{T}_{q\bar{q},NNLO,N_l}$ converges  to the real-virtual contributions $\ds^{RV}_{q\bar{q},\NNLO,N_l}$ as expected.  Similarly, figure 3(b) demonstrates the consistency of the cancellation in the collinear region, parametrised by the control variable $x=s_{15}/s$.

\begin{figure}[t]
  \begin{center}
    \subfigure[]{
      \label{fig.sketchsoftrv}
      \resizebox{0.35\linewidth}{!}{\includegraphics{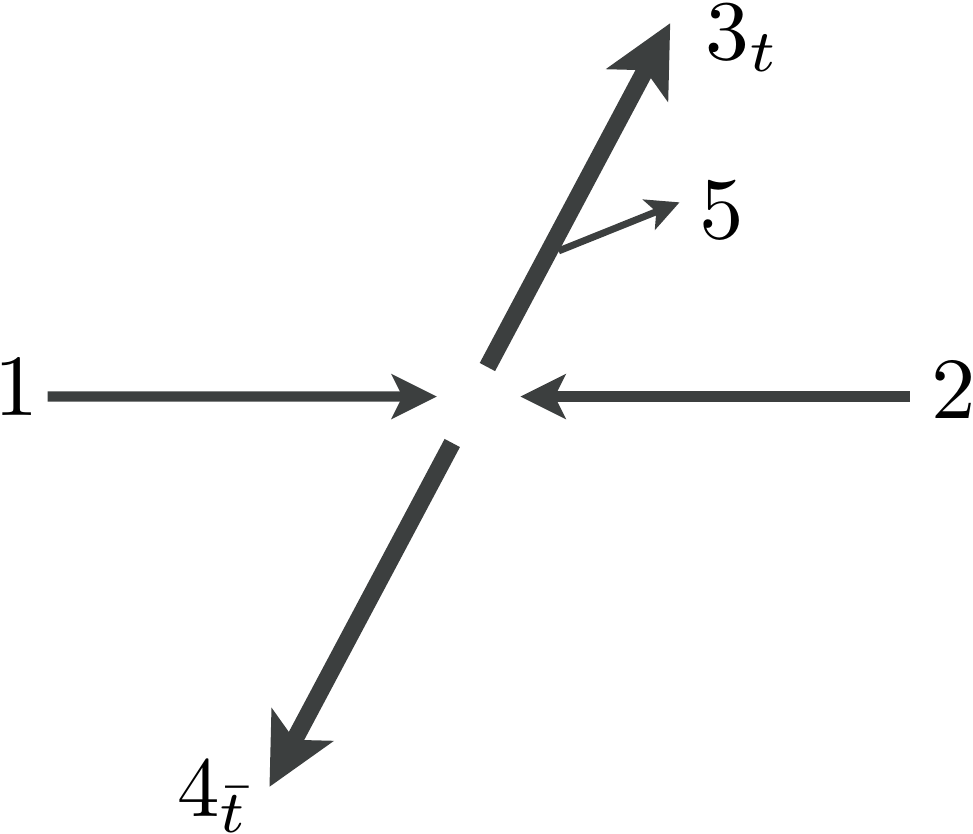}}
    }
    \subfigure[]{
      \label{fig.SoftRV}
      \resizebox{0.6\linewidth}{!}{\includegraphics{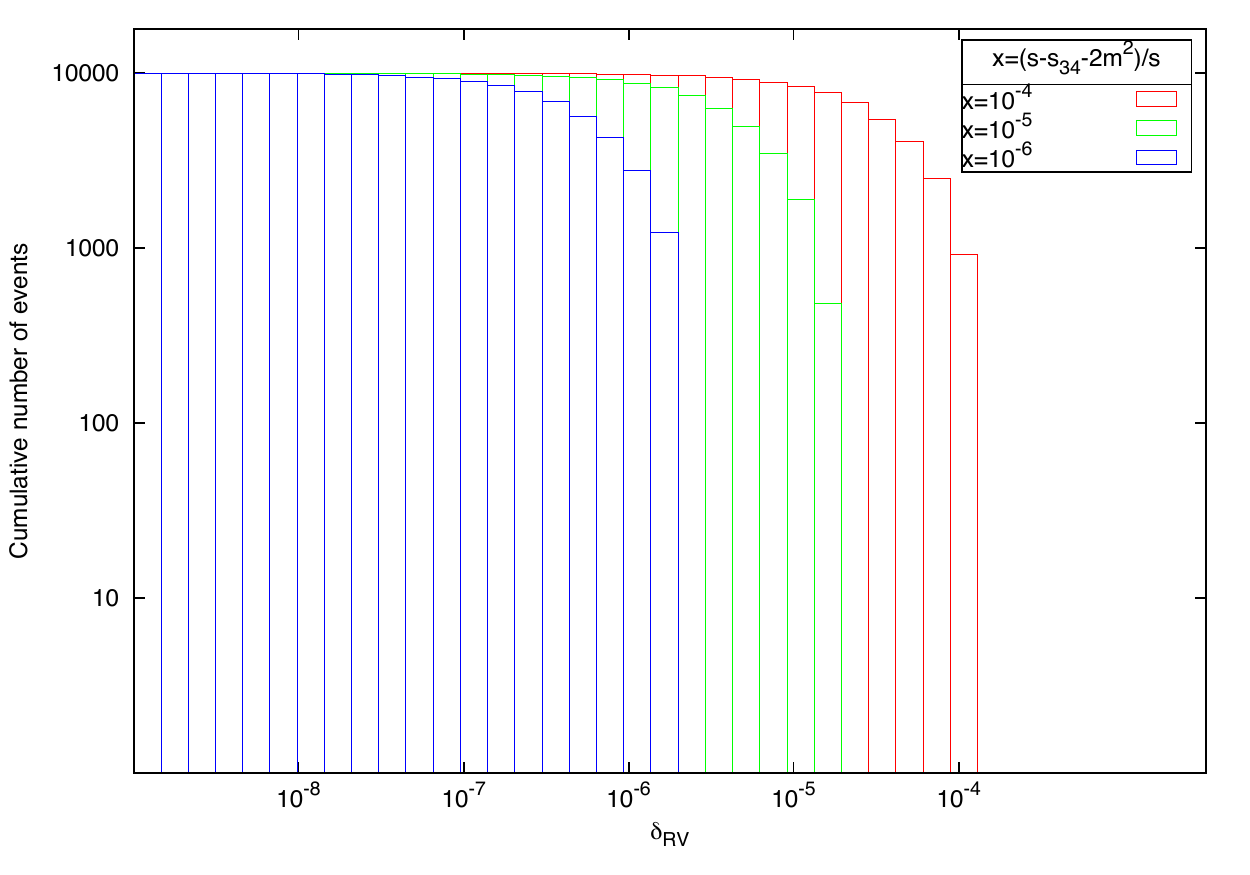}}
    }
    \caption{(a) Sketch of soft event limit. (b) Distribution of $R$ for $10^4$
             soft phase space points with three different values of $x$}
  \end{center}
\end{figure}

\begin{figure}[t]
  \begin{center}
    \subfigure[]{
      \label{fig.sketchcollrv}
      \resizebox{0.35\linewidth}{!}{\includegraphics{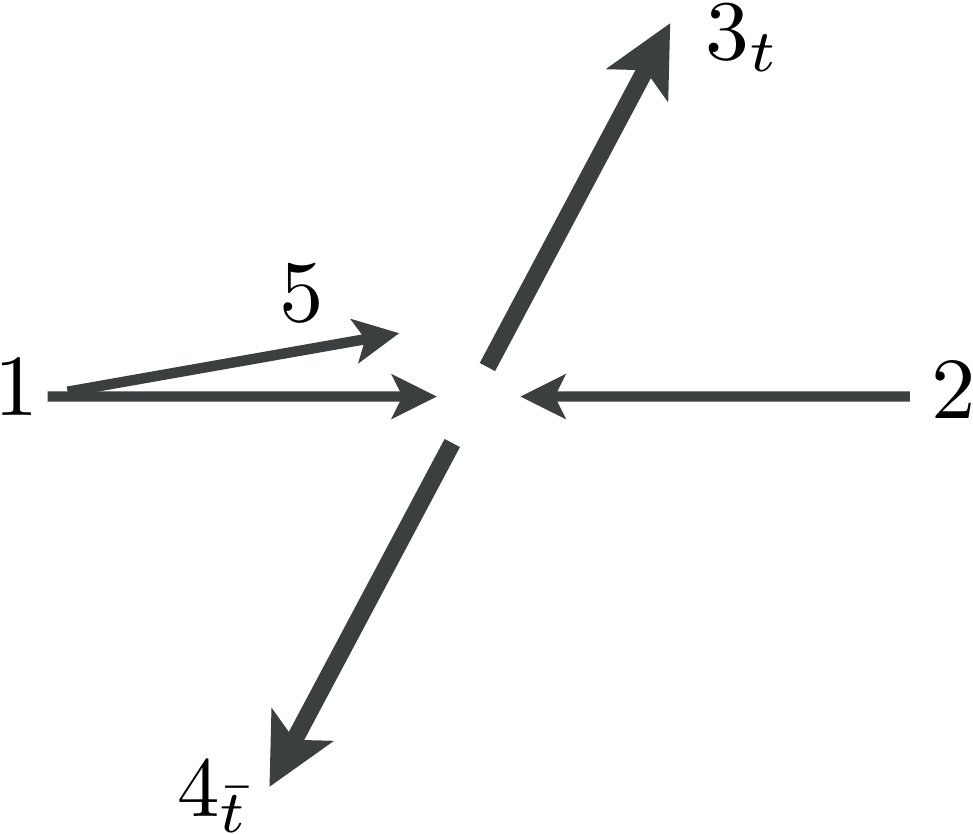}}
    }
    \subfigure[]{
      \label{fig.CollRV}
      \resizebox{0.6\linewidth}{!}{\includegraphics{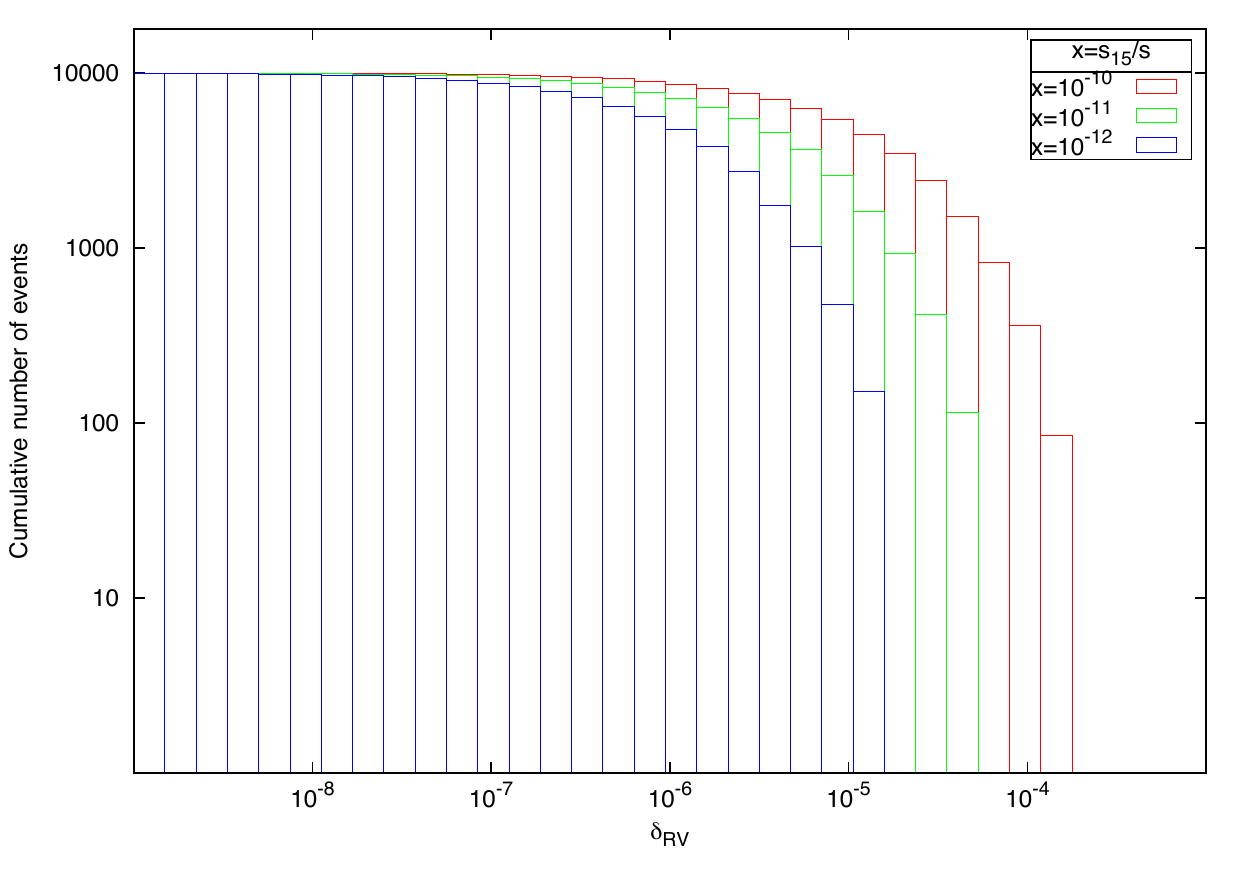}}
    }
    \caption{(a) Sketch of collinear event limit. (b) Distribution of $R$ for
             $10^4$ collinear phase space points with three different values of $x$}
  \end{center}
\end{figure}

%%%%%%%%%%%%%%%%%%%%%%%%%%%%%%%%%%%%%%%%%%%%%%%%%%%%%%
%%%%%%%%%%%%%%%%%%%%%%%%%%%%%%%%%%%%%%%%%%%%%%%%%%%%%%
%%%%%%%%%%%%%%%%%%%%%%%%%%%%%%%%%%%%%%%%%%%%%%%%%%%%%%
%               SECTION: Virtual-virtual contributions to $q\bar{q} \rightarrow t \bar{t}$: the $N_l$ part
%%%%%%%%%%%%%%%%%%%%%%%%%%%%%%%%%%%%%%%%%%%%%%%%%%%%%%

\section{Virtual-virtual contributions to $q\bar{q} \rightarrow t\bar{t}$: the $N_l$ part} 
\label{sec:virtualvirtual}
In this section we present the double virtual contributions to $q\bar{q} \to t\bar{t}$ proportional to $N_l$. We shall focus in particular in the double-virtual counter-term $\ds_{q \bar{q},NNLO,N_l}^{U}$ with which we construct the two-parton contribution 
\beqa
\int_{{\rm{d}}\Phi_{2\phantom{+1}}}\left[\ds_{q\bar{q},NNLO,N_l}^{VV}-\ds_{q \bar{q},NNLO,N_l}^{U}\right].
\eeqa

%%%%%%%%%%%%%%%%%%%%%%%%%%%%%%%%%%%%%%%%%%%%%%%%%%%%%%
%                                         SUBSECTION: The virtual-virtual contributions
%%%%%%%%%%%%%%%%%%%%%%%%%%%%%%%%%%%%%%%%%%%%%%%%%%%%%%

\subsection{The virtual-virtual contributions}
Following the general structure given in eq.(\ref{eq.setup.sigmannloVV}), the virtual-virtual contributions can be written as 
\beqa\label{eq:nnlovv}
\lefteqn{\ds^{VV}_{q \bar{q},NNLO,N_l}= \norm_{NNLO}^{VV}N_l \sum_{\textrm{perms}}\int\frac{{\rm d}x_1}{x_1}\frac{{\rm d}x_2}{x_2}\dphi_{2}(p_3,\ldots,p_{4};x_1p_1,x_2p_2)}\nonumber\\
&&\hspace{0.2in}\times\delta(1-x_1)\delta(1-x_2)|{\cal M}^2_{4}(\Q{3},\Qb{4},\qbi{2},\qi{1})|^2\;J_{2}^{(2)}(p_3,p_{4})\;,
\eeqa
where we have abbreviated
\beqa\label{eq:m2def}
|{\cal M}^2_{4}(\Q{3},\Qb{4},\qbi{2},\qi{1})|^2=
\bigg[ 2\re \left({\cal M}^2_{q_1\bar{q}_2\rightarrow t_3 \bar{t}_4} 
{\cal M}^{0\,\dagger}_{q_1\bar{q}_2\rightarrow t_3 \bar{t}_4}\right)
+|{\cal M}^1_{q_1\bar{q}_2\rightarrow t_3 \bar{t}_4}|^2\bigg]\Bigg|_{N_l},
\eeqa
and the normalisation factor $\norm_{NNLO}^{VV}$ was given in eq.(\ref{eq:NnnloVV}).

For the $N_l$ part of the two-loop matrix element in eq.(\ref{eq:m2def}) we employ the analytic results of \cite{Bonciani:2008az}. The ``one-loop squared'' term has been computed analytically in \cite{Korner:2008bn}. We re-derived it ourselves, also analytically, and use our own result in our event generator. We further compared our one-loop amplitude squared with results provided by Roberto Bonciani and found full agreement.

In general, NNLO double virtual contributions contain poles of up to order four, all of which originate from the loop integrations. However, in the colour factors that we are considering in this paper, i.e. $N_l\,N_c$ and $N_l/N_c$, the deepest poles are of order three. Since at this level no partons can become unresolved, the phase space integration of $\ds^{VV}_{q \bar{q},NNLO,N_l}$ does not yield any additional singularities. As we shall see below, all explicit poles of the double virtual contributions are captured and cancelled by those in the subtraction term $\ds^{U}_{q \bar{q},NNLO,N_l}$, which, as shown in eq.(\ref{eq.Udef}), contains mass factorisation counter-terms as well as double real and real-virtual integrated subtraction terms.

%%%%%%%%%%%%%%%%%%%%%%%%%%%%%%%%%%%%%%%%%%%%%%%%%%%%%%
%                    SUBSECTION: The mass factorisation counterterm $\ds_{NNLO}^{MF,2}$
%%%%%%%%%%%%%%%%%%%%%%%%%%%%%%%%%%%%%%%%%%%%%%%%%%%%%%

\subsection{The mass factorisation counterterm $\ds^{MF,2}$}
We start the construction of the virtual-virtual subtraction term $\ds^{U}_{q\bar{q},NNLO,N_l}$ by deriving the mass factorisation counter-term $\ds_{q\bar{q},NNLO,N_l}^{MF,2}$.

In general, for a given partonic process initiated by two partons $i$ and $j$ with momenta $p_1$ and $p_2$, the mass factorisation counterterm $\ds_{ij,NNLO}^{MF,2}$ to be included at NNLO together with the double virtual matrix elements contributions reads \cite{Currie:2013vh,GehrmannDeRidder:2011aa,GehrmannDeRidder:2012dg},
\beqa
\label{eq.mfnnlo2}
&&\hspace{-0.1in}\ds_{ij,NNLO}^{MF,2}(p_1,p_2)=-\cepb^2\bigg(\frac{\asmu}{2\pi} \bigg)^2\int\frac{{\rm d}x_1}{x_1}\frac{{\rm d}x_2}{x_2}\sum_{k,l} {\bf \Gamma}^{(2)}_{ij,kl}(x_1,x_2) \ds_{kl,LO}(x_1p_1,x_2p_2)\nonumber\\
&&\hspace{-0.05in}-\cepb\frac{\asmu}{2\pi}\hspace{-0.05in}\int\frac{{\rm d}x_1}{x_1}\frac{{\rm d}x_2}{x_2}\sum_{k,l} {\bf \Gamma}^{(1)}_{ij,kl}(x_1,x_2) \Big[ \ds_{kl,NLO}^V\hspace{-0.01in}+\hspace{-0.01in}\ds_{kl,NLO}^{MF}\hspace{-0.01in}+\hspace{-0.01in} \int_1\ds_{kl,NLO}^S \Big](x_1 p_1,x_2p_2), \nonumber\\
\eeqa
with the kernels ${\bf \Gamma}^{(1)}_{ij,kl}$ and ${\bf \Gamma}^{(2)}_{ij,kl}$ given by
\beqa
\label{eq.kernels11}
&&\hspace{-0.2in}{\bf \Gamma}^{(1)}_{ij,kl}(x_1,x_2)=\delta_{ki}\delta(1-x_1){\bf \Gamma}^{(1)}_{lj}(x_2)+\delta_{lj}\delta(1-x_2){\bf \Gamma}^{(1)}_{ki}(x_1)\\
&&\hspace{-0.2in}{\bf \Gamma}^{(2)}_{ij,kl}(x_1,x_2)=\delta_{ki}\delta(1-x_1){\bf \Gamma}^{(2)}_{lj}(x_2)+\delta_{lj}\delta(1-x_2){\bf \Gamma}^{(2)}_{ki}(x_1)+{\bf \Gamma}^{(1)}_{ki}(x_1){\bf \Gamma}^{(1)}_{lj}(x_2).
\eeqa

To simplify the construction of the double virtual subtraction term $\ds_{NNLO}^{U}$, the splitting kernel ${\bf \Gamma}^{(2)}_{ij,kl}$ can be written as suggested in \cite{Currie:2013vh} as 
\beq
\label{eq:bargam2}
{\bf \Gamma}_{ij;kl}^{(2)}(x_1,x_2)=\overline{{\bf \Gamma}}_{ij;kl}^{(2)}(x_1,x_2)-\frac{\beta_{0}}{\e}{\bf \Gamma}_{ij;kl}^{(1)}(x_1,x_2)+\frac{1}{2}\big[{\bf \Gamma}_{ij;ab}^{(1)}\otimes{\bf \Gamma}_{ab;kl}^{(1)}\big](x_1,x_2),
\eeq
such that,
\beq
\overline{{\bf \Gamma}}_{ij;kl}^{(2)}(x_1,x_2)=\overline{{\bf \Gamma}}_{ik}^{(2)}(x_1)\delta_{jl}\delta(1-x_2)+\overline{{\bf \Gamma}}_{jl}^{(2)}(x_2)\delta_{ik}\delta(1-x_1).
\eeq
The kernel $\overline{{\bf \Gamma}}_{ij}^{(2)}(z)$ is related to the usual Altarelli-Parisi splitting functions \cite{Altarelli:1977zs} as
\beqa
\overline{{\bf \Gamma}}_{ij}^{(2)}(x)&=&-\frac{1}{2\e}\bigg({\bf P}_{ij}^{1}(x)+\frac{\beta_{0}}{\e}{\bf P}_{ij}^{0}(x).\bigg).
\eeqa

For the partonic process $q\bar{q}\to t\bar{t}$, the part of the mass factorisation counter-term $\ds^{MF,2}_{q\bar{q},NNLO}$ that is proportional to $N_l$ reads
\beqa
&&\hspace{-0.2in}\ds_{q\bar{q},NNLO,N_l}^{MF,2}(p_1,p_2)= \nonumber\\
&&\hspace{0.22in}N_l \bigg[ - \cepb\frac{\asmu}{2\pi} \int \frac{{\rm d}x_1}{x_1} \frac{{\rm d}x_2}{x_2}{\bf \Gamma}_{qq;qq}^{(1)}(x_1,x_2)\ds_{q\bar{q},NLO,N_l}^{V} (x_1p_1,x_2 p_2)\nonumber \\
&&\hspace{0.5in}-\cepb ^2 \bigg(\frac{\asmu}{2\pi}\bigg)^2 \int\frac{{\rm d}x_1}{x_1}\frac{{\rm d}x_2}{x_2}{\bf \Gamma}_{qq;qq}^{(2),[N_l]}(x_1,x_2)\ds_{q\bar{q},LO} (x_1p_1,x_2 p_2)\bigg], \nonumber\\
\eeqa
where ${\bf \Gamma}_{qq;qq}^{(2),[N_l]}$ and $\ds_{q\bar{q},NLO,N_l}^{V}$ denote the pieces proportional to $N_l$ in ${\bf \Gamma}_{qq;qq}^{(2)}$ and $\ds_{q\bar{q},NLO}^{V}$ respectively. Moreover, $\ds_{q\bar{q},LO}$ is the leading order partonic cross section for the process $ q\bar{q}  \to t\bar{t}$.

Trading ${\bf \Gamma}^{(2)}_{qq;qq}$ for $\overline{\bf \Gamma}^{(2)}_{qq;qq}$ in eq.(\ref{eq:bargam2}), $\ds^{MF,2}_{q\bar{q},NNLO}$ can be compactly written as the sum of 
\beqa
&&\hspace{-0.2in}\ds^{MF,2a}_{q\bar{q},NNLO}(p_1,p_2)=- N_l\,\cepb\,\frac{\asmu}{2\pi} \int \frac{{\rm d}x_1}{x_1} \frac{{\rm d}x_2}{x_2}{\bf \Gamma}_{qq;qq}^{(1)}(x_1,x_2)\nonumber\\
&&\hspace{0.3in}\times\bigg[\ds_{q\bar{q},NLO,N_l}^{V}(x_1p_1,x_2 p_2) -\cepb\frac{\asmu}{2\pi}\frac{b_{0,F}} {\e} \ds_{qq,LO}(x_1p_1,x_2 p_2)\bigg]
\eeqa
and
\beq
\ds^{MF,2c}_{q\bar{q},NNLO}(p_1,p_2)= -N_l\,\cepb^2\bigg(\frac{\asmu}{2\pi}\bigg)^2 \int \frac{{\rm d}x_1}{x_1} \frac{{\rm d}x_2}{x_2}\overline{\bf \Gamma}_{qq;qq}^{(2),[N_l]}(x_1,x_2) \ds_{q\bar{q},LO}(x_1p_1,x_2 p_2),
\eeq
with the kernels ${\bf \Gamma}_{qq;qq}^{(1)}$ and $\overline{\bf \Gamma}_{qq;qq}^{(2),[N_l]}$ constructed as in eqs.(\ref{eq.kernels11}-\ref{eq:bargam2}) using the expressions given in the appendix of \cite{Currie:2013vh}.

Note finally that the mass factorisation counter-term denoted in \cite{Currie:2013vh} as $\ds^{MF,2b}$, which is necessary in the most general cases, is absent here. This is due to the fact that the one-loop kernel ${\bf \Gamma}_{qq;qq}^{(1)}$ does not contain any terms proportional to $N_l$.

%%%%%%%%%%%%%%%%%%%%%%%%%%%%%%%%%%%%%%%%%%%%%%%%%%%%%%
%                    SUBSECTION: The virtual-virtual subtraction term $\ds^{U}$
%%%%%%%%%%%%%%%%%%%%%%%%%%%%%%%%%%%%%%%%%%%%%%%%%%%%%%

\subsection{The virtual-virtual subtraction term $\ds^{U}$}
In addition to the mass factorisation counter-term derived above, the virtual-virtual subtraction term $\ds_{NNLO}^{U}$ contains in general contributions originated from the integrated double real and real-virtual subtraction terms. In the present case, it is given by
\beq
\ds^U_{q\bar{q},NNLO,N_l}=-\int_2\ds^{S,b\:4}_{q\bar{q},NNLO,N_l} -\ds^{MF,2}_{q\bar{q},NNLO,N_l}-\int_1 \ds^{VS,a}_{q\bar{q},NNLO,N_l}-\int_1 \ds^{VS,d}_{q\bar{q},NNLO,N_l}.
\eeq

The integrated double real subtraction term $\int_2\ds^{S,b\:4}_{q\bar{q},NNLO,N_l}$ in the equation above is obtained by integrating the four-parton B-type antennae in the subtraction term \linebreak
$\ds^{S,b\:4}_{q\bar{q},NNLO,N_l}$ given in eq.(\ref{eq.subtermqqttqq,b}). Doing so we find 
\beqa
&&\hspace{-0.1in}\int_2\ds^{S,b\:4}_{q\bar{q},NNLO,N_l}= {\cal N}_{NNLO}^{VV} N_l \int \frac{{\rm d}x_1}{x_1}\frac{{\rm d}x_2}{x_2}{\rm d}\Phi_2(p_3,p_4; x_1p_1,x_2p_2)\nonumber \\
&& \times \bigg\{  N_c \bigg [ {\cal B}^0_{q,Q q' \bar{q'}}(\e,s_{\bar{1}3},x_1,x_2)  +  {\cal B}^0_{q,Q q' \bar{q'}}(\e,s_{\bar{2}4},x_2,x_1)  \bigg] \nonumber\\
&&\hspace{0.08in}+\frac{1}{N_c}\bigg[ 2 {\cal B}^0_{q,Q q' \bar{q'}}(\e,s_{\bar{2}3},x_1,x_2) +  2 {\cal B}^0_{q,Q q' \bar{q'}}(\e,s_{\bar{1}4},x_2,x_1)\nonumber \\
&&\hspace{0.375in}   -2 {\cal B}^0_{q,Q q' \bar{q'}}(\e,s_{\bar{1}3},x_1,x_2)  -  2 {\cal B}^0_{q,Q q' \bar{q'}}(\e,s_{\bar{2}4},x_2,x_1) \phantom{\bigg[}\nonumber \\
&&\hspace{0.375in}- {\cal B}^0_{Q q\bar{q} \hat{Q}}(\e,s_{34},x_1,x_2)  -  {\cal B}^0_{q \bar{q},q' \bar{q'}}(\e,s_{\bar{1}\bar{2}},x_1,x_2) \bigg] \bigg\} |\cm_4(\Q{3},\Qb{4},\hat{\bar{2}}_{\bar{q}},\hat{\bar{1}}_{q})|^2 J_2^{(2)}(p_3,p_4) \nonumber \\
\eeqa 

Similarly, the integrated real-virtual subtraction terms $\int_1\ds^{VS,a}_{q\bar{q},NNLO,N_l}$ and \linebreak
$\int_1 \ds^{VS,d}_{q\bar{q},NNLO,N_l}$ are obtained from eqs.(\ref{eq.dsqqbvsanl}) and (\ref{eq.dsqqbvsdnl}) by integrating the tree-level and one-loop three parton antennae over their corresponding phase space. We obtain
\beqa
&&\hspace{-0.1in}\int_{1} \ds_{q\bar{q},NNLO,N_l}^{VS,a}={\cal N}_{NNLO}^{VV}\,N_{l} \int \frac{{\rm d}x_1}{x_1}  \frac{{\rm d}x_2}{x_2} {\rm d}\Phi_2(p_3,p_4; x_1p_1,x_2p_2) \nonumber \\
&& \times \bigg\{  N_c \bigg [ \bigg ({\cal A}^{1,l}_{q,Q g}(\e,s_{\bar{1}3},x_1,x_2) +{\cal A}^{1,l}_{q,Q g}(\e,s_{\bar{2}4},x_2,x_1) \bigg )|\cm_4(\Q{3},\Qb{4},\hat{\bar{2}}_{\bar{q}},\hat{\bar{1}}_{q})|^2  \nonumber \\
&&\hspace{0.375in}  + \bigg ({\cal A}^{0}_{q,Q g}(\e,s_{\bar{1}3},x_1,x_2) +   {\cal A}^{0}_{q,Q g}(\e,s_{\bar{2}4},x_2,x_1) \bigg )|\cmb_4^{[l]}(\Q{3},\Qb{4},\hat{\bar{2}}_{\bar{q}},\hat{\bar{1}}_{q})|^2  \bigg]  \nonumber\\ 
&&\hspace{0.08in}+\frac{1}{N_c}\bigg[ \bigg ( 2{\cal A}^{1,l}_{q,Q g}(\e,s_{\bar{2}3},x_2,x_1) +  2{\cal A}^{1,l}_{q,Q g}(\e,s_{\bar{1}4},x_1,x_2) \nonumber \\
&&\hspace{0.5in}-2{\cal A}^{1,l}_{q,Q g}(\e,s_{\bar{1}3},x_1,x_2) -  2{\cal A}^{1,l}_{q,Q g}(\e,s_{\bar{2}4},x_2,x_1) \phantom{\bigg(}   \nonumber \\
&&\hspace{0.5in} -2{\cal A}^{1,l}_{Q g \bar{Q}}(\e,s_{34},x_1,x_2) -2 {\cal A}^{1,l}_{q\bar{q}, g}(\e,s_{\bar{1}\bar{2}},x_1,x_2) \bigg )|\cm_4(\Q{3},\Qb{4},\hat{\bar{2}}_{\bar{q}},\hat{\bar{1}}_{q})|^2 \nonumber \\
&&\hspace{0.38in} +\bigg ( 2{\cal A}^{0}_{q,Q g}(\e,s_{\bar{2}3},x_2,x_1) +  2{\cal A}^{0}_{q,Q g}(\e,s_{\bar{1}4},x_1,x_2) \nonumber \\
&&\hspace{0.5in}-2{\cal A}^{0}_{q,Q g}(\e,s_{\bar{1}3},x_1,x_2) -  2{\cal A}^{0}_{q,Q g}(\e,s_{\bar{2}4},x_2,x_1)\phantom{\bigg(} \nonumber \\
&&\hspace{0.5in} -2{\cal A}^{0}_{Q g \bar{Q}}(\e,s_{34},x_1,x_2) -2 {\cal A}^{0}_{q\bar{q}, g}(\e,s_{\bar{1}\bar{2}},x_1,x_2) \bigg )|\cmb_4^{[l]}(\Q{3},\Qb{4},\hat{\bar{2}}_{\bar{q}},\hat{\bar{1}}_{q})|^2 \bigg] \bigg\}J_2^{(2)}(p_3,p_4), \nonumber\\
\eeqa
while for the integrated form of renormalisation scale dependent subtraction term we get
\beqa
&&\hspace{-0.2in}
\int_{1} \ds_{q \bar{q},NNLO,N_l}^{VS,d}= {\cal N}_{NNLO}^{VV}\,N_{l}\,b_{0,F} \int \frac{{\rm d}x_1}{x_1} \frac{{\rm d}x_2}{x_2} {\rm d}\Phi_2(p_3,p_4; x_1p_1,x_2p_2) |\cm_4(\Q{3},\Qb{4},\hat{\bar{2}}_{\bar{q}},\hat{\bar{1}}_{q})|^2  \nonumber \\
&& \times \bigg\{  N_c \bigg [ \bigg (\log \left(\frac{\mu^2}{s_{\bar{1}3}}\right){\cal A}^{0}_{\hat{q},Q g}(\e,s_{\bar{1}3},x_1,x_2)  + \log \left(\frac{\mu^2}{s_{\bar{2}4}}\right){\cal A}^{0}_{\hat{q},Q g}(\e,s_{\bar{2}4},x_2,x_1)\bigg) \bigg ] \nonumber \\ 
&&\hspace{0.075in}+\frac{1}{N_c}\bigg[ \bigg ( 2 \log \left(\frac{\mu^2}{s_{\bar{2}3}}\right){\cal A}^{0}_{\hat{q},Q g}(\e,s_{\bar{2}3},x_2,x_1) +2 \log \left(\frac{\mu^2}{s_{\bar{1}4}}\right){\cal A}^{0}_{\hat{q},Q g}(\e,s_{\bar{1}4},x_1,x_2) \nonumber \\
&&\hspace{0.475in} -2 \log \left(\frac{\mu^2}{s_{\bar{1}3}}\right){\cal A}^{0}_{\hat{q},Q g}(\e,s_{\bar{1}3},x_1,x_2) -2 \log \left(\frac{\mu^2}{s_{\bar{2}4}}\right){\cal A}^{0}_{\hat{q},Q g}(\e,s_{\bar{2}4},x_2,x_1) \nonumber \\
&&\hspace{0.475in} - \log \left(\frac{\mu^2}{s_{34}}\right){\cal A}^{0}_{Q\bar{Q} g}(\e,s_{34},x_1,x_2) -\log \left(\frac{\mu^2}{s_{\bar{1}\bar{2}}}\right){\cal A}^{0}_{\hat{q}\hat{\bar{q}}, g}(\e,s_{\bar{1}\bar{2}},x_2,x_1)\bigg ) \bigg] \bigg \}J_2^{(2)}(p_3,p_4). \nonumber\\
\eeqa

With the explicit expressions obtained for the integrated antennae, one can show analytically that 
\beq\label{eq.poles2}
\poles\left(\ds_{NNLO}^{VV}-\ds_{NNLO}^{U}\  \right)=0
\eeq
demonstrating that we have correctly implemented all the subtractions terms at real-real, real-virtual and virtual-virtual levels in leading ($N_l N_c$) and subleading-colour ($N_l/N_c$) contributions for the parton-level process $q \bar{q} \to t \bar{t}$.
   
This pole cancellation provides furthermore a crucial check on the integrated forms of the massless and massive four-parton B-type antennae derived in \cite{Abelof:2012he,Bernreuther:2011jt,Boughezal:2010mc} and of the correct implementation of the one-and two-loop massive amplitudes. Finally, it also provides a proof that the massive extension of the NNLO antenna formalism can be used to extract and cancel the infrared singularities of the top-antitop pair production cross section evaluated at the NNLO level, enabling the construction of a fully differential event generator. 

%%%%%%%%%%%%%%%%%%%%%%%%%%%%%%%%%%%%%%%%%%%%%%%%%%%%%%
%%%%%%%%%%%%%%%%%%%%%%%%%%%%%%%%%%%%%%%%%%%%%%%%%%%%%%
%%%%%%%%%%%%%%%%%%%%%%%%%%%%%%%%%%%%%%%%%%%%%%%%%%%%%%
%                           SECTION: Numerical results: NNLO differential distributions
%%%%%%%%%%%%%%%%%%%%%%%%%%%%%%%%%%%%%%%%%%%%%%%%%%%%%%

\section{Numerical results: NNLO differential distributions}
\label{sec:numerics}
With the double real counter-terms derived in \cite{Abelof:2011ap} and recalled in appendix \ref{sec:dss} together with the real-virtual and virtual-virtual subtraction terms presented above in sections \ref{sec:realvirtual} and \ref{sec:virtualvirtual}, we developed a Monte Carlo parton-level event generator based on the set up of eq.(\ref{eq.subnnlo}). This program, written in {\tt Fortran} following the structure of the program developed in \cite{GehrmannDeRidder:2013mf} for the calculation of the di-jet hadronic cross section at NNLO, allows us to produce the first differential distributions for top pair production including exact NNLO results, namely the $\order{\alpha_s^4}$ corrections to the $q\bar{q}\rightarrow t\bar{t}$ process proportional to the number of light quark flavors $N_l$.  

In this section we present differential distributions in the top quark transverse momentum, the top quark rapidity, as well as in the rapidity and invariants mass of the top-antitop system. We consider the $q\bar{q}$ initiated process only, convoluted with parton distribution functions (PDFs) corresponding to proton-proton collisions, at leading order, next-to-leading order, and next-to-next-to-leading order. The next-to-next-to leading order results denoted as NNLO ($N_l$) in the figures 4-7 presented in this section are the sum of the full NLO corrections and the purely fermionic NNLO corrections to $q\bar{q}\rightarrow t\bar{t}$ discussed in this paper. 

We employ a hadronic center-of-mass energy $\sqrt{s}=7$ TeV, a value of the top quark mass given by $m_t=173.5$ GeV, and the PDF sets MSTW2008LO90CL, MSTW2008NLO90CL and MSTW2008NNLO90CL for the computations at LO, NLO and NNLO respectively \cite{Martin:2009iq}. We set the factorisation and renormalisation scales to be equal to the top quark mass $\mu_R=\mu_F=m_t$.

\begin{figure}[t]
  \begin{center}
     \resizebox{0.9\linewidth}{!}{\includegraphics{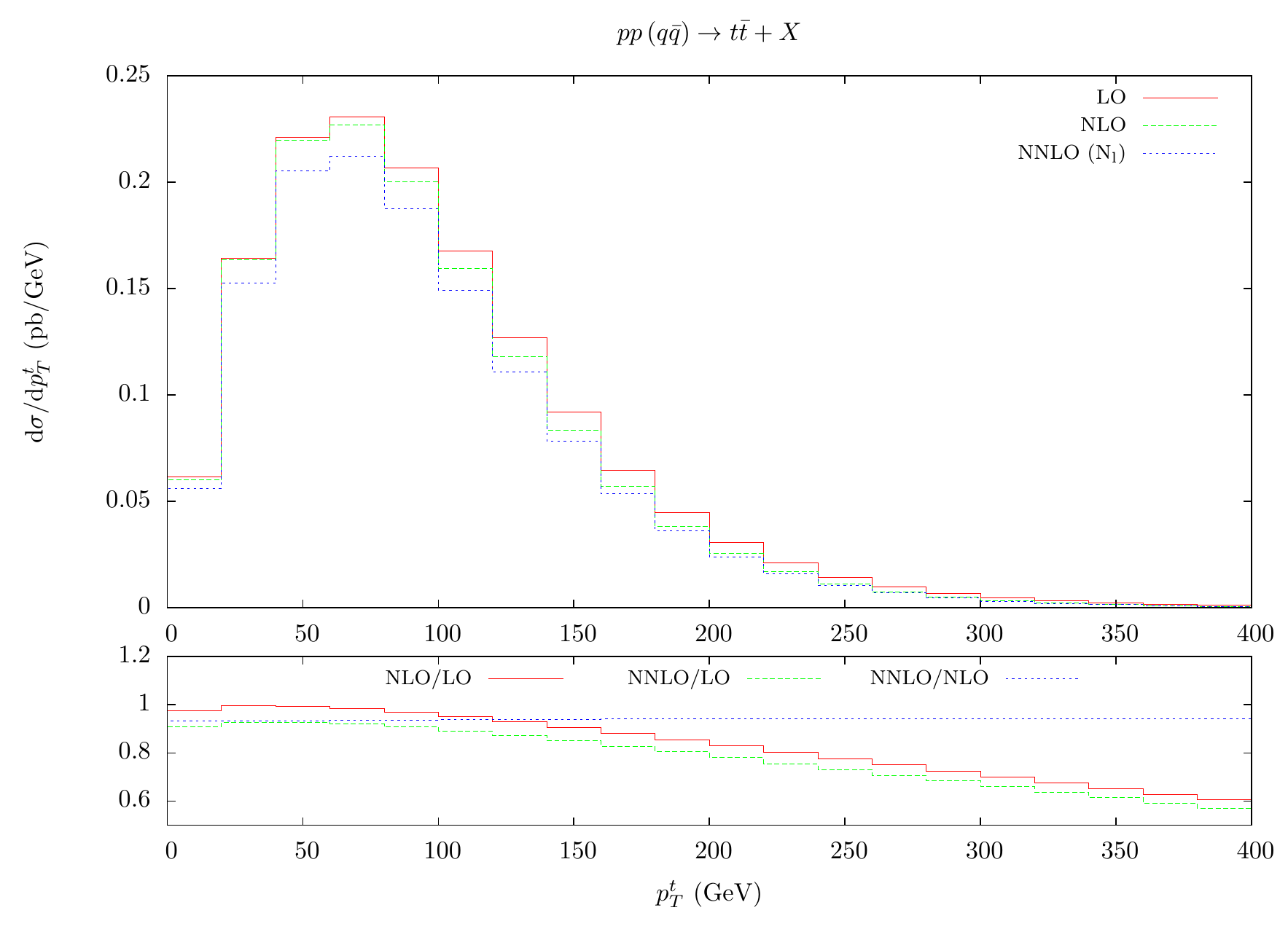}}
     \caption{Transverse momentum distribution of a single top quark ${\rm d}\sigma / {\rm d} p_T^t$ for $\sqrt{s}=7$ TeV at LO (red), NLO (green), and NNLO (blue). The lower panel shows the ratios of LO, NLO and NNLO cross sections.}
     \label{fig.ptdist}
  \end{center}
\end{figure}

In figure 4 we present the inclusive $t\bar{t}$ cross section as a function of the top quark transverse momentum $p_T^t$ at LO, NLO, and NNLO together with the corresponding $k$-factors. From the NNLO/NLO ratio we see that the NNLO contributions proportional to $N_l$ decrease the cross section by $6-7\%$ depending on the value of the transverse momentum $p_T^t$ compared to the NLO result.

\begin{figure}[t]
  \begin{center}
     \resizebox{0.9\linewidth}{!}{\includegraphics{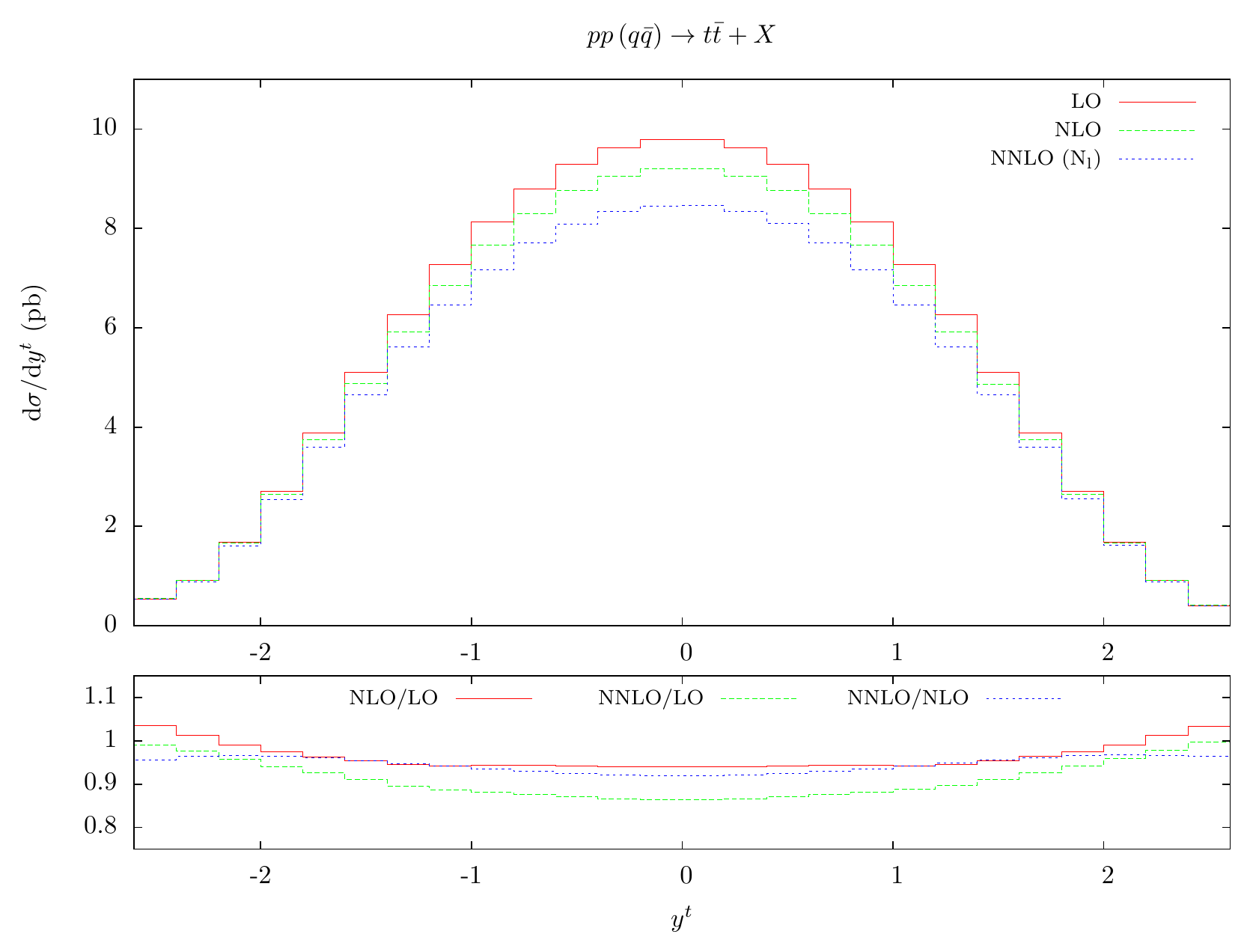}}
     \caption{Rapidity distribution of a single top quark ${\rm d}\sigma / {\rm d} y^t$ for $\sqrt{s}=7$ TeV at LO (red), NLO (green), and NNLO (blue). The lower panel shows the ratios of LO, NLO and NNLO cross sections.}
  \end{center}
  \label{fig.etatdist}
\end{figure}

In figure 5 we show the inclusive $t\bar{t}$ cross section as a function of the top quark rapidity $y^t$ at LO, NLO, and NNLO with the corresponding $k$-factors in the lower panel. The NNLO/NLO ratio shows that the NNLO contributions proportional to $N_l$ decrease the cross section by $8\%$ in the central region and $5\%$ in the forward and backward regions.

\begin{figure}[t]
  \begin{center}
     \resizebox{0.9\linewidth}{!}{\includegraphics{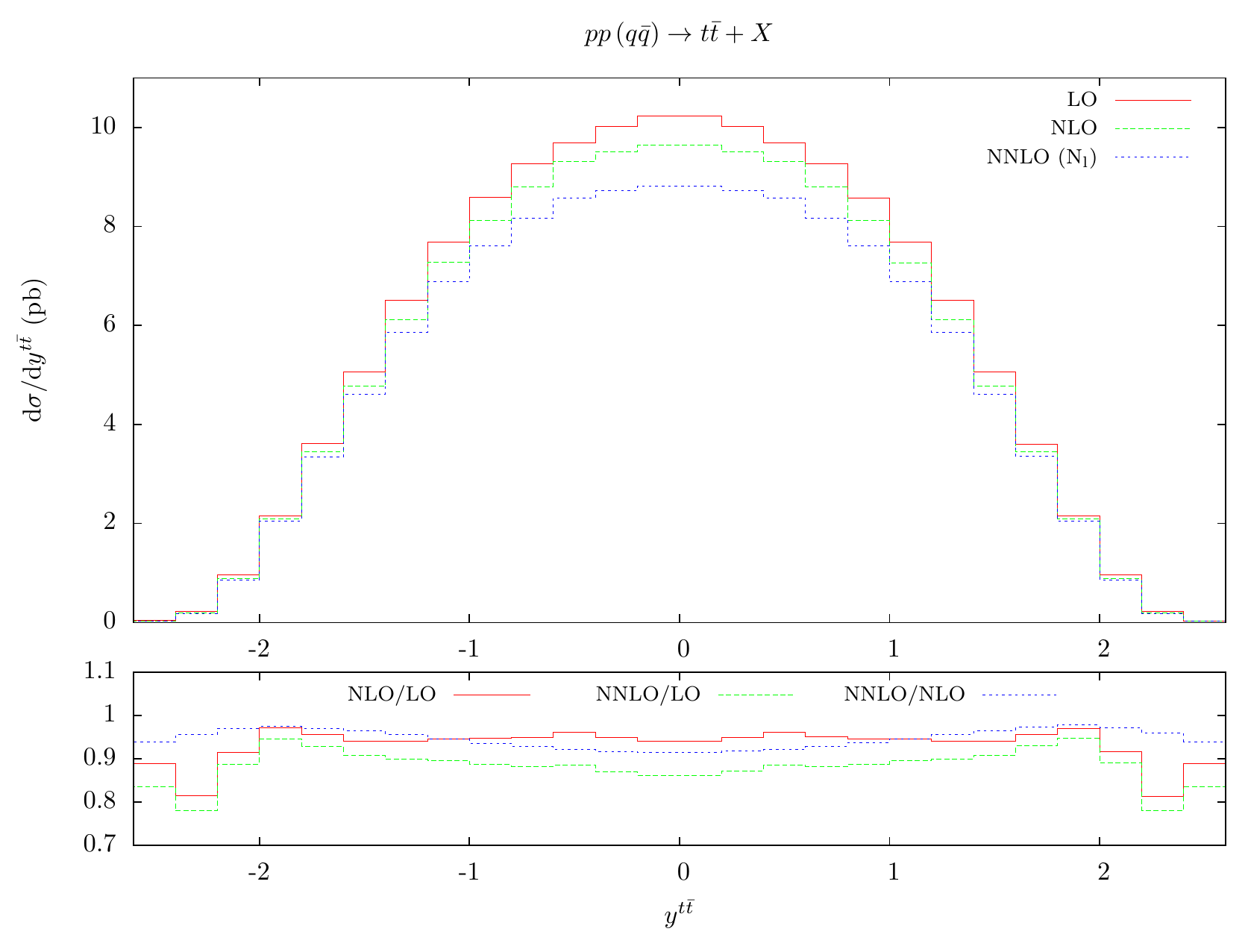}}
     \caption{Rapidity distribution of the top-antitop system ${\rm d}\sigma / {\rm d} y^{t\bar{t}}$ for $\sqrt{s}=7$ TeV at LO (red), NLO (green), and NNLO (blue). The lower panel shows the ratios of LO, NLO and NNLO cross sections.}
  \end{center}
  \label{fig.etattdist}
\end{figure}

Figure 6 contains the inclusive $t\bar{t}$ cross section as a function of the rapidity of the top-antitop system $y^{t\bar{t}}$ at LO, NLO and NNLO together with their corresponding $k$-factors. Once again we note an overall decrease in the cross section at NNLO, ranging from $2.5$ to $8.5\%$ with respect to the NLO result depending on the rapidity region considered.

\begin{figure}[t]
  \begin{center}
     \resizebox{0.9\linewidth}{!}{\includegraphics{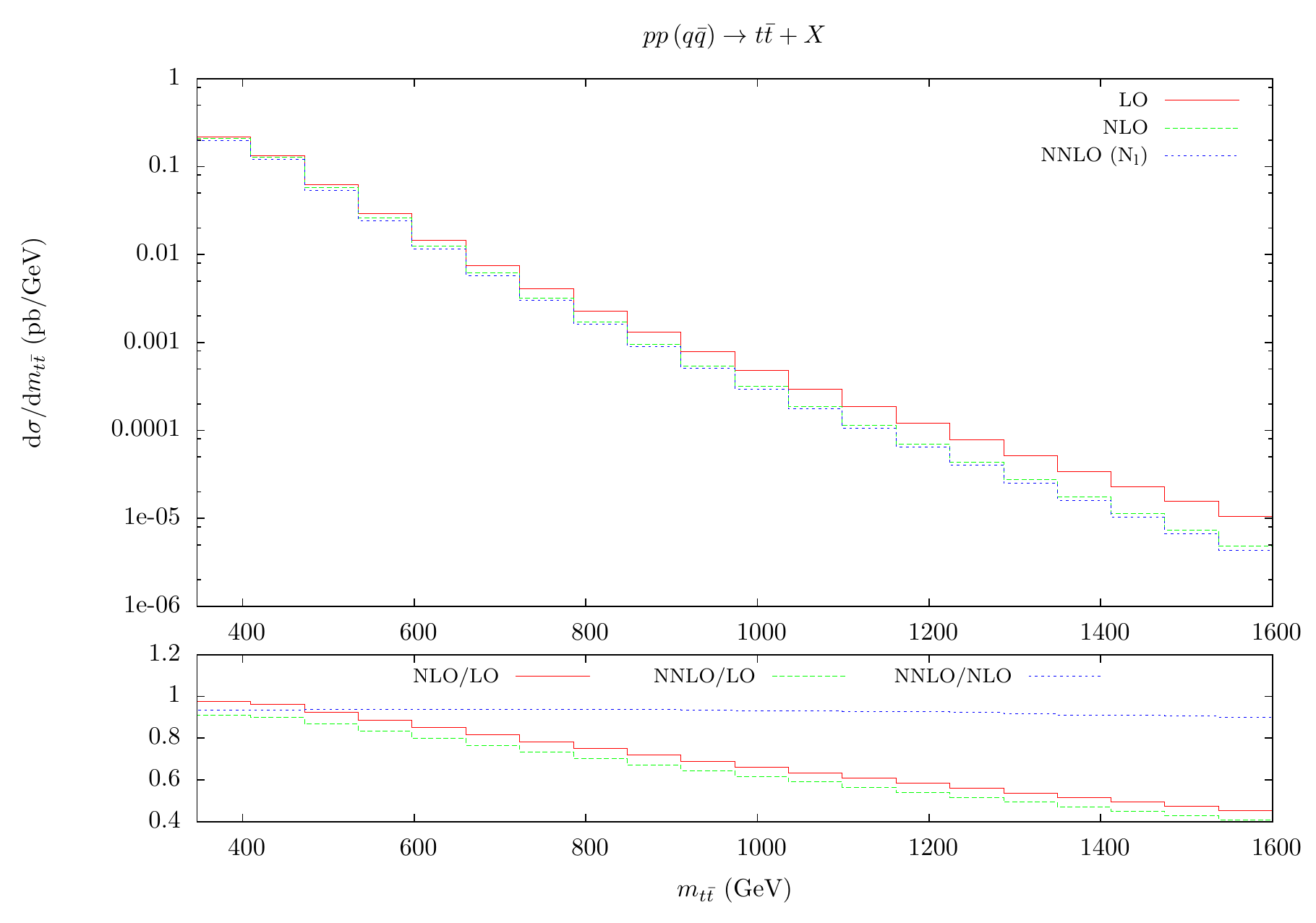}}
     \caption{Invariant mass distribution of the top-antitop system ${\rm d}\sigma / {\rm d} m_{t\bar{t}}$ for $\sqrt{s}=7$ TeV at LO (red), NLO (green), and NNLO (blue). The lower panel shows the ratios of LO, NLO and NNLO cross sections.}
  \end{center}
  \label{fig.mttdist}
\end{figure}

In figure 7 we present the cross section as a function of the invariant mass of the $t\bar{t}$ system at LO, NLO and NNLO, with all three ratios in the lower panel. As can be seen from the NNLO/NLO ratio, the NNLO corrections proportional to the number of light quark flavours $N_l$ decrease the cross section over the entire $m_{t\bar{t}}$ spectrum considered. The decrease ranges from $6.5\%$ in the low invariant mass region to almost $10\%$ in the high invariant mass region.

\begin{figure}[t]
  \begin{center}
    \subfigure[]{
      \label{fig.pttlcslc}
      \resizebox{0.475\linewidth}{!}{\includegraphics{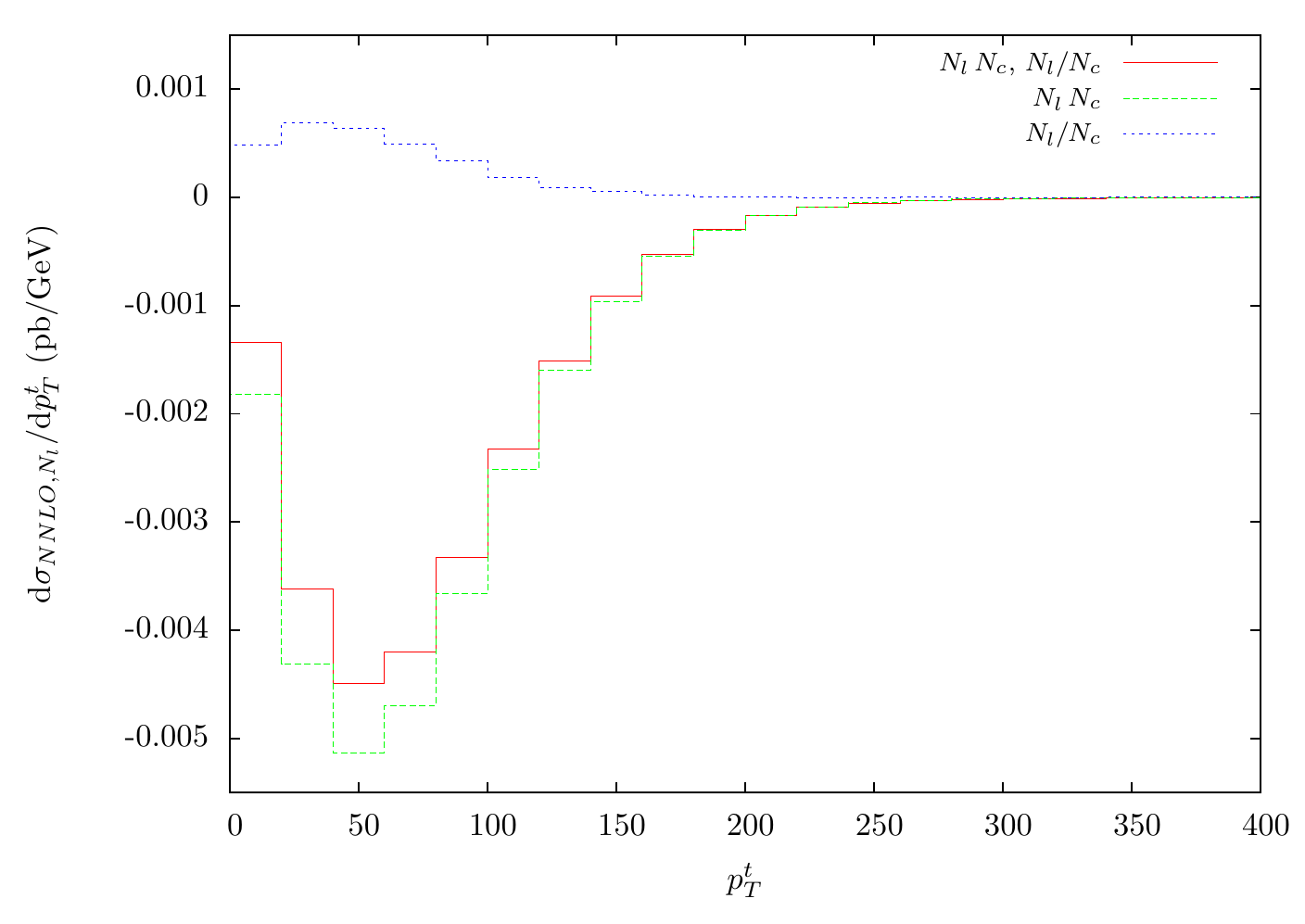}}
    }
    \subfigure[]{
      \label{fig.ytlcslc}
      \resizebox{0.475\linewidth}{!}{\includegraphics{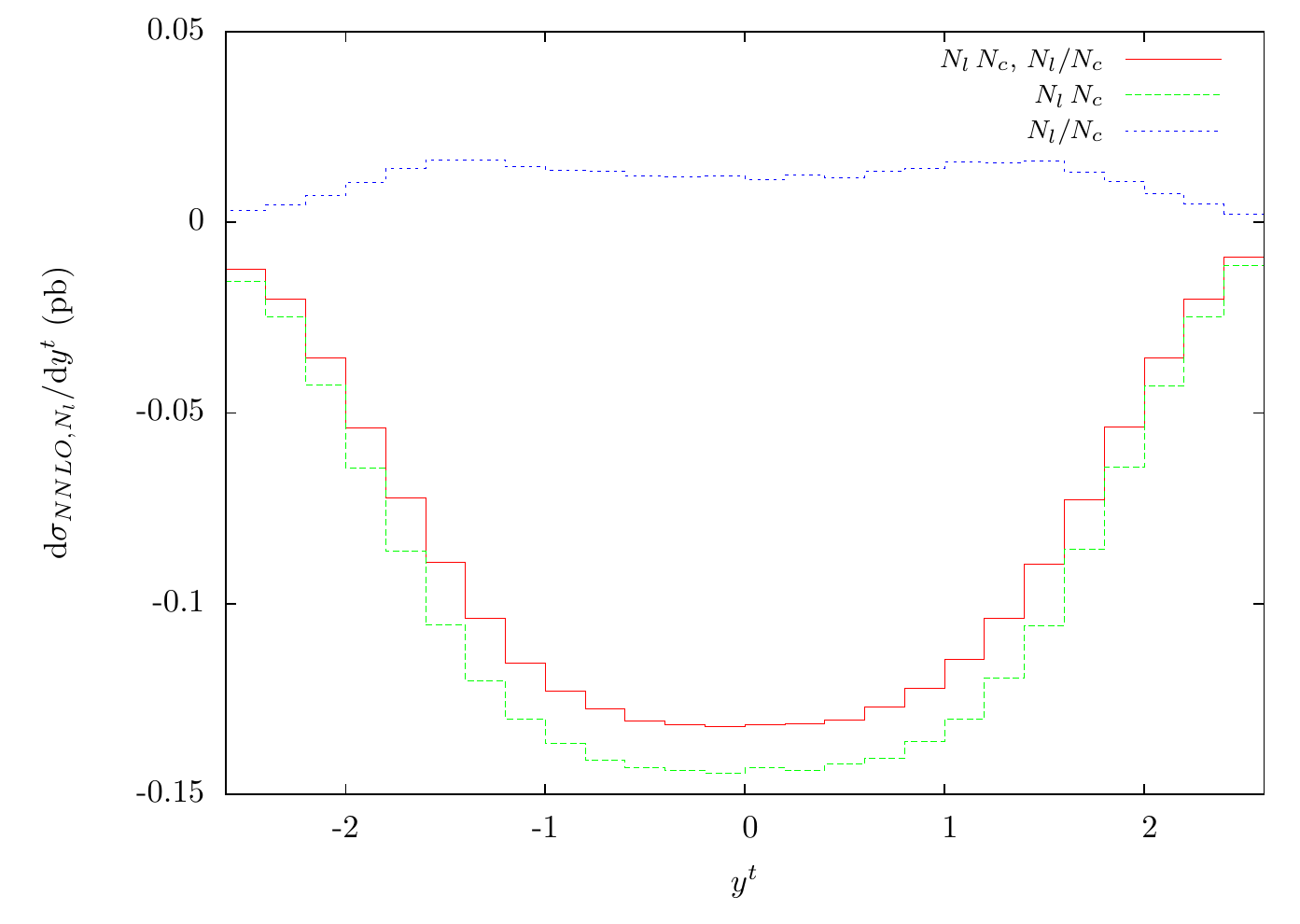}}
    }
    \caption{Individual NNLO contributions due to the colour factors $N_l\,N_c$ and $N_l/N_c$ as a function of (a) the top quark transverse momentum, (b) the top quark rapidity.}
  \end{center}
\end{figure}

Na\"ive power counting suggests that the leading-colour NNLO contributions considered in this paper, i.e. the pieces proportional to $N_l\,N_c$ should approximately account for $90\%$ of the NNLO corrections, with the remainder being given by the subleading-colour contributions, proportional to $N_l/N_c$. At the level of the total cross section we find that the subleading-colour pieces are opposite in sign to the leading-colour and represent $13\%$ of the full result, in agreement with this na\"ive power counting expectation.

In order to assess the impact of the subleading-colour contributions, we studied the NNLO corrections proportional to $N_lN_c$ and $N_l/N_c$ separately as a function of the top quark transverse momentum and rapidity. The results of these studies are shown in figures 8(a) and 8(b). We find that the subleading-colour piece has a different shape and sign than the leading-colour part, and contributes more significantly in the forward and backwards regions and less when the top quark is more centrally produced. In particular, in the $p_T^t$ spectrum of figure 8(a) we find that at low transverse momenta, the size of the $N_l/N_c$ colour factor is of approximately $30\%$ of the full colour result, while at high transverse momenta its contribution is negligible.

As a check of our numerical results, we compared the value of the total NNLO correction proportional to $N_l$ with the result presented in \cite{Baernreuther:2012ws}. In order to perform this check we had to incorporate to our calculation the NNLO contributions proportional to $N_hN_l$, with $N_h$ the number of heavy quark flavours, since those are included in the coefficient function $F_1(\beta)$ of \cite{Baernreuther:2012ws}. These contributions are rather simple, as they only enter at the double virtual level, and they are finite. We employed the two-loop matrix elements derived in \cite{Bonciani:2008az}, and computed the square of the one-loop matrix elements ourselves. With the same input parameters and PDF sets used throughout the present section, we find that the total NNLO correction to top pair production proportional to $N_l$ in the $q\bar{q}$ channel is $-0.5780$ pb, whereas with the coefficient function $F_1(\beta)$ of \cite{Baernreuther:2012ws} we obtain $-0.5822$ pb. These two results agree within less that $1\%$, which represents a very strong check of both completely independent calculations. 

%%%%%%%%%%%%%%%%%%%%%%%%%%%%%%%%%%%%%%%%%%%%%%%%%%%%%%
%%%%%%%%%%%%%%%%%%%%%%%%%%%%%%%%%%%%%%%%%%%%%%%%%%%%%%
%%%%%%%%%%%%%%%%%%%%%%%%%%%%%%%%%%%%%%%%%%%%%%%%%%%%%%
%                                                               SECTION: Conclusions
%%%%%%%%%%%%%%%%%%%%%%%%%%%%%%%%%%%%%%%%%%%%%%%%%%%%%%

\section{Conclusions}
\label{sec:conclusions}
In this paper we presented the ${\cal O}(\alpha_s^4)$ light fermionic corrections to the partonic process $q\bar{q} \to t\bar{t}$. Those corrections are of three types: virtual-virtual, real-virtual and double real, with two, three and four particles in the final state respectively. While the ultraviolet divergencies are removed by renormalisation, the infrared singularities are treated using the massive extension of the NNLO antenna formalism.   

The double real corrections had been derived in \cite{Abelof:2011ap}. In this paper we presented the real-virtual and virtual-virtual contributions and derived their related subtraction terms explicitly. We showed that the real-virtual subtraction terms correctly approximate the real-virtual matrix elements in all their singular limits. Furthermore, combined with the integrated forms of the subtraction terms, we show analytically that all explicit infrared singularities are cancelled both at real-virtual and at the virtual-virtual levels. This analytic pole cancellation provides us with a very strong check on the correctness of our result and on the applicability of the antenna subtraction method to reactions involving massive final states.  

The NNLO results of this paper are implemented in a Monte Carlo parton-level generator providing full kinematical information on an event-by event basis. This program, written in {\tt Fortran}, allowed us to produce NNLO differential distributions for top pair production in the partonic channel under consideration for the first time. We studied the impact of the NNLO corrections proportional to $N_l$ for distributions in $p_T$ and rapidity of the top quark as well as in the rapidity and invariant mass of the $t\bar{t}$ system and found that those corrections are substantial. They reduce the cross section compared to the NLO results, with the reduction varying considerably over the phase space. We also assessed the separate sizes of the leading and subleading-colour $N_l$ contributions finding that they have opposite signs, and that, at the level of the total cross section, the subleading-colour part contributes with $13\%$. Its importance, however, varies substantially over different regions of phase space. It can be as high as $30\%$ for low transverse momenta of the top quark, and negligible for high transverse momenta. 

The results presented in this paper can be regarded as a major step towards the computation of the complete cross section of top pair production including all partonic channels in fully differential form. For the first time, we have presented exact NNLO ($N_l$) corrections to one partonic channel in this process and produced NNLO differential distributions. Further work will include the computation of all remaining partonic channels. 

%%%%%%%%%%%%%%%%%%%%%%%%%%%%%%%%%%%%%%%%%%%%%%%%%%%%%%
%%%%%%%%%%%%%%%%%%%%%%%%%%%%%%%%%%%%%%%%%%%%%%%%%%%%%%

\acknowledgments
We would like to thank Joao Pires for many useful and stimulating discussions, Roberto Bonciani for providing us with the numerical code to evaluate the virtual contributions \cite{Bonciani:2008az}, and Thomas Gehrmann for adapting the phase space generator used in the di-jet calculation of \cite{GehrmannDeRidder:2013mf} to top pair production. This research was supported by the Munich Institute for Astro-and Particle Physics (MIAPP) of the DFG cluster of excellence ``Origin and Structure of the Universe'', where A.G. was a workshop participant in July 2014. G.A. is very grateful to the Galileo Galilei Institute for Theoretical Physics and the Institute for Theoretical Physics at ETH Z\"urich for their hospitality. We acknowledge support from the Swiss National Science Foundation under contract PBEZP2-145917, and from EuropeanCommission through the 'LHCPhenoNet' Initial Training Network PITN-GA-2010-264564' and through the ERC Advanced Grant  'MC@NNLO' (340983).

\appendix

%%%%%%%%%%%%%%%%%%%%%%%%%%%%%%%%%%%%%%%%%%%%%%%%%%%%%%
%%%%%%%%%%%%%%%%%%%%%%%%%%%%%%%%%%%%%%%%%%%%%%%%%%%%%%
%%%%%%%%%%%%%%%%%%%%%%%%%%%%%%%%%%%%%%%%%%%%%%%%%%%%%%
%         SECTION: The double real subtraction term for $q\bar{q} \rightarrow t \bar{t}$ : the $N_l$ part}
%%%%%%%%%%%%%%%%%%%%%%%%%%%%%%%%%%%%%%%%%%%%%%%%%%%%%%

\section{The double real subtraction term for $q\bar{q} \rightarrow t \bar{t}$: the $N_l$ part}
\label{sec:dss}
The double real corrections to the process $q\bar{q} \rightarrow t \bar{t}$ proportional to the number of light quark flavours are due to the tree-level partonic process $q\bar{q} \rightarrow t \bar{t}q'\bar{q}'$. We derived the corresponding subtraction terms in \cite{Abelof:2011ap} and found that only two pieces are needed, namely $ds^{S,a}_{q\bar{q},NNLO,N_l}$ and $\ds^{S,b}_{q\bar{q},NNLO,N_l}$. The subtraction term $ds^{S,a}_{q\bar{q},NNLO,N_l}$ capturing the single unresolved limits of the real-real contributions read
\beqa\label{eq.subtermqqttqq,a}
\lefteqn{\ds^{S,a }_{q\bar{q},NNLO,N_l}=\norm_{NNLO}^{RR,q\bar{q}}\,N_l\,\dphi_4(p_3,p_4,p_5,p_6;p_1,p_2)}\nonumber\\
&&\times\bigg\{ N_c\bigg[ \frac{1}{2} E_3^0(\Q{3},\qp{5},\qpb{6})\bigg( |\cm_5(\Q{(\wt{35})},\qi{1};;\qbi{2},\gl{(\wt{56})},\Qb{4})|^2\nonumber\\
&&\:\:\:\:\:\:\:\:\:\:\:\:\:\:\:\:\:\:\:\:\:\:\:\:\:\:\:\:\: +|\cm_5(\Q{(\wt{35})},\gl{(\wt{56})},\qi{1};;\qbi{2},\Qb{4})|^2 \bigg)J_2^{(3)}(p_{\wt{35}},p_4,p_{\wt{56}})\nonumber\\
&&\:\:\:\:\:\:\:\:\:\:\:+\frac{1}{2} E_3^0(\Qb{4},\qp{5},\qpb{6})\bigg( |\cm_5(\Q{3},\qi{1};;\qbi{2},\gl{(\wt{56})},\Qb{(\wt{45})})|^2\nonumber\\
&&\:\:\:\:\:\:\:\:\:\:\:\:\:\:\:\:\:\:\:\:\:\:\:\:\:\:\:\:\:+|\cm_5(\Q{3},\gl{(\wt{56})},\qi{1};;\qbi{2},\Qb{(\wt{45})})|^2\bigg)J_2^{(3)}(p_3,p_{\wt{45}},p_{\wt{56}})\bigg]\nonumber\\
&&\:\:+\frac{1}{N_c}\bigg[\frac{1}{2} E_3^0(\Q{3},\qp{5},\qpb{6})\bigg( |\cm_5(\Q{(\wt{35})},\Qb{4};;\qbi{2},\gl{(\wt{56})},\qi{1})|^2\nonumber\\
&&\:\:\:\:\:\:\:\:\:\:\:\:\:\:\:\:\:\:\:\:\:\:\:\:\:\:\:\:\:\:\:\:\:\:\:\:\:\:\:\:\:\:\:\:\:\:+|\cm_5(\Q{(\wt{35})},\gl{(\wt{56})},\Qb{4};;\qbi{2},\qi{1})|^2\nonumber\\
&&\:\:\:\:\:\:\:\:\:\:\:\:\:\:\:\:\:\:\:\:\:\:\:\:\:\:\:\:\:\:\:\:\:\:\:\:\:\:\:\:\:\:\:\:\:\:-2|\cm_5(\Q{(\wt{35})},\Qb{4},\qbi{2},\qi{1},\ph{(\wt{56})})|^2\bigg)J_2^{(3)}(p_{\wt{35}},p_4,p_{\wt{56}})\nonumber\\
&&\:\:\:\:\:\:\:\:\:\:\:+\frac{1}{2}E_3^0(\Qb{4},\qp{5},\qpb{6})\bigg(|\cm_5(\Q{3},\Qb{(\wt{45})};;\qbi{2},\gl{(\wt{56})},\qi{1})|^2\nonumber\\
&&\:\:\:\:\:\:\:\:\:\:\:\:\:\:\:\:\:\:\:\:\:\:\:\:\:\:\:\:\:\:\:\:\:\:\:\:\:\:\:\:\:\:\:\:\:\:+|\cm_5(\Q{3},\gl{(\wt{56})},\Qb{(\wt{45})};;\qbi{2},\qi{1})|^2\nonumber\\
&&\:\:\:\:\:\:\:\:\:\:\:\:\:\:\:\:\:\:\:\:\:\:\:\:\:\:\:\:\:\:\:\:\:\:\:\:\:\:\:\:\:\:\:\:\:\:-2|\cm_5(\Q{3},\Qb{(\wt{45})},\qbi{2},\qi{1},\ph{(\wt{56})})|^2\bigg)J_2^{(3)}(p_3,p_{\wt{45}},p_{\wt{56}})\bigg]\bigg\}.\nonumber\\ 
\eeqa
$\ds^{S,b}_{q\bar{q},NNLO,N_l}$ captures the double unresolved limit. It reads
\beqa\label{eq.subtermqqttqq,b}
\lefteqn{\ds^{S,b}_{q\bar{q},NNLO,N_l}=\norm_{NNLO}^{RR,q\bar{q}}\,N_l\,\dphi_4(p_3,p_4,p_5,p_6;p_1,p_2)}\nonumber\\
&&\times\bigg\{ N_c\bigg[ \bigg( B_4^0(\Q{3},\qpb{6},\qp{5},\qi{1})- \frac{1}{2}E_3^0(\Q{3},\qp{5},\qpb{6})A_3^0(\Q{(\wt{35})},\gl{(\wt{56})},\qi{1})\nonumber\\
&&\:\:\:\:\:\:\:\:\:\:\:\:\:\:\: -\frac{1}{2} E_3^0(\Qb{4},\qp{5},\qpb{6})A_3^0(\Q{3},\gl{(\wt{56})},\qi{1})\bigg) |\cm_4(\Q{(\wt{356})},\Qb{4},\qbi{2},\qi{\bar{1}})|^2 J_2^{(2)}(p_{\wt{356}},p_4)\nonumber\\
&&\:\:\:\:\:\:\:\:\:\:\:+\bigg( B_4^0(\Qb{4},\qp{5},\qpb{6},\qbi{2}) -\frac{1}{2}E_3^0(\Q{3},\qp{5},\qpb{6})A_3^0(\Qb{4},\gl{(\wt{56})},\qbi{2})\nonumber\\
&&\:\:\:\:\:\:\:\:\:\:\:\:\:\:\:-\frac{1}{2} E_3^0(\Qb{4},\qp{5},\qpb{6})A_3^0(\Qb{(\wt{45})},\gl{(\wt{56})},\qbi{2})\bigg) |\cm_4(\Q{3},\Qb{(\wt{456})},\qbi{\bar{2}},\qi{1})|^2 J_2^{(2)}(p_3,p_{\wt{456}})\bigg] \nonumber \\ 
&& -\frac{1}{N_c}\bigg[\bigg( B_4^0(\Q{3},\qpb{6},\qp{5},\Qb{4}) - \frac{1}{2}E_3^0(\Q{3},\qp{5},\qpb{6})A_3^0(\Q{(\wt{35})},\gl{(\wt{56})},\Qb{4})\nonumber\\
&&\:\:\:\:\:\:\:\:\:\:\:\:\:\:-\frac{1}{2} E_3^0(\Qb{4},\qp{5},\qpb{6})A_3^0(\Q{3},\gl{(\wt{56})},\Qb{(\wt{45})})\bigg)|\cm_4(\Q{(\wt{356})},\Qb{(\wt{456})},\qbi{2},\qi{1})|^2 J_2^{(2)}(p_{\wt{356}},p_{\wt{456}})\nonumber\\
&&\:\:\:\:\:\:\:\:\:+\bigg( B_4^0(\qbi{2},\qpb{6},\qp{5},\qi{1})-\frac{1}{2}E_3^0(\Q{3},\qp{5},\qpb{6})A_3^0(\qbi{2},\gl{(\wt{56})},\qi{1})\nonumber\\
&&\:\:\:\:\:\:\:\:\:\:\:\:\:\:-\frac{1}{2} E_3^0(\Qb{4},\qp{5},\qpb{6})A_3^0(\qbi{2},\gl{(\wt{56})},\qi{1})\bigg)|\cm_4(\Q{\tilde{3}},\Qb{\tilde{4}},\qbi{\bar{2}},\qi{\bar{1}})|^2 J_2^{(2)}(\tilde{p}_3,\tilde{p}_4)\nonumber\\
&&\:\:\:\:\:\:\:\:\:+2\bigg( B_4^0(\Q{3},\qpb{6},\qp{5},\qi{1})- \frac{1}{2}E_3^0(\Q{3},\qp{5},\qpb{6})A_3^0(\Q{(\wt{35})},\gl{(\wt{56})},\qi{1})\nonumber\\
&&\:\:\:\:\:\:\:\:\:\:\:\:\:\:-\frac{1}{2} E_3^0(\Qb{4},\qp{5},\qpb{6})A_3^0(\Q{3},\gl{(\wt{56})},\qi{1})\bigg)|\cm_4(\Q{(\wt{356})},\Qb{4},\qbi{2},\qi{\bar{1}})|^2 J_2^{(2)}(p_{\wt{356}},p_4)\nonumber\\
&&\:\:\:\:\:\:\:\:\:+2\bigg( B_4^0(\Qb{4},\qp{5},\qpb{6},\qbi{2}) -\frac{1}{2}E_3^0(\Q{3},\qp{5},\qpb{6})A_3^0(\Qb{4},\gl{(\wt{56})},\qbi{2})\nonumber\\
&&\:\:\:\:\:\:\:\:\:\:\:\:\:\:-\frac{1}{2} E_3^0(\Qb{4},\qp{5},\qpb{6})A_3^0(\Qb{(\wt{45})},\gl{(\wt{56})},\qbi{2})\bigg) |\cm_4(\Q{3},\Qb{(\wt{456})},\qbi{\bar{2}},\qi{1})|^2 J_2^{(2)}(p_1,p_{\wt{456}})\nonumber\\
&&\:\:\:\:\:\:\:\:\:-2\bigg( B_4^0(\Q{3},\qpb{6},\qp{5},\qbi{2})-\frac{1}{2}E_3^0(\Q{3},\qp{5},\qpb{6})A_3^0(\Q{(\wt{35})},\gl{(\wt{56})},\qbi{2})\nonumber\\
&&\:\:\:\:\:\:\:\:\:\:\:\:\:\:
-\frac{1}{2} E_3^0(\Qb{4},\qp{5},\qpb{6})A_3^0(\Q{3},\gl{(\wt{56})},\qbi{2})\bigg)
|\cm_4(\Q{(\wt{356})},\Qb{4},\qbi{\bar{2}},\qi{1})|^2 J_2^{(2)}(p_{\wt{356}},p_4)\nonumber\\
&&\:\:\:\:\:\:\:\:\:-2\bigg( B_4^0(\Qb{4},\qp{5},\qpb{6},\qi{1}) -\frac{1}{2}E_3^0(\Q{3},\qp{5},\qpb{6})A_3^0(\Qb{4},\gl{(\wt{56})},\qi{1})\nonumber\\
&&\:\:\:\:\:\:\:\:\:\:\:\:\:\:-\frac{1}{2} E_3^0(\Qb{4},\qp{5},\qpb{6})A_3^0(\Qb{(\wt{45})},\gl{(\wt{56})},\qi{1})\bigg)|\cm_4(\Q{3},\Qb{(\wt{456})},\qbi{2},\qi{\bar{1}})|^2 J_2^{(2)}(p_1,p_{\wt{456}})\bigg]\bigg\},\nonumber\\ 
\eeqa
where the pieces containing $B_4^0$ antennae subtract the double soft and triple collinear limits, while those of the form $E_3^0\times A_3^0$ subtract the single collinear limits of the B-type antennae, as well as spurious double unresolved limits of the $\ds^{S,a }_{q\bar{q},NNLO,N_l}$ subtraction term.

%%%%%%%%%%%%%%%%%%%%%%%%%%%%%%%%%%%%%%%%%%%%%%%%%%%%%%
%%%%%%%%%%%%%%%%%%%%%%%%%%%%%%%%%%%%%%%%%%%%%%%%%%%%%%
%%%%%%%%%%%%%%%%%%%%%%%%%%%%%%%%%%%%%%%%%%%%%%%%%%%%%%
%                                  SECTION: Colour-ordered infrared singularity operators
%%%%%%%%%%%%%%%%%%%%%%%%%%%%%%%%%%%%%%%%%%%%%%%%%%%%%%

\section{Colour-ordered infrared singularity operators}
\label{sec.iones}
The explicit pole structure of colour-ordered one-loop matrix elements can be written in terms of colour-ordered infrared singularity operators $\ione{i}{j}$. Within the antenna subtraction method, the pole part of one-loop antennae as well as that of integrated three-parton antennae can be also captured by these operators.

If only massless particles are involved, the following set of operators is sufficient (in addition to the one-loop splitting kernels $\Gamma^{(1)}_{ij}(x)$) to express the pole structure of a QCD one-loop amplitude as well as that of a one-particle inclusive integral of a tree-level amplitude 
\cite{Daleo:2006xa,GehrmannDeRidder:2005cm}:
\beqa
&&\ione{q}{\bar{q}}(\e,s_{q\bar{q}})=-\frac{e^{\e\gamma_E}}{2\Gamma(1-\e)}\left(\frac{|s_{q\bar{q}}|}{\mu^2}\right)^{-\e}\left[  \frac{1}{\e^2}+\frac{3}{2\e} \right]\\
&&\ione{q}{g}(\e,s_{qg})=-\frac{e^{\e\gamma_E}}{2\Gamma(1-\e)}\left(\frac{|s_{qg}|}{\mu^2}\right)^{-\e}\left[  \frac{1}{\e^2}+\frac{5}{3\e} \right]\\
&&\ione{g}{g}(\e,s_{gg})=-\frac{e^{\e\gamma_E}}{2\Gamma(1-\e)}\left(\frac{|s_{gg}|}{\mu^2}\right)^{-\e}\left[  \frac{1}{\e^2}+\frac{11}{6\e} \right]\\
&&\ione{q}{g,F}(\e,s_{qg})=\frac{e^{\e\gamma_E}}{2\Gamma(1-\e)}\left(\frac{|s_{qg}|}{\mu^2}\right)^{-\e}\frac{1}{6\e}\label{eq.ioneqgf}\\ 
&&\ione{g}{g,F}(\e,s_{gg})=\frac{e^{\e\gamma_E}}{2\Gamma(1-\e)}\left(\frac{|s_{gg}|}{\mu^2}\right)^{-\e}\frac{1}{3\e}.
\eeqa
When massive fermions are involved the following operators must also be considered \cite{Abelof:2011jv}
\beqa
&&\ione{Q}{\bar{Q}}(\e,s_{Q\bar{Q}})=-\frac{e^{\e\gamma_E}}{2\Gamma(1-\e)}\left(\frac{|s_{Q\bar{Q}}|}{\mu^2}\right)^{-\e}\left[ \frac{1}{\e}\left(1-\frac{1+r_0}{2\sqrt{r_0}}\ln\left(\frac{1+\sqrt{r_0}}{1-\sqrt{r_0}} \right)\right)\right]\\
&&\ione{Q}{\bar{q}}(\e,s_{Q\bar{q}})=-\frac{e^{\e\gamma_E}}{2\Gamma(1-\e)}\left(\frac{|s_{Q\bar{q}}|}{\mu^2}\right)^{-\e}\left[ \frac{1}{2\e^2}+\frac{5}{4\e}+\frac{1}{2\e}\ln\left( \frac{m_Q^2}{|s_{Q\bar{q}}|}\right)\right]\\
&&\ione{Q}{g}(\e,s_{Qg})=-\frac{e^{\e\gamma_E}}{2\Gamma(1-\e)}\left(\frac{|s_{Qg}|}{\mu^2}\right)^{-\e}\left[ \frac{1}{2\e^2}+\frac{17}{12\e}+\frac{1}{2\e}\ln\left( \frac{m_Q^2}{|s_{Q\bar{g}}|}\right)\right]\\
&&\ione{Q}{g,F}(\e,s_{Qg})=\frac{e^{\e\gamma_E}}{2\Gamma(1-\e)}\left(\frac{|s_{Qg}|}{\mu^2}\right)^{-\e} \frac{1}{6\e},\phantom{ \left( \frac{m_Q^2}{|s_{Q\bar{g}}|}\right)}\label{eq.ioneQgf}
\eeqa
with
\beq
r_0=1-\frac{4m_Q^2}{s_{Q\bar{Q}}+2m_Q^2}.
\eeq

%%%%%%%%%%%%%%%%%%%%%%%%%%%%%%%%%%%%%%%%%%%%%%%%%%%%%%
%%%%%%%%%%%%%%%%%%%%%%%%%%%%%%%%%%%%%%%%%%%%%%%%%%%%%%
%%%%%%%%%%%%%%%%%%%%%%%%%%%%%%%%%%%%%%%%%%%%%%%%%%%%%%
%                             SECTION: Infrared properties of tree-level and one-loop amplitudes
%%%%%%%%%%%%%%%%%%%%%%%%%%%%%%%%%%%%%%%%%%%%%%%%%%%%%%

\section{Infrared properties of tree-level and one-loop amplitudes}
\label{sec:infraredfact}
In this section we collect the single unresolved tree and one-loop factors needed in the context of the computation presented in this paper. 

%%%%%%%%%%%%%%%%%%%%%%%%%%%%%%%%%%%%%%%%%%%%%%%%%%%%%%
%                                      SUBSECTION: Infrared limits of tree-level amplitudes
%%%%%%%%%%%%%%%%%%%%%%%%%%%%%%%%%%%%%%%%%%%%%%%%%%%%%%

\subsection{Infrared limits of tree-level amplitudes}
The factorisation properties of colour-ordered tree-level amplitudes in single soft and collinear limits are well known \cite{Campbell:1997hg,Catani:1999ss}. They involve universal unresolved factors that are also encountered in the single unresolved limits of tree-level antennae. In the following we shall recall the general factorisation of tree-level matrix elements in single soft and collinear limits and present only those unresolved factors that are needed in the context of this paper.

%%%%%%%%%%%%%%%%%%%%%%%%%%%%%%%%%%%%%%%%%%%%%%%%%%%%%%
%                                                 SUBSUBSECTION: Collinear limits
%%%%%%%%%%%%%%%%%%%%%%%%%%%%%%%%%%%%%%%%%%%%%%%%%%%%%%

\subsubsection{Collinear limits}
When two massless colour-connected partons $i$ and $j$ become collinear and cluster into a parent particle $k$, a sub-amplitude squared factorises as 
\beq\label{eq.collfact}
|\cm_m(\ldots,i,j,\ldots)|^2
\stackrel{^{p_i||p_j}}{\longrightarrow}\frac{1}{s_{ij}}
P_{ij\rightarrow k}(z) |\cm_{m-1}(\ldots,k,\ldots)|^2.
\eeq
The spin averaged Altarelli-Parisi splitting function $P_{ij\rightarrow k}(z)$ is different for the different parton-parton splittings. The definition of the momentum fraction $z$ in eq.(\ref{eq.collfact}) depends on whether both collinear particles are in the final state, or if one of them is in the initial state. In the final-final case we have
\beq\label{eq.zff}
p_i\rightarrow z\,p_k\hspace{1in} p_j\rightarrow (1-z)p_k,
\eeq
whereas if $p_i$ is in the initial state and $p_j$ in the final state, we have
\beq\label{eq.zif}
p_j\rightarrow z\,p_i\hspace{1in}p_k\rightarrow (1-z)p_i.
\eeq

In the calculation presented in this paper, only the collinear limit of an initial state (anti)quark and a final state gluon must be considered. The corresponding splitting function is given by
\beq\label{eq.splittree}
P_{\hat{q}g\rightarrow \hat{q}}(z)=\frac{1+(1-z)^2-\e z^2}{z(1-z)}.
\eeq

%%%%%%%%%%%%%%%%%%%%%%%%%%%%%%%%%%%%%%%%%%%%%%%%%%%%%%
%                                                        SUBSUBSECTION: Soft limits
%%%%%%%%%%%%%%%%%%%%%%%%%%%%%%%%%%%%%%%%%%%%%%%%%%%%%%
 
 \subsubsection{Soft limits}
Colour-ordered tree-level amplitudes develop an implicit soft singularity when the four-momentum of a final state gluon vanishes. When a gluon with momentum $p_j$ becomes soft in a colour-ordered amplitude where it is colour-connected to two hard particles with momenta $p_i$ and $p_k$, the amplitude factorises as,  
\beq\label{eq.factampg}
\cm_{m}(\ldots,i,j,k,\ldots) 
\stackrel{^{p_j \rightarrow 0}}{\longrightarrow}
\e^{\mu}(p_j,\lambda)J_{\mu}(p_i,p_j,p_k)\cm_{m-1}(\ldots,i,k,\ldots)
\eeq
where $\e^{\mu}(p_j,\lambda)$ is the soft gluon's polarisation vector, and the soft currents are given by
\beq\label{eq.currentg}
J_{\mu}(p_i,p_j,p_k)=e_{\mu}(p_i,p_j)-e_{\mu}(p_k,p_j)
\eeq
with
\beq
e^{\mu}(p_a,p_b)=\frac{p_a^{\mu}}{\sqrt{2}p_a\cdot p_b}.
\eeq
After squaring eq.(\ref{eq.factampg}) we obtain the well-known formula
\beq\label{eq.wellknownfact}
|\cm_{m}(\ldots,i,j,k,\ldots)|^2
\stackrel{^{p_j \rightarrow 0}}{\longrightarrow}
\ssoft{i}{j}{k}(m_i,m_k)|\cm_{m-1}(\ldots,i,k,\ldots)|^2,
\eeq
with the massive soft eikonal factor given by
\beq
\label{eq.ssoft}
\ssoft{i}{j}{k}(m_i,m_k)=\frac{2s_{ik}}{s_{ij}s_{jk}}-\frac{2m_i^2}{s_{ij}^2}-\frac{2m_k^2}{s_{jk}^2}.
\eeq

In multi-parton processes one often encounters certain subleading-colour contributions that, in addition to colour-ordered amplitudes squared, contain interferences of different colour-ordered amplitudes. Using eq.(\ref{eq.factampg}), we find that in their soft gluon limits, these types of interferences factorise as
\beqa
&&\hspace{-0.3in}\cm_{m}(\ldots,i,j,k,\ldots)\cm_{m}(\ldots,l,j,m,\ldots)^{\dagger}
\stackrel{^{p_j \rightarrow 0}}{\longrightarrow}\phantom{\frac{2s_{il}}{s_{ij}s_{jl}}}\nonumber\\
&&\hspace{0.2in}\frac{1}{2}\left(\ssoft{i}{j}{m}(m_i,m_m)
+\ssoft{k}{j}{l}(m_k,m_l)
-\ssoft{i}{j}{l}(m_i,m_l)-\ssoft{k}{j}{m}(m_k,m_m)\right)\nonumber\\
&&\hspace{0.6in}\times
\cm_{m}(\ldots,i,k,\ldots)\cm_{m}(\ldots,l,m,\ldots)^{\dagger}.\phantom{\frac{2s_{il}}{s_{ij}s_{jl}}}
\label{eq.factinterferenceg2}
\eeqa
In some cases we can employ decoupling identities and replace the subleading-colour interference terms by sub-amplitudes squared in which a gluon is $U(1)$-like, and does not have any non-abelian couplings. In these cases, the factorisation in the soft limit at the amplitude level is analogous to the soft photon factorisation of QED matrix elements. It reads
\beq\label{eq.factabeliang}
\cm_m(1,\ldots,m;;j_{\gamma})
\stackrel{^{p_j\rightarrow 0}}{\longrightarrow}
\e^{\mu}(p_j,\lambda) \Bigg( \sum_{i\in\{q\}}e_{\mu}(p_i,p_j)
-\sum_{k\in\{\bar{q}\}}e_{\mu}(p_k,p_j) \Bigg) \cm_{m-1}(1,\ldots,m),
\eeq
where $\{q\}$ is the set of all final state quarks and initial state antiquarks in the process, and $\{ \bar{q} \}$ stands for all final state antiquarks and initial state quarks. Squaring eq.(\ref{eq.factabeliang}) and rearranging the result, we find the following factorisation for the amplitude squared:
\beqa\label{eq.factabeliang2}
&&\hspace{-5mm}|\cm_m(\ldots,m;;j_{\gamma})|^2
\stackrel{^{p_j\rightarrow 0}}{\longrightarrow} 
\Bigg( \sum_{\substack{i\in\{q\} \\ k\in\{\bar{q}\}}}\ssoft{i}{j}{k}(m_i,m_k)
-\frac{1}{2}\sum_{\substack{(i,k) \in\{q\}\\ i\neq k }}\ssoft{i}{j}{k}(m_i,m_k)\nonumber\\
&&\hspace{55mm}-\frac{1}{2}\sum_{\substack{(i,k) \in\{\bar{q}\}\\ i\neq k}}
\ssoft{i}{j}{k}(m_i,m_k) \Bigg)|\cm_{m-1}(\ldots,m)|^2.
\eeqa
We have employed this factorisation in the construction of the real-virtual subtraction term for the colour-ordered amplitude squared $|\cmb_5^{[l]}(\Q{3},\Qb{4},\qbi{2},\qi{1},\ph{5})|^2$ in eq.(\ref{eq.qqbNlRV}). 
 
%%%%%%%%%%%%%%%%%%%%%%%%%%%%%%%%%%%%%%%%%%%%%%%%%%%%%%
%                                      SUBSECTION: Infrared limits of one-loop amplitudes
%%%%%%%%%%%%%%%%%%%%%%%%%%%%%%%%%%%%%%%%%%%%%%%%%%%%%%

\subsection{Infrared limits of one-loop amplitudes}
\label{sec:infraredoneloop}
Like at tree level, in their infrared limits the one-loop colour-ordered amplitudes undergo well-known factorisations involving universal one-loop and tree-level soft and collinear factors \cite{Weinzierl:2003ra,Bern:1994zx,Bern:1998sc,Kosower:1999xi,Kosower:1999rx,Bern:1999ry,Catani:2000pi,Kosower:2002su,Kosower:2003cz,Catani:2003vu,Bern:2004cz,Badger:2004uk,Bierenbaum:2011gg}. Those singular factors are also found in the single unresolved limits of one-loop antennae.

Introducing the following notation
\beq\label{eq.int}
|\cmb^{1}_m|^2 \equiv 2\re(\cmb^{1}_m \cmb^{0\,\,\dagger}_m),
\eeq
the factorisation of $|\cmb^{1}_m|^2$ in single soft and collinear limits can be generically written as
\beq\label{eq.square}
|\cmb_m^{1}|^2\rightarrow \text{Sing}^{(0)}_1 |\cmb^1_{m-1}|^2+\text{Sing}^{(1)}_1 |\cmb^0_{m-1}|^2,
\eeq
where $\text{Sing}^{(1)}_1$ is a one-loop single unresolved factor and $ |\cmb^1_{m-1}|^2$ is a one-loop reduced sub-amplitude squared. Following the decomposition of the one-loop colour-ordered amplitudes into primitives, the one-loop unresolved factor can be decomposed as
\beq
\text{Sing}^{(1)}_1=
N_c \,\text{Sing}^{(1),[lc]}_1
+N_l \, \text{Sing}^{(1),[l]}_1+N_h \, \text{Sing}^{(1),[h]}_1
-\frac{1}{N_c} \, \text{Sing}^{(1),[slc]}_1.
\eeq
In subleading-colour contributions involving interferences of one-loop and tree-level amplitudes with different colour-orderings the factorisation in eq.(\ref{eq.square}) does not hold for the soft limit. As was the case at tree level, multiple soft factors arise.

In the following, we shall explicitly present the one-loop singular factors that we encountered in the calculation presented in this paper. 

%%%%%%%%%%%%%%%%%%%%%%%%%%%%%%%%%%%%%%%%%%%%%%%%%%%%%%
%                                        SUBSUBSECTION: One-loop collinear factors
%%%%%%%%%%%%%%%%%%%%%%%%%%%%%%%%%%%%%%%%%%%%%%%%%%%%%%

\subsubsection{One-loop collinear factors}
In single collinear limits, the interference of a one-loop and a tree-level colour-ordered matrix element factorises as in eq.(\ref{eq.square}), with $\text{Sing}^{(0)}_1$ and $\text{Sing}^{(1)}_1$ given by a tree-level and a one-loop splitting function respectively. For the calculation presented in this paper only the $N_l$ part of the one-loop initial-final splitting function denoted by $P^{1,[N_l]}_{\hat{q}g\rightarrow \hat{q}}(z)$ is needed. It is proportional to its tree-level counterpart, which was given in eq.(\ref{eq.splittree}), and reads
\beq
\label{eq.split1lif}
P^{(1),[l]}_{\hat{q}g\rightarrow \hat{q}}(z)=-2\frac{b_{0,F}}{\e}P_{\hat{q}g\rightarrow \hat{q}}(z)
\eeq
with $z$ defined as in eq.(\ref{eq.zif}). 

%%%%%%%%%%%%%%%%%%%%%%%%%%%%%%%%%%%%%%%%%%%%%%%%%%%%%%
%                                        SUBSUBSECTION: One-loop soft factors
%%%%%%%%%%%%%%%%%%%%%%%%%%%%%%%%%%%%%%%%%%%%%%%%%%%%%%

\subsubsection{One-loop soft factors}
As was the case at tree level, when a soft gluon is emitted between massive fermions in the colour chain, the soft one-loop factor contains mass dependant terms. While at tree level the massless soft factor can be obtained from the massive one by setting the massess of the hard radiators to zero, this is no longer the case at the one-loop level: masses are present in the arguments of logarithms and polylogarithms that diverge in the massless limit. We must therefore consider separately the soft factors with: (a) two massless hard radiators, (b) one massless and one massive hard radiator, (c) two massive hard radiators. In the subleading-colour contributions to the real-virtual cross section and subtraction terms presented in section \ref{sec:realvirtual} all these three cases arise.

When a gluon $j$ becomes soft in a one-loop primitive amplitude where it is colour-connected to the hard particles $i$ and $k$, the amplitude factorises as
\beqa
&&\hspace{-0.5in}\cmb^{1,[X]}_{m}(\ldots,i,j,k,\ldots)\stackrel{^{p_j\rightarrow 0}}{\longrightarrow}\,\e^{\mu}(p_j,\lambda)J_{\mu}(p_i,p_j,p_k)\cmb^{1,[X]}_{m-1}(\ldots,i,k,\ldots)\phantom{\Big(}\nonumber\\
&&\hspace{-0.62in}\phantom{\cmb^{1,[X]}_{m}(\ldots,i,j,k,\ldots)\stackrel{^{p_j\rightarrow 0}}{\longrightarrow}}+\e^{\mu}(p_j,\lambda)J_{\mu}^{(1),[X]}(p_i,p_j,p_k;m_i,m_k)\cm_{m-1}(\ldots,i,k,\ldots),
\eeqa
where $X=lc,l,h,slc$. The tree-level current $J_{\mu}(p_i,p_j,p_k)$ was given in eq.(\ref{eq.currentg}), and the primitive one-loop currents $J_{\mu}^{(1),[X]}(p_i,p_j,p_k;m_i,m_k)$ take a different form depending on whether $m_i$ and/or $m_k$ vanish. Restricting ourselves to the light quark currents $J_{\mu}^{(1),[l]}$ we have
\beqa
&&\hspace{-0.2in}J_{\mu}^{(1),[l]}(p_i,p_j,p_k;0,0)=-\frac{b_{0,F}}{\e}J_{\mu}(p_i,p_j,p_k)\\
&&\hspace{-0.2in}J_{\mu}^{(1),[l]}(p_i,p_j,p_k;m_i,0)=-\frac{b_{0,F}}{\e}J_{\mu}(p_i,p_j,p_k)\\
&&\hspace{-0.2in}J_{\mu}^{(1),[l]}(p_i,p_j,p_k;m_i,m_k)=-\frac{b_{0,F}}{\e}J_{\mu}(p_i,p_j,p_k)
\eeqa
with $b_{0,F}=-1/3$.

The reason why the light-fermion one-loop currents $J_{\mu}^{(1),[l]}$ are proportional to the tree-level currents is that these contributions are absent in the bare currents, and only enter in the renormalised ones, which are obtained as
\beq
J_{\mu,\,ren}^{(1),[l]}(p_i,p_j,p_k;m_i,m_k) = 
\frac{1}{\cepb}J_{\mu,\,bare}^{(1),[l]}(p_i,p_j,p_k;m_i,m_k)-\frac{b_{0,F}}{\e}J_{\mu}(p_i,p_j,p_k;m_i,m_k).
\eeq
The fact that the current $J_{\mu}^{(1),[l]}$  is proportional to the tree-level current is crucial for the construction of our real-virtual subtraction terms presented in section \ref{sec:realvirtual}. Indeed, it allows us to use tree-level relations like those given in eqs.(\ref{eq.factinterferenceg2}) and (\ref{eq.factabeliang2}) for the one-loop $N_l$ subtraction term at subleading colour.

Using these one-loop currents, the one-loop soft factors read, 
\beq
S^{(1),[X]}_{ijk}(m_i,m_k)=-2\,\eta^{\mu\nu}\,\re\left(J_{\mu}^{(1),[X]}(p_i,p_j,p_k;m_i,m_k)J_{\nu}(p_i,p_j,p_k)\right),
\eeq
which, for the $N_l$ parts, yields
\beqa
&&\hspace{-0.2in}S^{(1),[l]}_{ijk}(0,0)=-\frac{2b_{0,F}}{\e}\ssoft{i}{j}{k}(0,0)\\
&&\hspace{-0.2in}S^{(1),[l]}_{ijk}(m_i,0)=-\frac{2b_{0,F}}{\e}\ssoft{i}{j}{k}(m_i,0)\\
&&\hspace{-0.2in}S^{(1),[l]}_{ijk}(m_i,m_j)=-\frac{2b_{0,F}}{\e}\ssoft{i}{j}{k}(m_i,m_j)
\eeqa
with, $\ssoft{i}{j}{k}(m_i,m_j)$ given eq.(\ref{eq.ssoft}). 

%%%%%%%%%%%%%%%%%%%%%%%%%%%%%%%%%%%%%%%%%%%%%%%%%%%%%%
%%%%%%%%%%%%%%%%%%%%%%%%%%%%%%%%%%%%%%%%%%%%%%%%%%%%%%
%%%%%%%%%%%%%%%%%%%%%%%%%%%%%%%%%%%%%%%%%%%%%%%%%%%%%%
%                          SECTION: Infrared properties of the massive one-loop antennae $A^1_3$
%%%%%%%%%%%%%%%%%%%%%%%%%%%%%%%%%%%%%%%%%%%%%%%%%%%%%%

\section{Infrared properties of the massive one-loop antennae $A^1_3$} 
\label{sec:a13}
Before presenting the infrared limits of the one-loop antenna functions appearing in the real-virtual subtraction term $\ds^{T}_{q \bar{q},NNLO,N_l}$, derived in section \ref{sec:realvirtual} and required to subtract the single unresolved behaviour of the real-virtual contributions $\ds^{RV}_{q \bar{q},NNLO,N_l}$, let us first recall how one-loop antennae are defined in general.

%%%%%%%%%%%%%%%%%%%%%%%%%%%%%%%%%%%%%%%%%%%%%%%%%%%%%%
%                                             SUBSECTION: One-loop antenna functions
%%%%%%%%%%%%%%%%%%%%%%%%%%%%%%%%%%%%%%%%%%%%%%%%%%%%%%

\subsection{One-loop antenna functions}
Within the antenna formalism, the infrared limits of the real-virtual contributions are captured by three-parton one-loop antennae \cite{GehrmannDeRidder:2005cm,GehrmannDeRidder:2011aa}. These are generally denoted as $X^1_3(i,j,k)$ and depend on the antenna momenta $p_i,p_j,p_k$ as well as on the masses of the hard radiators in the massive case. Those one-loop antenna functions are constructed with colour-ordered one-loop three-parton and two-parton matrix elements as
\beq\label{eq:X1def}
X_{3}^1(i,j,k) = S_{ijk,IK}\, \frac{|{\cal M}^1_{3}(i,j,k)|^2}{|{\cal M}^0_2(I,K)|^2} - X_3^0(i,j,k)\, \frac{|{\cal M}^1_{2}(I,K)|^2}{|{\cal M}^0_2(IK)|^2} \;,
\eeq
where the tree-level antenna function, denoted by $X_3^0(i,j,k)$, is given by
\beq
X_3^0(i,j,k) = S_{ijk,IK}\, \frac{|{\cal M}^0_{ijk}|^2}{|{\cal M}^0_{IK}|^2}. 
\eeq
$S_{ijk,IK}$ denotes the symmetry factor associated with the antenna, which accounts both for potential identical particle symmetries and for the presence of more than one antenna in the basic two-parton process. Initial-final and initial-initial antennae can be obtained from their final-final counterparts by the appropriate crossing of partons to the initial-state. This procedure is straightforward at tree level but requires some care in the one-loop case, since one-loop antennae contain polylogarithms or hypergeometric functions that must be analytically continued to the appropriate kinematical region \cite{Daleo:2009yj,Gehrmann:2011wi}.

In any of the three kinematical configurations, one-loop antenna functions like one-loop amplitudes, can be conveniently decomposed into primitives according to their colour factors as follows
\beq\label{eq.antennadec}
X_3^1(i,j,k)=N_c X_3^{1,lc}(i,j,k) + N_l X_3^{1,l}(i,j,k)+ N_h X_3^{1,h}(i,j,k)-\frac{1}{N_c}X_3^{1,slc}(i,j,k).
\eeq

The one-loop matrix elements in eq.(\ref{eq:X1def}) are renormalised in the scheme of \cite{Bonciani:2008az,Bonciani:2009nb,Bonciani:2010mn,Bonciani:2013ywa} which we recalled in section \ref{sec:realvirtual}. The renormalisation of the antennae is performed at the scale $\mu^2=|s_{ijk}|$. To ensure the correct subtraction of terms arising from renormalisation and to ensure that the antennae and real-virtual matrix elements are computed at the same renormalisation scale, for the $N_l$ contribution we must substitute
\beq
X_{ijk}^{1,l}\rightarrow X_{ijk}^{1,l} + \frac{b_{0,F}}{\e} X_{ijk}^0 \left( (|s_{ijk}|)^{-\e}-(\mu^2)^{-\e}\right).
\eeq
As explained in section \ref{sec:realvirtual}, this substitution gives rise to the real-virtual subtraction term of type $\ds^{VS,(d)}$.

After renormalisation, one-loop antennae contain explicit and implicit infrared singularities. The structure of the former can be entirely captured by colour-ordered infrared singularity operators, whereas the latter occur when massless partons in the antenna become soft or collinear.

The one-loop three parton antennae encountered in the context of this paper are of A-type. They involve either no masses, as the initial-initial antenna $A_3^{1,l}(\qi{1},\gl{3},\qbi{2})$, one mass as the initial-final antenna $A_3^{1,l}(\Q{1},\gl{3},\qi{2})$, or two masses as the final-final antenna $A_3^{1,l}(\Q{1},\gl{3},\Qb{2})$.

%%%%%%%%%%%%%%%%%%%%%%%%%%%%%%%%%%%%%%%%%%%%%%%%%%%%%%
%           SUBSECTION: Explicit infrared pole structure of the massive one-loop antennae $A^{1,l}_3$
%%%%%%%%%%%%%%%%%%%%%%%%%%%%%%%%%%%%%%%%%%%%%%%%%%%%%%

\subsection{Explicit infrared pole structure of $A^{1,l}_3$ one-loop antennae}
In general, the explicit infrared poles of one-loop antennae can be written in terms of massless and massive $\ione{i}{j}$ operators, which were all recalled in the appendix \ref{sec.iones}. The poles of the one-loop antennae required in the calculation presented in this paper are proportional to their tree-level counterparts as 
\beqa
\label{eq.polesA13ffnf} &&\hspace{-0.2in}
\poles \left( A_3^{1,l}(\Q{1},\gl{3},\Qb{2})\right)=2\, b_{0,F} A_3^0(\Q{1},\gl{3},\Qb{2})\\
\label{eq.polesA13ifnf} &&\hspace{-0.2in}
\poles \left( A_3^{1,l}(\Q{1},\gl{3},\qi{2})\right)=2 b_{0,F} A_3^0(\Q{1},\gl{3},\qi{2})\\
\label{eq.polesA13iinf} &&\hspace{-0.2in}
\poles \left( A_3^{1,l}(\qi{1},\gl{3},\qbi{2})\right)=2\, b_{0,F} A_3^0(\qi{1},\gl{3},\qbi{2}).
\eeqa

%%%%%%%%%%%%%%%%%%%%%%%%%%%%%%%%%%%%%%%%%%%%%%%%%%%%%%
%                    SUBSECTION: Infrared limits of the massive one-loop antennae $A^{1,l}_3$
%%%%%%%%%%%%%%%%%%%%%%%%%%%%%%%%%%%%%%%%%%%%%%%%%%%%%%

\subsection{Infrared limits of $A^{1,l}_3$ one-loop antennae}
We conclude this section on one-loop antennae by listing the infrared limits of the three different $A^{1,l}_3$ antenna functions employed in this paper.

The infrared limits of of the one-loop initial-initial antenna $A_3^{1,l}(\hat{1}_q,3_g,\hat{2}_{\bar{q}})$ are given by
\beqa
&&A_3^{1,l}(\hat{1}_q,3_g,\hat{2}_{\bar{q}})\mathop{\longrightarrow}^{p_3\rightarrow 0}S^{(1),[l]}_{132}(0,0)\\
&&A_3^{1,l}(\hat{1}_q,3_g,\hat{2}_{\bar{q}})\mathop{\longrightarrow}^{p_1||p_3}\frac{1}{s_{13}}P^{1,[l]}_{qg\leftarrow Q}(z)\\
&&A_3^{1,l}(\hat{1}_q,3_g,\hat{2}_{\bar{q}})\mathop{\longrightarrow}^{p_2||p_3}\frac{1}{s_{23}}P^{1,[l]}_{qg\leftarrow Q}(z).
\eeqa

The limits of the flavour-violating initial-final antenna $A_3^{1,l}(1_Q,3_g,\hat{2}_{\bar{q}})$ are
\beqa
&&A_3^{1,l}(1_Q,3_g,\hat{2}_{\bar{q}})\mathop{\longrightarrow}^{p_3\rightarrow 0}S^{(1),[l]}_{132}(m_Q^2,0)\\
&&A_3^{1,l}(1_Q,3_g,\hat{2}_{\bar{q}})\mathop{\longrightarrow}^{p_2||p_3}\frac{1}{s_{23}}P^{1,[l]}_{qg\leftarrow Q}(z).
\eeqa

Finally, the massive final-final antennae $A_3^{1,l}(1_Q,3_g,2_{\bar{Q}})$ only have soft limits
\beqa
&&A_3^{1,l}(1_Q,3_g,2_{\bar{Q}})\mathop{\longrightarrow}^{p_3\rightarrow 0}S^{(1),[l]}_{132}(m_Q^2,m_Q^2).
\eeqa

The one-loop soft factors $S^{(1),[l]}$ and $P^{1,[l]}_{qg\leftarrow Q}(z)$ have been given in appendix \ref{sec:infraredoneloop}.

%%%%%%%%%%%%%%%%%%%%%%%%%%%%%%%%%%%%%%%%%%%%%%%%%%%%%%
%%%%%%%%%%%%%%%%%%%%%%%%%%%%%%%%%%%%%%%%%%%%%%%%%%%%%%
%%%%%%%%%%%%%%%%%%%%%%%%%%%%%%%%%%%%%%%%%%%%%%%%%%%%%%
%                  SECTION: The integrated massive initial-final antenna ${\cal B}_{q,Qq'\bar{q'}}$
%%%%%%%%%%%%%%%%%%%%%%%%%%%%%%%%%%%%%%%%%%%%%%%%%%%%%%

\section{The integrated massive initial-final antenna ${\cal B}_{q,Qq'\bar{q'}}$}
\label{sec.b04}
In \cite{Abelof:2012he} we have derived the integrated tree-level four-parton massive antenna ${\cal B}_{q,Qq'\bar{q'}}$ expressing our results in terms of the following variables
\beq
y= 1- \frac{Q^2 +m_{Q}^2}{2 p_i\cdot q}\hspace{1.5in}z=\frac{m_{Q}^2}{E^2_{cm}}
\eeq
with $E^2_{cm}=y/(1-y-z)Q^2$. In the present paper, this integrated antenna is more easily employed and combined with other integrated subtraction terms at the virtual-virtual level if it is expressed in terms of
\beq
x_1=\frac{Q^2 +m_Q^2}{2 p_i\cdot q}\hspace{1.5in}x_0=\frac{Q^2}{Q^2 +m_Q^2}
\eeq
In terms of those variables, the pole part of the integrated four-parton antenna function denoted by ${\cal B}_{q,Qq'\bar{q'}}$ reads
\beqa 
\lefteqn{{\cal B}_{q,Qq'\bar{q'}}(\e,s,x_1,x_2)=\delta(1-x_2)\,\bigg[-\frac{1}{24\e^3}\delta(1-x_1)
-\frac{1}{\e^2}\bigg(\frac{11}{72}\delta(1-x_1)}\nonumber\\
&&+\frac{6}{72}\delta(1-x_1){\rm G}(1;x_0)-\frac{1}{6}{\cal D}_0(x_1)+\frac{1}{12}(1+x_1)\bigg)
+\frac{1}{\e}\bigg(\frac{65}{108}\delta(1-x_1)-\frac{7}{144}\delta(1-x_1)\nonumber\\
&&-\frac{11}{36}\delta(1-x_1){\rm G}(1;x_0)-\frac{1}{6}\delta(1-x_1){\rm G}(1,1;x_0)
+\frac{11}{18}{\cal D}_0(x_1)+\frac{1}{3}{\cal D}_0(x_1){\rm G}(1;x_0)-\frac{2}{3}{\cal D}_1(x_1)\nonumber\\
&&-\frac{17}{36}-\frac{5}{36}x_1-\frac{(1-x_1)}{12(1-xx_0)^2}+\frac{(1+x_1^2)}{4(1-x_1)}{\rm G}(0;x_1)
+\frac{1+x_1}{3}{\rm G}(1;x_1)-\frac{1}{3(1-x_1)}{\rm G}(1;x_0)\nonumber\\
&&+\frac{(1+x_1^2)}{6(1-x_1)}{\rm G}\bigg(\frac{1}{x_1};x_0\bigg)+\order{\e^0}\bigg],\nonumber\\
\eeqa
where the functions denoted as ${\rm G}$ are two dimensional harmonic polylogarithms \cite{Gehrmann:2001jv}, and, as usual,
\beq
{\cal D}_n(x)=\bigg(\frac{\ln^n(1-x)}{(1-x)}\bigg)_{+}.
\eeq

%%%%%%%%%%%%%%%%%%%%%%%%%%%%%%%%%%%%%%%%%%%%%%%%%%%%%%
%%%%%%%%%%%%%%%%%%%%%%%%%%%%%%%%%%%%%%%%%%%%%%%%%%%%%%

\bibliography{bibliography}

\end{document}